\documentclass[12pt]{article}
\pdfoutput=1
\usepackage{amsmath, amsthm,comment}
\usepackage{amsfonts}
\usepackage{amssymb,graphics,psfrag}
\usepackage{array,epsfig,stmaryrd,graphicx}
\usepackage{cite}
\usepackage{graphicx}
\usepackage{makeidx}
\usepackage{multicol}
\usepackage{geometry}
\usepackage{amsfonts}
\usepackage{mathrsfs}
\usepackage{amssymb}
\usepackage{amsmath}
\usepackage{slashed}
\usepackage{heck}%
\providecommand{\U}[1]{\protect\rule{.1in}{.1in}}
\newcommand{\beq}{\begin{equation}}
\newcommand{\eeq}{\end{equation}}
\newcommand{\be}{\begin{equation}}
\newcommand{\ee}{\end{equation}}
\newcommand{\bea}{\begin{eqnarray}}
\newcommand{\eea}{\end{eqnarray}}
\newcommand{\ben}{\begin{eqnarray*}}
\newcommand{\een}{\end{eqnarray*}}
\newcommand{\ba}{\begin{aligned}}
\newcommand{\ea}{\end{aligned}}
\newcommand{\bt}{\begin{tabular}}
\newcommand{\et}{\end{tabular}}
\newcommand{\bc}{\begin{center}}
\newcommand{\ec}{\end{center}}

\newcommand{\ds}{\displaystyle}

\newcommand{\cref}{{\bf [check ref]}}

\newcommand{\bs}{\begin{subarray}{c}}
\newcommand{\es}{\end{subarray}}

\begin{document}

\date{March, 2009}
\title{The Footprint of F-theory at the LHC}

\preprint{arXiv:0903.3609 \\ MCTP-09-08}

\institution{HarvardU}{\centerline{${}^{1}$Jefferson Physical
Laboratory, Harvard University, Cambridge, MA 02138, USA}}

\institution{MCTP}{\centerline{${}^{2}$Michigan Center for Theoretical
Physics, University of Michigan, Ann Arbor, MI 48109, USA}}%

\authors{Jonathan J. Heckman\worksat{\HarvardU}%
\footnote{e-mail: \texttt{jheckman@fas.harvard.edu}%
}, Gordon L. Kane\worksat{\MCTP}%
\footnote{e-mail: \texttt{gkane@umich.edu}%
}, \\[2mm] Jing Shao\worksat{\MCTP}%
\footnote{e-mail: \texttt{jingshao@umich.edu}%
} and Cumrun Vafa\worksat{\HarvardU}%
\footnote{e-mail: \texttt{vafa@physics.harvard.edu}%
}}

\abstract{Recent work has shown that compactifications of F-theory
provide a potentially attractive phenomenological scenario. The low
energy characteristics of F-theory GUTs consist of a deformation away
from a minimal gauge mediation scenario with a high messenger scale.
The soft scalar masses of the theory are all shifted by a stringy
effect which survives to low energies. This effect can range from $0$ GeV
up to $\sim 500$ GeV. In this paper we study potential
collider signatures of F-theory GUTs, focussing in particular on ways
to distinguish this class of models from other theories with an MSSM
spectrum. To accomplish this, we have adapted the general footprint
method developed recently for distinguishing broad classes of string
vacua to the specific case of F-theory GUTs. We show that with only
$5$~fb$^{-1}$ of simulated LHC data, it is possible to distinguish
many mSUGRA models and low messenger scale gauge mediation models from F-theory GUTs.
Moreover, we find that at $5$~fb$^{-1}$, the stringy deformation
away from minimal gauge mediation produces observable consequences
which can also be detected to a level of order $\sim \pm 80$ GeV.
In this way, it is possible to distinguish between models with
a large and small stringy deformation. At $50$~fb$^{-1}$,
this improves to $\sim \pm 10$~GeV.}

%TCIMACRO{\TeXButton{Maketitle}{\maketitle}}%
%BeginExpansion
\maketitle
%EndExpansion

\enlargethispage{\baselineskip}
\tableofcontents

\section{Introduction \label{INTRO}}

Bridging the gap between string theory and experiment would at first
appear to require enormous energy scales to probe more intrinsically
\textquotedblleft stringy phenomena\textquotedblright. This is compounded by the
fact that a given low energy theory may possess several completions
at higher energy scales. Indeed, at low energy scales string theory
can always be represented by an effective field theory.

On the other hand, there is no guarantee that a given effective field theory will
have a UV completion in string theory. Thus, while a direct confirmation of the theory may not be possible,
indirect manifestations of the theory are likely to be present at
lower energy scales. In particular, although a given string compactification
may reduce to a well-defined effective field theory, the specific
choice of the field content, mass scales and parameters will very
much depend on the details of the compactification. Seemingly contrived
field theories may have a very natural stringy origin. In this way, a class
of string theory compactifications can provide a preferred set
of effective field theories, each with a specific class of potential
observable signatures.

From the perspective of the string theorist, the primary challenge
is then to determine a set of criteria which select a class of
UV completions of the Standard Model of particle physics. One well-motivated
possibility is to assume the existence of low scale supersymmetry
with the spectrum of the MSSM, and the presence of a Grand Unified
Theory (GUT) at high energy scales $M_{GUT}\sim10^{16}$~GeV. This
has typically been taken as evidence for the existence of an extra
unification structure near the Planck scale, which is indeed very
suggestive of the potential relevance of stringy physics. Thus, in
an indirect way, the requirement of GUT scale physics and the existence
of supersymmetry manifested at low scales provides a first criterion
for vacuum selection.

Although the GUT scale is very close to the Planck scale, there
is still a small hierarchy in that $M_{GUT}/M_{pl}\sim10^{-3}$. In
\cite{BHVII} it was proposed that the smallness of this parameter
be promoted to the additional selection criterion that a candidate vacuum should admit
a limit where the dynamics of quantum gravity can in principle decouple (as $M_{pl}\rightarrow\infty$).
A surprising consequence of the existence of both a GUT and
a decoupling limit is that it imposes strong restrictions on the content
of the low energy theory. In the particular context of compactifications
of a strongly coupled formulation of type IIB\ string theory known
as F-theory, this question has been studied. See \cite{BHVI,BHVII,HMSSNV,HVGMSB,HVLHC,HVFLAV,FGUTSCosmo,DonagiWijnholt,WatariTATARHETF,DonagiWijnholtBreak,MarsanoGMSB,MarsanoToolbox,Wijnholt:2008db,Font:2008id,Blumenhagen:2008zz,Blumenhagen:2008aw,Bourjaily:2009vf,Hayashi:2009ge,Andreas:2009uf,Chen:2009me} for recent work on F-theory based models of Grand Unified Theories (F-theory
GUTs). Returning to the form of the low energy theory, bottom up considerations
also serve to constrain the details of the compactification. Proceeding
iteratively, it was shown in \cite{HVGMSB,HVLHC} that the sparticle
spectrum is constrained to a remarkable degree.

From the perspective of the low energy theory, there are two novel
features associated with the supersymmetry breaking sector of F-theory
GUTs. First, because there exists a limit where gravity decouples,
F-theory GUTs are incompatible with gravity mediated supersymmetry
breaking, but rather, most naturally accommodate gauge mediated supersymmetry
breaking (GMSB). F-theory GUTs constitute a deformation away from
the minimal GMSB scenario both in terms of the most natural input energy scales,
and in the form of additional contributions to the soft mass terms
of the theory.

Recall that in minimal GMSB, the effects of supersymmetry breaking
are parameterized by the number of vector-like pairs of messengers
in the $5\oplus\overline{5}$, $N_{5}$, and the characteristic mass
scale for gauginos, $\Lambda$, such that the mass of the gaugino at
a low energy scale is:
\begin{equation}
m_{\text{gaugino}}\sim N_{5}\frac{\alpha}{4\pi}\Lambda\sim N_{5}\frac{\alpha}{4\pi}\frac{F}{M_{\text{mess}}}\text{,}
\end{equation}
where $\alpha$ is shorthand for the fine structure constants of the various Standard Model gauge groups,
$\sqrt{F}$ is the scale of supersymmetry breaking, and $M_{\text{mess}}$
is the mass of the messenger fields communicating supersymmetry breaking
to our sector. From the low energy point of view, it is most natural
to take the messenger scale as low as possible without running into
conflict with experiments so that $\sqrt{F}\sim M_{\text{mess}}\sim10^{5}$~GeV.
By contrast, in F-theory GUTs the scale of supersymmetry breaking
is typically much higher, and is given by $\sqrt{F}\sim10^{8}-10^{9}$~GeV,
and the messenger scale is $M_{\text{mess}}\sim10^{12}$~GeV. Even
though such high values are allowed from the viewpoint of mGMSB, there
is nothing distinguished about such energy scales based on low energy
considerations. Rather, the requirement that this model admit a UV
completion within F-theory requires an increase in the scale of supersymmetry
breaking. This is because the $\mu$ term is generated by GUT scale
dynamics so that only a small range of scales for supersymmetry breaking
will generate a value for the $\mu$ parameter near the scale of electroweak symmetry
breaking \cite{HVGMSB}.

Besides motivating a specific high energy scale for the masses of the
messenger fields, there are also more stringy manifestations of F-theory
GUTs which survive to low energies. These effects constitute a \textit{predictive
and measurable} shift away from the soft supersymmetry breaking terms
of the minimal gauge mediation scenario. This deformation is due to
the fact that in any quantum theory of gravity, global symmetries
must be gauged, or will be violated by Planck scale effects. One such
gauge symmetry which is broken at the string scale persists as a global
$U(1)_{PQ}$ Peccei-Quinn symmetry at lower energies. In the low energy
theory, this global $U(1)$ symmetry is anomalous and so would seemingly
lead to an inconsistent theory if gauged. In string theory, however, such anomalies
are cancelled via the generalized Green-Schwarz mechanism. What is
particularly interesting is that because the fields of the MSSM and
supersymmetry breaking sector are both charged under this symmetry,
heavy $U(1)_{PQ}$ gauge boson exchange generates additional contributions
to the soft scalar masses. Because the charges of all of the MSSM
fields are constrained by the existence of higher GUT symmetries,
this amounts to a \textit{predictive, and stringy prediction} for
the expected shifts in the mass spectrum away from the minimal gauge
mediation scenario. Typically, the size of these mass shifts are comparable
to the total mass of the sleptons generated by gauge mediation effects.
\textit{Therefore this stringy \textquotedblleft PQ deformation\textquotedblright\ will have measurable consequences.}

The aim of the present paper is to study collider signatures of F-theory
GUTs, and in particular, to establish whether it is possible to distinguish
between other models with an MSSM spectrum, but with different input
Lagrangians at the TeV scale. In this regard, our goal is to view
the LHC\ as a tool by which one can differentiate between distinct
extensions of the Standard Model.

One particularly promising way to distinguish between distinct models
with only limited integrated luminosity is based on the general footprint
method developed in \cite{Kane:2006yi,KaneFootprint}. This consists of creating
two-dimensional plots of various candidate signatures and scanning
over the parameters in a given class of models. Performing such a
scan for two classes of models, it is then possible to determine a
set of signatures which can distinguish between distinct models. This
can then be supplemented by more quantitative measures such as chi-square
like fits to establish the distinguishability of two models. To minimize
the effects of Standard Model background, we have typically selected
events which contain either a hard jet, or some other hard process
which is difficult to replicate by purely Standard Model effects.

In this paper, we show that with $5$~fb$^{-1}$ of simulated LHC\ data
(which is lower than the expected annual luminosity for the LHC in the
first three years), it is possible to distinguish F-theory GUTs from
other models with an MSSM\ sparticle spectrum. In F-theory GUTs,
the bino or the lightest stau typically corresponds to a quasi-stable
NLSP which decays outside the detector. Because reconstruction of
the charged track left by a stau is a relatively easy signature to
detect, in this paper we will primarily focus on the case of a bino
NLSP. Within this class of F-theory GUT models, we scan over the various
parameters of the theory, and compare the signals obtained with those
of mSUGRA\ models with mass spectra similar to those of F-theory
GUTs. Scanning over many such mSUGRA\ models, we show that there
are indeed signatures which can reliably distinguish between F-theory
GUTs and such models.\footnote{For mSUGRA models with large A-terms, only
F-theory GUTs with one messenger can be distinguished given the
limited signatures and integrated luminosity.} In a certain sense, this is to be expected,
because when the squark (resp. slepton) mass spectra are similar to
those of F-theory GUTs, the slepton (resp. squark) masses are typically
different. Moreover, in those cases where both the squark and slepton
spectra are similar, the corresponding branching fractions between
the two classes of models are different enough that it is still possible
to develop a class of signatures which can distinguish between F-theory
GUTs and mSUGRA models. We also find that is possible to distinguish
F-theory GUTs from minimal gauge mediation scenarios with a low messenger scale.

At the next stage of analysis, we show that with the same integrated
luminosity, it is possible to distinguish between high scale minimal
gauge mediation models, and F-theory GUTs. This amounts to showing
that the effects of the F-theoretic PQ-deformation are observable at the LHC. To this
end, we first show that it is possible to determine a class of signatures
which are sensitive to $N_{5}$ and $\Lambda$. Having fixed these
values, we next show that in the case of single messenger models,
it is indeed possible to roughly distinguish between models with distinct
values of the PQ deformation up to a level of $\sim \pm 80$~GeV with $5$~fb$^{-1}$
of integrated luminosity. We find that this sensitivity improves to
an uncertainty of $\sim \pm 10$~GeV with $50$~fb$^{-1}$ of simulated
LHC\ data. In the case of multiple messenger models, the effects
of the PQ\ deformation are less pronounced in the regime where the
bino is the NLSP. As a consequence, distinguishing between models
in this range of models appears more challenging.

The organization of the rest of this paper is as follows. In section
\ref{FATLHC} we provide additional background on F-theory GUTs. This
includes a more detailed description of the parameter space of F-theory
GUTs, and the defining characteristics of F-theory GUTs which are
relevant for collider studies. This is followed in section \ref{DISTINGUISH}
by an analysis of signatures which can distinguish between F-theory
GUTs, mSUGRA\ models, and low messenger scale minimal gauge mediation models.
Next, in section \ref{DETFth} we study the extent to which it is possible
to distinguish between distinct F-theory GUTs. Section
\ref{CONCLUSIONS} contains our conclusions.

\section{F-theory at the LHC \label{FATLHC}}

In this section we define the class of models which correspond to
F-theory GUTs, and in particular, the low energy content of relevance
for collider studies. We refer the interested reader to \cite{HVGMSB,HVLHC}
for further details on the reasoning by which a narrow and predictive
range of parameters is determined. For further details of F-theory
GUTs, we refer the interested reader to \cite{BHVI,BHVII,HMSSNV,HVGMSB,HVLHC,HVFLAV,FGUTSCosmo,DonagiWijnholt,WatariTATARHETF,DonagiWijnholtBreak,MarsanoGMSB,MarsanoToolbox,Wijnholt:2008db,Font:2008id,Blumenhagen:2008zz,Blumenhagen:2008aw,Bourjaily:2009vf,Hayashi:2009ge,Andreas:2009uf,Chen:2009me}.

At low energies, an F-theory GUT corresponds to a deformation away
from a minimal gauge mediation scenario. The relatively high messenger
scale required to generate a viable $\mu$ term implies that in this
class of models, the LSP\ is the gravitino with a mass of $\sim 10-100$
MeV, and the NLSP is either a quasi-stable bino-like lightest neutralino,
or a quasi-stable stau. In scenarios with a stau NLSP the decay of
$\widetilde{\chi}_{1}^{0}\rightarrow\widetilde{\tau}_{1}^{\pm}\tau^{\mp}$
will produce staus which will leave a charged track which is relatively
easy to detect. In the similar \textquotedblleft sweet spot\textquotedblright\ model
of gauge mediation with a high messenger scale, an analysis of a quasi-stable
stau NLSP\ scenario was recently considered \cite{KitanoIbeSweetSpot}. See for example, references
\cite{Martin:1998vb,Feng:1997zr,Ambrosanio:2000zu,Ellis:2006vu} for additional
information on collider studies of quasi-stable stau NLSP scenarios.
Indeed, if this parameter range for F-theory GUTs is realized in nature,
it will show up as a striking signal at the LHC, the essential point
being that because the stau leaves a charged track, it is then possible
to reconstruct the mass of the stau and then the associated decay
products. Proceeding up the decay chain, it is then possible to reconstruct
more detailed information regarding the mass spectrum of such a scenario.
Due to the fact that this analysis has already been performed, e.g. in
\cite{Ellis:2006vu,KitanoIbeSweetSpot}, and is itself based on earlier analysis
of quasi-stable stau NLSP\ scenarios, in this paper we shall focus on the
more challenging case of F-theory GUTs with a bino NLSP.

The remainder of this section is organized as follows. After reviewing
the main features of the low energy theory, we next describe the scan
over parameter space of F-theory GUT models. This is followed by a
discussion of the characteristic features of the mass spectrum, cross
sections, and branching fractions, and in particular, the dependence
of these quantities on the inputs of F-theory GUTs.

\subsection{Review of F-theory GUTs}

In F-theory GUTs, the defining features of the GUT model are determined
by the worldvolume theory of a seven-brane which fills our spacetime
and wraps four internal directions of the six hidden dimensions of
string theory. The chiral matter of the MSSM\ localizes on Riemann
surfaces in the seven-brane, and interaction terms between chiral
matter localize at points in the geometry. As argued in \cite{HVGMSB},
crude considerations based on the existence of a limit where the effects
of gravity can decouple imposes sharp restrictions on the low energy
content of the effective field theory. In particular, because such
models admit a limit where the effects of gravity can decouple, they
are incompatible with mechanisms such as gravity mediation. Rather,
in F-theory GUTs the effects of supersymmetry breaking are communicated
to the MSSM\ via gauge mediation.

From the perspective of the low energy effective theory, the defining
characteristic of F-theory GUTs is that it constitutes a deformation
away from a high scale minimal gauge mediation scenario. Letting $X$
denote the GUT\ singlet chiral superfield which develops a supersymmetry
breaking vev:
\begin{equation}
\left\langle X\right\rangle =x+\theta^{2}F_{X}\text{,}
\end{equation}
the characteristic mass scales associated with the sparticle spectrum
in minimal gauge mediation are controlled by the parameter:
\begin{equation}
\Lambda=\frac{F_{X}}{x}\text{.}
\end{equation}
For example, in a model with $N_{5}$ vector-like pairs of messenger
fields in the $5\oplus\overline{5}$ of $SU(5)$, the masses of the
gauginos and scalar sparticles scale as:
\begin{align}
m_{\text{gaugino}} & \sim N_{5}\frac{\alpha}{4\pi}\Lambda\\
m_{\text{scalar}} & \sim\sqrt{N_{5}}\frac{\alpha}{4\pi}\Lambda
\end{align}
where in the above, $\alpha$ is shorthand for the contribution from
the various fine structure constants of the Standard Model gauge groups.

In the specific context of F-theory GUTs, the $\mu$ term is roughly
given as:
\begin{equation}
\mu\sim\frac{F_{X}}{M_{X}^{KK}}\text{,}
\end{equation}
where $M_{X}^{KK}\sim10^{15}$ GeV is a Kaluza-Klein mass scale of
a GUT singlet in the compactification. Thus, obtaining the correct value
of $\mu$ requires:
\begin{equation}
F_{X}\sim10^{16}-10^{18}\text{ GeV}^{2}\text{.}
\end{equation}

This range of values for $F_{X}$ implies that the mass of the gravitino
is $\sim 10-100$~MeV. Moreover, the fact that the scale of supersymmetry
breaking is relatively high compared to other models of gauge mediation
implies that the NLSP will decay outside the detector due to its long
lifetime.

The rough range of values for $\Lambda$ extends from $\Lambda\sim 10^{4}$
to $\Lambda\sim 10^{6}$. Beyond this range, the mini-hierarchy problem
is exacerbated. In fact, we shall typically consider a smaller range
on the order of:
\begin{equation}
10^{4}\text{ GeV}\lesssim\Lambda\lesssim2\times10^{5}\text{ GeV,}
\end{equation}
because for larger values of $\Lambda$, the masses of the gluinos
and squarks would be too heavy to be produced at the LHC. Finally,
in the context of F-theory GUTs, the $B\mu$ term and the A-terms
all vanish at the messenger scale. Thus, in this class of models,
$B\mu$ and the A-terms are radiatively generated, and $\tan\beta$ is typically in
the range of $20-40$.

\begin{figure}
[ptb]
\begin{center}
\includegraphics[
height=0.8925in,
width=4.5558in
]%
{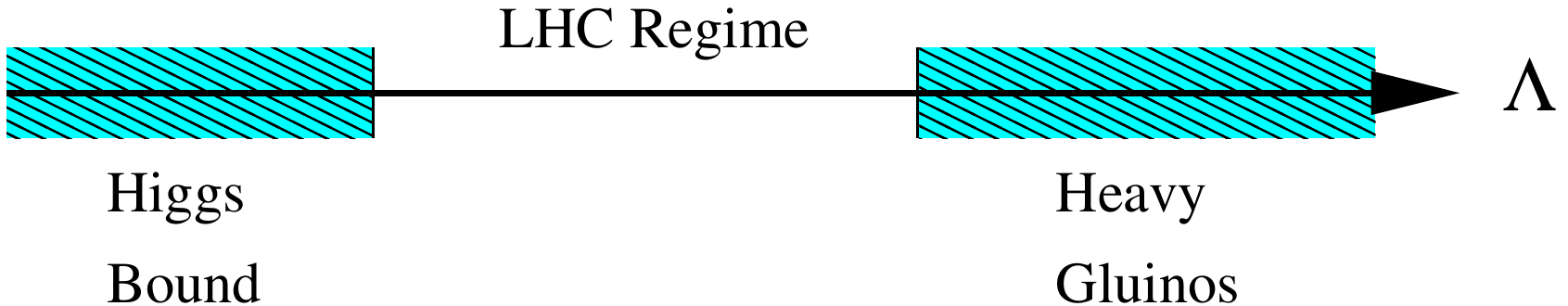}%
\caption{The range of values for the F-theory GUT\ parameter $\Lambda$ extends
from a lower bound set by the current limits on the mass of the Higgs, to an
upper bound determined by the requirement that the gluinos be light enough to
be produced at the LHC.}%
\label{lambda}%
\end{center}
\end{figure}

The mass spectrum of F-theory GUTs corresponds to a deformation away
from the minimal gauge mediation scenario. This is due to the fact
that the theory contains an anomalous $U(1)_{PQ}$ gauge symmetry. This
anomaly is cancelled via the generalized Green-Schwarz mechanism. The
essential point is that this introduces additional higher dimension
operators into the theory which have the effect of shifting by a universal amount
the soft scalar masses:
\begin{equation}
m_{\text{soft}}^{2}=\widehat{m}^{2}+e_{\Phi}\Delta_{PQ}^{2}\text{,}
\end{equation}
where $\widehat{m}$ denotes the mass in the absence of the PQ deformation,
and the charge $e_{\Phi}$ is defined as:
\begin{align}
\text{Chiral Matter} & \text{: }e_{\Phi}=-1\\
\text{Higgs} & \text{: }e_{\Phi}=+2\text{.}
\end{align}
To leading order, the gaugino masses and trilinear couplings are unchanged by this deformation.

At a more fundamental level, the deformation parameter originates
from integrating out a heavy gauge boson of mass $M_{U(1)_{PQ}}$
so that:
\begin{equation}
\Delta_{PQ}^{2}\sim4\pi\alpha_{PQ}\frac{F_{X}^{2}}{M_{U(1)_{PQ}}^{2}}\text{,}
\end{equation}
where $\alpha_{PQ}$ denotes the fine structure constant of the $U(1)_{PQ}$ gauge theory.
A priori, the value of this gauge boson mass can be on the order
of the string scale, GUT scale, or even somewhat lower. The phenomenologically
most interesting region of course corresponds to the case of lower
$M_{U(1)_{PQ}}$, or higher $\Delta_{PQ}$.

In fact, the cosmology of F-theory GUTs suggest a lower bound on $\Delta_{PQ}$
on the order of \cite{FGUTSCosmo}:
\begin{equation}
\Delta_{PQ}\gtrsim50\text{ GeV.}\label{PQlowerBound}
\end{equation}
This comes from the fact that in F-theory GUTs, the Goldstone mode
associated with $U(1)_{PQ}$ symmetry breaking is the QCD\ axion
\cite{HVGMSB}. The other real degree of freedom in the associated
supermultiplet is the saxion, which has a mass proportional to $\Delta_{PQ}$.
As shown in \cite{FGUTSCosmo}, the oscillations of the saxion eventually
come to dominate the energy density of the Universe, and its decay
reheats the Universe provided the saxion is sufficiently heavy so that
a decay channel to a mode other than axions is available. This imposes
the kinematic constraint that the saxion be heavier than the Higgs,
which translates into the lower bound given by inequality (\ref{PQlowerBound}).

There is also an upper bound to the size of $\Delta_{PQ}$ because
increasing $\Delta_{PQ}$ decreases the soft masses of the squarks
and sleptons. Thus, for large enough values of $\Delta_{PQ}$ on the
order of $500$~GeV (the precise value of which depends on $\Lambda$ and the number
of messenger fields), the low energy spectrum will contain a tachyon.
Depending on the number of messengers as well as the size of the PQ\ deformation,
either a bino-like neutralino, or a lightest stau could be the NLSP.
Due to the fact that the scale of supersymmetry breaking is so high,
the NLSP\ decays outside the detector, effectively behaving as a
stable particle.

\begin{figure}
[ptb]
\begin{center}
\includegraphics[
height=0.8562in,
width=5.3618in
]%
{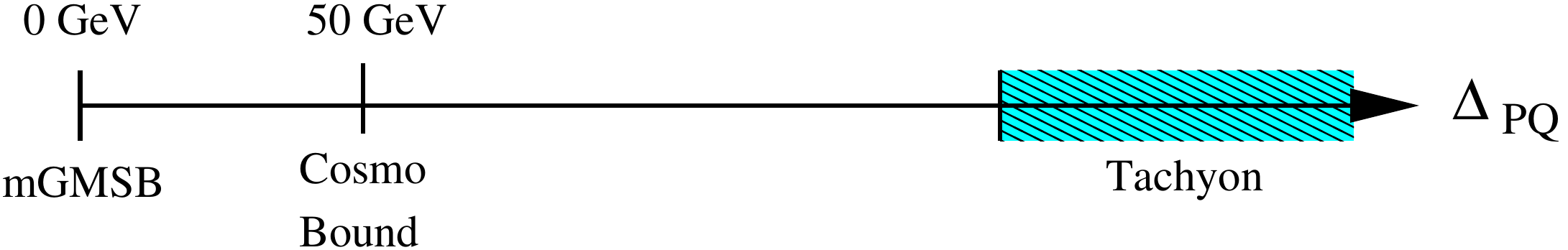}%
\caption{The PQ deformation parameter $\Delta_{PQ}$ of F-theory
GUTs lowers the squark and slepton soft scalar masses
in relation to the value expected from a high messenger scale
model of minimal gauge mediated supersymmetry breaking. At $\Delta_{PQ}=0$,
F-theory GUTs reduce to a high messenger scale mGMSB model. Cosmological
considerations impose a lower bound on $\Delta_{PQ}$ of order $50$ GeV.
Finally, there is also an upper bound on $\Delta_{PQ}$ which comes from the
requirement that the slepton sector not contain a tachyonic mode.}%
\label{pq}%
\end{center}
\end{figure}

It is quite exciting that a remnant of stringy physics in the form
of the PQ deformation has a measurable manifestation at low energy
scales. One of the aims of this paper is to study how the effect of
this deformation can be extracted from collider data.

\subsection{Parameter Space Scan}

Due to the fact that F-theory GUTs depend on one discrete parameter,
$N_{5}$, and two continuous parameters, $\Lambda$ and $\Delta_{PQ}$,
it is possible to perform a scan over much of the parameter space
of models. The range of the scan performed over the parameters $\Lambda$
and $\Delta_{PQ}$ depends on the number of messengers $N_{5}$, because
if the masses of the gluinos and squarks are too heavy, the LHC will
not be able to generate sparticles of the desired mass. For example,
in minimal GMSB, the masses of the gluinos and squarks respectively
scale as $N_{5}\Lambda$ and $\sqrt{N_{5}}\Lambda$. Thus, even a
factor of five increase in $\Lambda$ can jeapordize the production
of gluinos at the LHC. Further, increasing $\Lambda$ exacerbates
the fine tuning already present in the Higgs sector. For this reason,
we believe it is theoretically well-motivated to primarily consider
scenarios where $\Lambda$ is as small as possible, without violating
current experimental lower bounds on the masses in the sparticle spectrum.
As explained earlier, we will focus on bino NLSP\ scenarios because
the lightest stau NLSP\ scenario of similar models has been studied
elsewhere, such as in \cite{KitanoIbeSweetSpot}. Nevertheless, some
distinctive features of such models in the context of F-theory GUTs
are discussed briefly in subsection \ref{STAUDISCUSS}.

Restricting to the bino NLSP\ case, since increasing the number of
messengers lowers the stau mass relative to the gaugino masses, the
condition that the masses satisfy the relation:
\begin{equation}
m_{\widetilde{B}}<m_{\widetilde{\tau_{1}}}
\end{equation}
translates into the requirement that:
\begin{equation}
1\leq N_{5}\leq3\text{.}
\end{equation}

Moreover, as $\Lambda$ decreases, the Higgs mass also decreases.
Thus, the bound on the Higgs mass obtained from LEP II\ puts a lower
bound on $\Lambda$, which we denote by $\Lambda^{min}(N_{5})$. The
maximum value of $\Lambda$ we consider, which we denote by $\Lambda^{max}(N_{5})$
is chosen so that the resulting sparticle masses are light enough
to generate enough events at the LHC after a few years. We note that
both $\Lambda^{min}(N_{5})$ and $\Lambda^{max}(N_{5})$ are fairly
insensitive to $\Delta_{PQ}$. For each scanned value of $\Lambda$,
we also scanned over $\Delta_{PQ}$ from $\Delta_{PQ}=0$ to a value
of $\Delta_{PQ}$ such that $m_{\widetilde{\tau_{1}}}-m_{\widetilde{B}}<10$
GeV. To summarize, in this paper we have scanned over F-theory GUTs
in the parameter range:
\begin{itemize}
\item {$1\leq N_{5}\leq3$}
\item {$\Lambda^{min}(N_{5})\leq\Lambda\leq\Lambda^{max}(N_{5})$}
\item {$0\leq{\Delta_{PQ}\leq}\Delta_{PQ}^{max}(N_{5},\Lambda)$.}
\end{itemize}

\subsection{Mass Spectrum}

Scanning over all regions of interest in F-theory parameter space,
we have generated the associated sparticle spectra using \texttt{SOFTSUSY}
\cite{SOFTSUSYAllanach} by imposing the boundary condition $B\mu=0$
at the messenger scale. Compatibility with electroweak symmetry breaking
then fixes $\tan\beta$ to a large value between $20-40$, the exact
value of which depends on the specifics of the model. The dependence
of the mass spectrum on $N_{5}$ and $\Lambda$ when $\Delta_{PQ}=0$
corresponds to the case of mGMSB with a high messenger scale $M_{\text{mess}}\sim10^{12}$~GeV.
See, for example, \cite{GiudiceSUSYReview} for a review of gauge
mediated supersymmetry breaking. In this section, we discuss the effect
of $\Delta_{PQ}$ on the mass spectrum.

The spectrum separates into those particles which are affected by
the PQ\ deformation, and those which are not. To leading order, the
masses of the gauginos are not affected by the PQ deformation. Just
as in minimal gauge mediation, the low scale gaugino masses $m_{i}$
of the three gauge group factors satisfy the relation:
\begin{equation}
m_{1}:m_{2}:m_{3}\sim 1:2:6\text{.}
\end{equation}
In the context of F-theory GUTs, the two lightest neutralinos and
the lightest chargino correspond to gauginos, with a primarily bino-like
lightest neutralino.
\begin{figure}[ptb]
\begin{center}
\includegraphics[
height=4.12in,
width=6.4091in
]{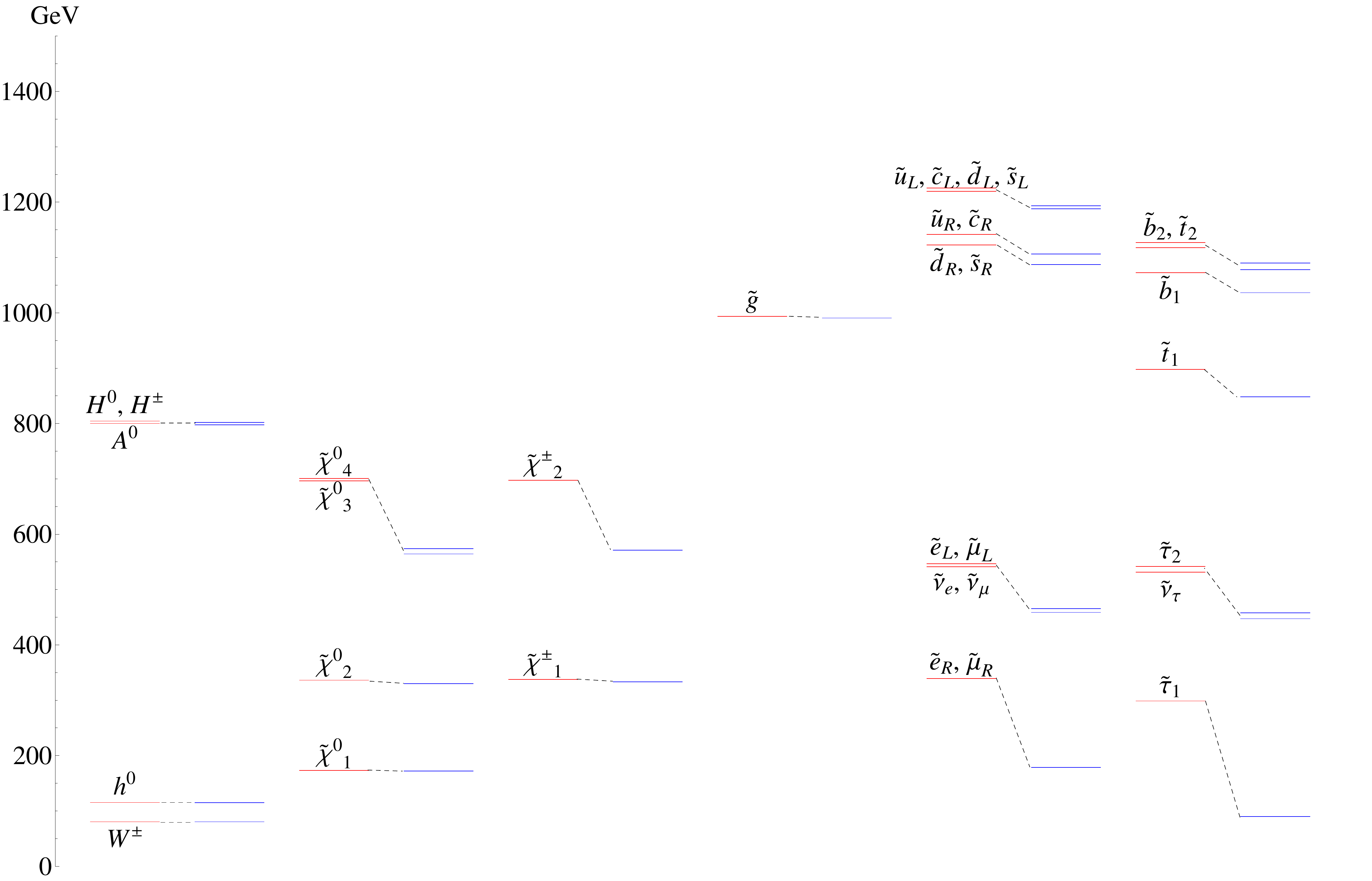}
\end{center}
\caption{Plot of the mass spectrum of F-theory GUTs with $N_{5}=1$,
$\Lambda=1.3\times10^{5}$ GeV, and minimal (red, left part of each column) and maximal (blue, right part of each column) PQ
deformation. See figure \ref{FIG:THRMESSSPEC} in Appendix F for the spectra of three messenger models.}%
\label{FIG:ONEMESSSPEC}%
\end{figure}
The spectrum of MSSM particles which are not affected by the PQ deformation
are therefore:
\begin{equation}
\text{NPQ: }\widetilde{\chi}_{1}^{0}\text{, }\widetilde{\chi}_{2}^{0}\text{,
}\widetilde{\chi}_{1}^{\pm}\text{, }\widetilde{g}\text{,}
\end{equation}
where we have ordered the sparticles from lightest on the left to
heaviest on the right.

\begin{figure}[ptb]
\begin{center}
\includegraphics[
height=3.4272in,
width=5.719in
]{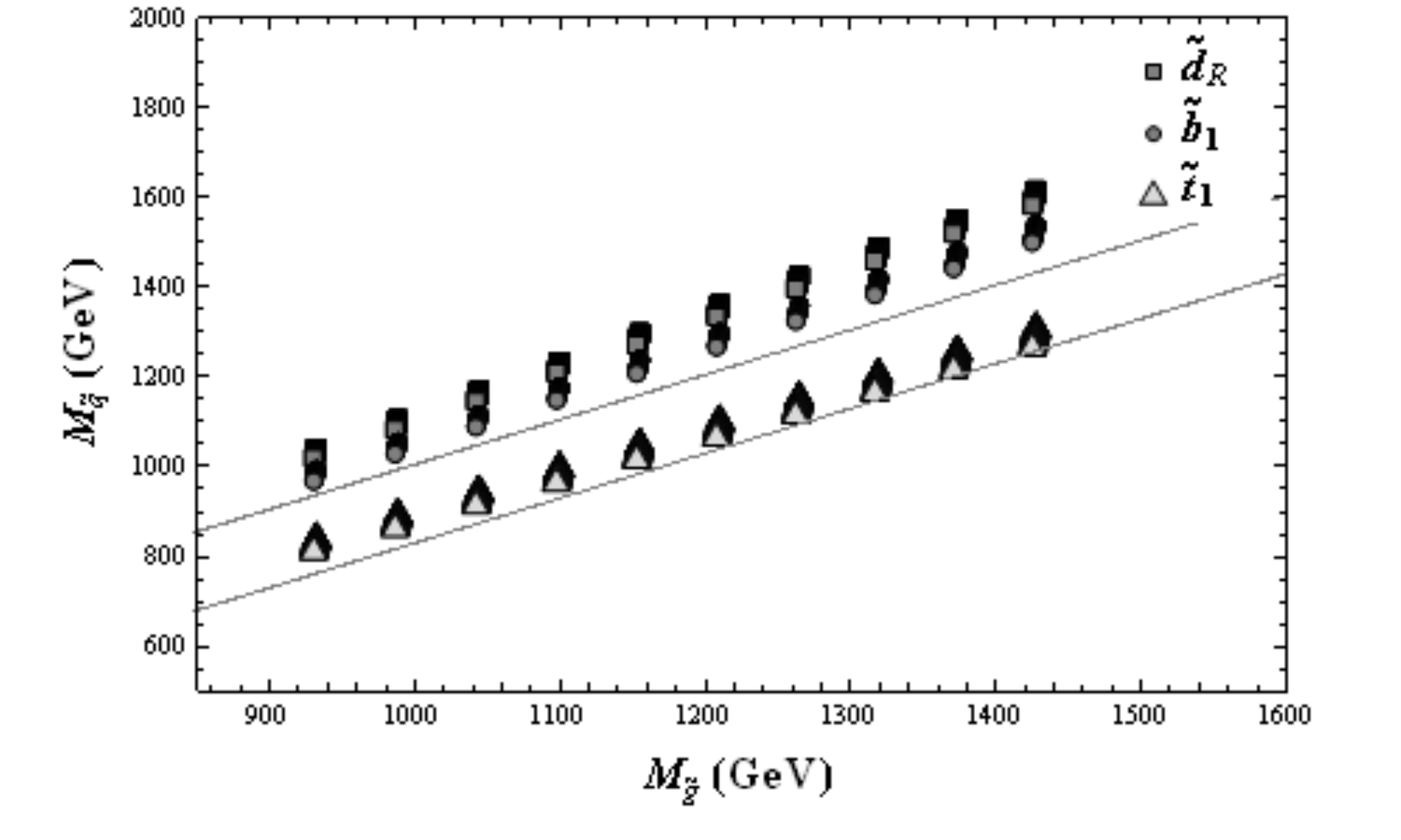}
\includegraphics[
height=3.4272in,
width=5.719in
]{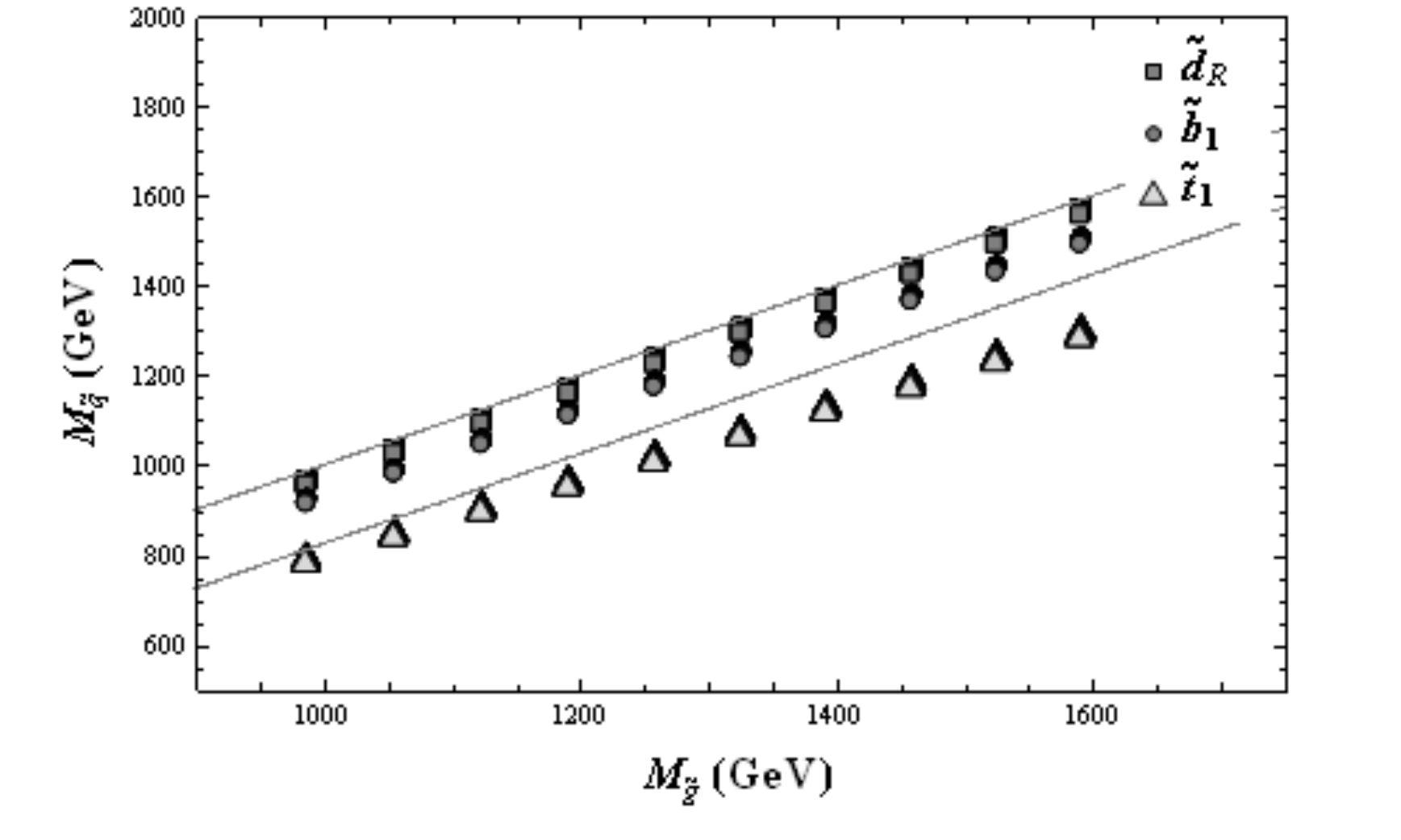}
\end{center}
\caption{Plots of squark masses versus the gluino mass in F-theory GUTs. The top and bottom plots are for
one and two messenger models respectively. The upper line in the plots
corresponds to $m_{\tilde{g}}$, while the lower one corresponds to
$m_{\tilde{g}}-m_{t}$. These figures imply that the decay of the gluino in one
messenger models proceeds via a 3-body process, but in two messenger
models it decays in a two-body one. The three messenger case is similar
to the two messenger case. The gluino mass is primarily determined by $\Lambda$, and
the variations of the squark masses in these figures are due to the change of $\Delta_{PQ}$.}
\label{FIG:GLUINO-SQUARK-FTH}
\end{figure}

The masses of the remaining sparticles of the MSSM all shift due to
the PQ deformation. This includes not just the squarks and sleptons,
but also the Higgsinos. This latter shift is more indirect, and can
be traced back to the fact that the PQ deformation alters the form
of the scalar Higgs potential. As a consequence, achieving proper
electroweak symmetry breaking leads to a shift in the value of the
$\mu$ parameter at the messenger scale. This in turn alters the masses
of the Higgsinos. The spectrum of MSSM\ particles which are affected
by the PQ deformation ordered by lowest mass sparticles on the left
to most massive on the right are:
\begin{equation}
\text{PQ: }\widetilde{l}\text{, }\widetilde{\chi}_{3}^{0}\text{,
}\widetilde{\chi}_{4}^{0}\text{, }\widetilde{\chi}_{2}^{\pm}\text{, }
\widetilde{q}\text{.}
\end{equation}

\begin{figure}[ptb]
\begin{center}
\includegraphics[
height=3.4272in,
width=5.719in
]{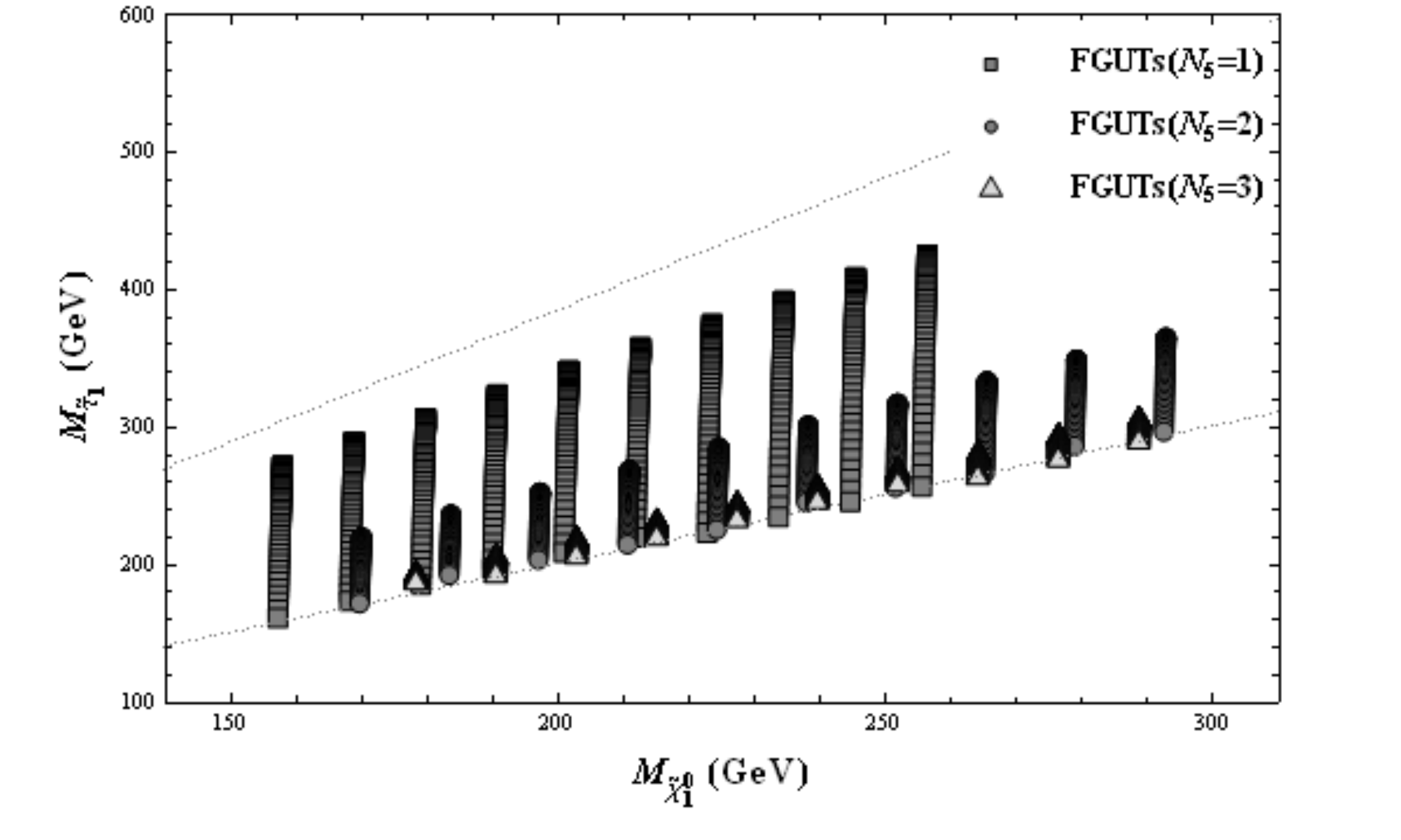}
\end{center}
\caption{Plot of the stau masses in F-theory GUTs versus the bino
NLSP mass for $N_5=1,2,3$ messengers. The upper line in the plot corresponds
to $m_{\tilde{\chi}_{2}^{0}}$, while the lower one corresponds to $m_{\tilde{\chi}_{1}^{0}}$.
This figure shows the typical hierarchy in the masses of $\tilde \tau_1$ and $\tilde \chi_{1,2}^{0}$.
The variation in the stau mass corresponds to the change of $\Delta_{PQ}$,
which becomes less significant in the bino NLSP regime as the number of messengers increases.}
\label{FIG:MLSP-MSTAU-FTH}
\end{figure}

The mass shift due to the PQ\ deformation is most prominent for lighter
sparticles. At the messenger scale, the mass shift for squarks and
sleptons is:
\begin{equation}
m=\widehat{m}\sqrt{1-\frac{\Delta_{PQ}^{2}}{\widehat{m}^{2}}}\text{,}
\label{massshift}
\end{equation}
where $\widehat{m}$ denotes the mass at the messenger scale in the
absence of the PQ deformation. Hence, when $\widehat{m}\gg\Delta_{PQ}$,
there is little change in the mass of the sparticle, so that the squarks
will shift by a comparably small amount. On the other hand, the masses
of the sleptons can shift significantly. Since the mass spectrum is
generated mainly by gauge mediation, the absence of an $SU(2)$ gauge
coupling implies that the right-handed selectron $\widetilde{e}_{R}$,
smuon  $\widetilde{\mu}_{R}$ and stau $\widetilde{\tau}_{R}$ will be
lighter, and thus more sensitive to the PQ\ deformation in comparison
with their left-handed counterparts. Depending on the range of
parameter space, the $\widetilde{e}_{R}$, $\widetilde{\mu}_{R}$ and
$\widetilde{\tau}_{R}$ mass can either be above or below the mass
of the $\widetilde{\chi}_{2}^{0}$. It is also possible in some cases
for $\widetilde{e}_{R}$, $\widetilde{\mu}_{R}$ and $\widetilde{\tau}_{R}$
to become comparable in mass to $\widetilde{\chi}_{1}^{0}$.

\begin{figure}[ptb]
\begin{center}
\includegraphics[
height=3.4272in,
width=5.719in
]{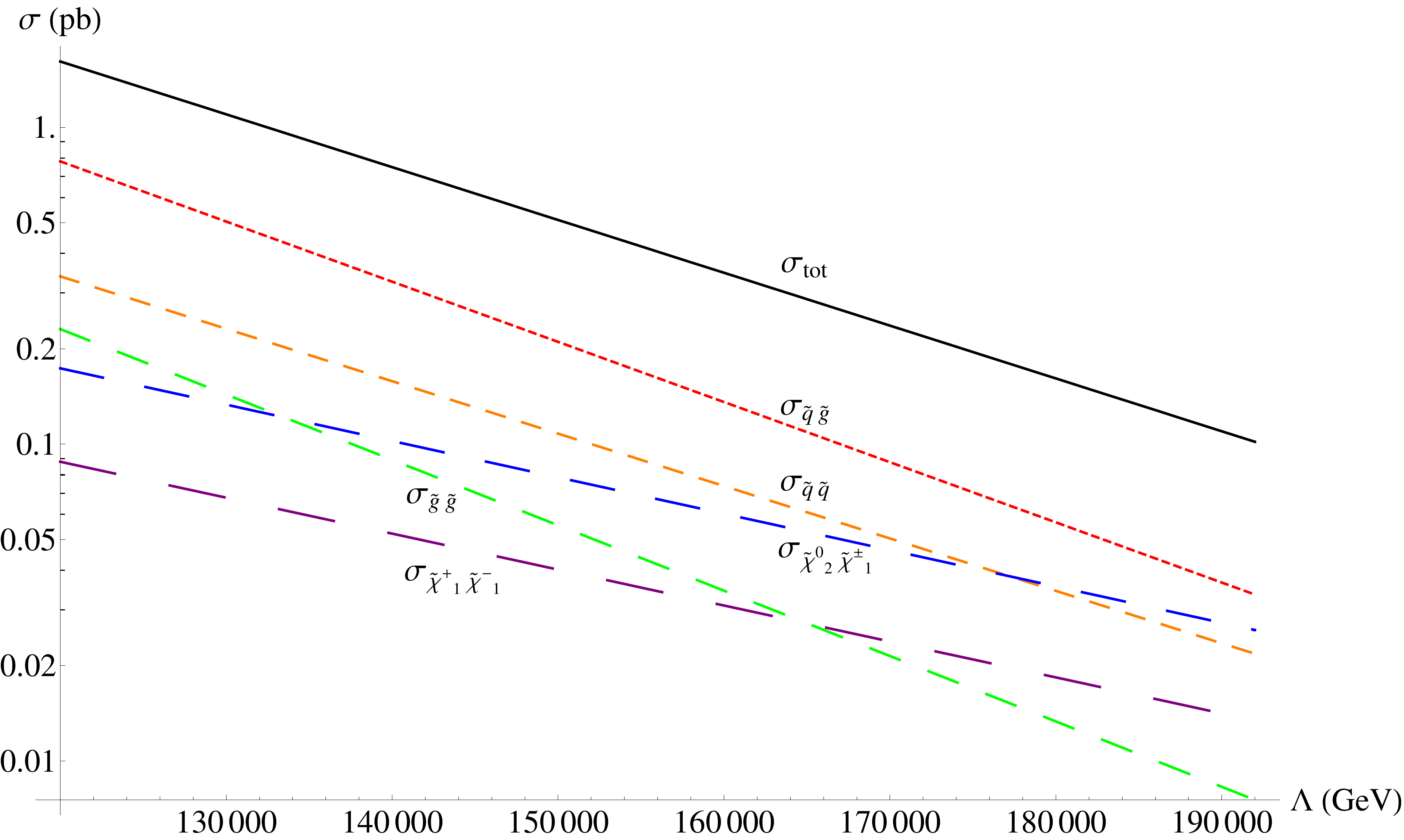}
\end{center}
\caption{Plot of the cross sections at $14$ TeV CM energy for the primary supersymmetric processes as a function of
$\Lambda$ for F-theory GUT models with $N_{5} = 1$ and $\Delta_{PQ} = 0$ GeV.
Note that as $\Lambda$ increases, the cross sections decrease by roughly
an order of magnitude. See figure \ref{FIG:ThrL} in Appendix F for a similar plot of the $N_{5} = 3$ case.}%
\label{FIG:OneL}%
\end{figure}

Due to the large Yukawa couplings present in the third generation,
RG flow will amplify the effects of the PQ deformation in the third
generation squarks and sleptons. The stop and sbottom can typically
become lighter than the gluino in such models, and the $\widetilde{\tau}_{1}$
is lighter than $\widetilde{\chi}_{2}^{0}$. A comparison of these
mass hierarchies for different number of messengers can be seen in
figures \ref{FIG:GLUINO-SQUARK-FTH} and \ref{FIG:MLSP-MSTAU-FTH}.
Further, for large enough
values of $\Delta_{PQ}$, the $\widetilde{\tau}_{1}$ can be lighter
than $\widetilde{\chi}_{1}^{0}$. Figure \ref{FIG:ONEMESSSPEC} and
figure \ref{FIG:THRMESSSPEC} in Appendix F illustrate the mass spectrum
of a single messenger and three messenger model at minimal and maximal
PQ deformation. By inspection of these figures, it follows that progressing
up in mass tends to diminish the effects of the PQ\ deformation.

The interplay between $\Delta_{PQ}$ and the mGMSB parameters $N_{5}$
and $\Lambda$ also influences the form of the low energy spectrum.
Returning to figures \ref{FIG:ONEMESSSPEC} and \ref{FIG:THRMESSSPEC},
we note that although it is a well-known aspect of minimal gauge mediation,
an important feature of these spectra is that as the number of messengers
increases, the scalar sparticles tend to decrease in mass faster than
their fermionic counterparts. This will have important consequences
when we discuss potential decay chains of interest.

\subsection{Cross Sections}

We now discuss the dependence of the associated cross sections on
the parameters $N_{5}$, $\Lambda$ and $\Delta_{PQ}$. Scanning over
F-theory parameter space, we have generated the (leading order) cross sections for
parton collisions using \texttt{PYTHIA} \cite{PYTHIA}. The dominant
supersymmetric processes are associated with events where parton collisions
generate either two gluinos ($\sigma_{\widetilde{g}\widetilde{g}}$),
a squark and a gluino ($\sigma_{\widetilde{q}\widetilde{g}}$), two
squarks ($\sigma_{\widetilde{q}\widetilde{q}}$), two lightest charginos
($\sigma_{\widetilde{\chi}_{1}^{+}\widetilde{\chi}_{1}^{-}}$), or a second neutralino and a
lightest chargino ($\sigma_{\widetilde{\chi}_{2}^{0}\widetilde{\chi}_{1}^{\pm}}$),
where here, $\widetilde{q}$ denotes a first or second generation squark.
We have also determined the total cross section generated by all dominant
processes, which we denote by $\sigma_{\text{tot}}$.

Figure \ref{FIG:OneL} and figure \ref{FIG:ThrL} in Appendix F show
plots of the largest cross sections as a function of $\Lambda$ with
$\Delta_{PQ}=0$ for $N_{5}=1$ and $N_{5}=3$. Note that as expected,
increasing $\Lambda$ tends to decrease the cross section. An interesting
feature of these plots is that although they are not colored particles,
there is a significant amount of chargino and second neutralino production.
In addition, we find that there is only very weak dependence on $\Delta_{PQ}$
in the dominant cross sections. Turning the discussion around, because
the cross section has little dependence on $\Delta_{PQ}$, we can
deduce that the primary effects from the PQ deformation must originate
from shifts in the masses, or branching ratios as a function of $\Delta_{PQ}$.

\subsection{Decay Channels}

In this subsection we discuss the primary decay channels which are
present in F-theory GUTs. Due to the fact that the cross section is
dominated by the production of gluinos, squarks, lightest charginos
and second lightest neutralinos, these
will also be the first particles present in the decay chain. We have
computed the relevant branching fractions by linking the output from
the mass spectrum calculated via \texttt{SOFTSUSY} \cite{SOFTSUSYAllanach}
with \texttt{SDECAY} \cite{SDECAY}. Due to the fact that the relative
masses of the squarks and the gluino heavily depends on whether $N_{5}=1$
or $N_{5}>1$, in the following discussion we separate our analysis
of these two cases.

\subsubsection{Single Messenger Case}

First consider single messenger F-theory GUT models. In this
case, only the stop and sbottom are typically lighter than the gluino
whereas for $N_{5}>1$, all of the squarks are lighter than the gluino.
Thus, the gluino decays via off shell squarks to three decay products
in the case of $N_{5}=1$ whereas for $N_{5}>1$, two body decays
are instead preferred. The dominant decay channels for the gluino
at small PQ deformation are presented below in the case of $N_{5}=1$
with $\Lambda=1.44\times10^{5}$ GeV:
\begin{equation}%
\begin{tabular}
[c]{|l|c|c|}\hline
Channel & $\Delta_{PQ}=0$ & $\Delta_{PQ}=250$ GeV\\\hline
$\widetilde{g}\rightarrow\widetilde{\chi}_{1}^{\pm}q_{1}^{\mp}q_{2}^{\pm}$ &
$27\%$ & $20\%$\\\hline
$\widetilde{g}\rightarrow\widetilde{\chi}_{1,2}^{0}q^{\mp}q^{\pm}$ & $26\%$ &
$20\%$\\\hline
$\widetilde{g}\rightarrow\widetilde{\chi}_{1,2}^{\pm}t^{\mp}b^{\pm}$ & $21\%$
& $30\%$\\\hline
$\widetilde{g}\rightarrow\widetilde{\chi}_{1,2,3,4}^{0}b^{\mp}b^{\pm}$ &
$20\%$ & $10\%$\\\hline
$\widetilde{g}\rightarrow\widetilde{\chi}_{1,2,3,4}^{0}t^{\mp}t^{\pm}$ &
$13\%$ & $17\%$\\\hline
\end{tabular}
\ \text{.}%
\end{equation}
In the above, we have adopted a condensed notation where $\pm$ and
$\mp$ refer to the charges of the particle and its anti-particle.
Also $q$ refers to the first two quark generations and $q_{1,2}$
refer to distinct components of an $SU(2)$ quark doublet. By inspection,
the stringy PQ deformation does produce a change in the branching
fractions. Note that the decay of a single $\tilde{g}$ to two tops
is a decay which is easy to see with small luminosity \cite{Acharya:2009gb}.

The decay of the right-handed squarks provides another channel which
is potentially sensitive to the effects of the PQ\ deformation. For
example, the branching ratios for the decay of the right-handed sdown
are:%
\begin{equation}
\text{%
\begin{tabular}
[c]{|l|c|c|}\hline
Channel & $\Delta_{PQ}=0$ & $\Delta_{PQ}=250$ GeV\\\hline
$\widetilde{d}_{R}\rightarrow\widetilde{\chi}_{1}^{0}d$ & $12\%$ &
$15\%$\\\hline
$\widetilde{d}_{R}\rightarrow\widetilde{g}d$ & $88\%$ & $84\%$\\\hline
\end{tabular}
,}%
\end{equation}
while for the right-handed sup, we find:
\begin{equation}
\text{%
\begin{tabular}
[c]{|l|c|c|}\hline
Channel & $\Delta_{PQ}=0$ & $\Delta_{PQ}=250$ GeV\\\hline
$\widetilde{u}_{R}\rightarrow\widetilde{\chi}_{1}^{0}u$ & $31\%$ &
$36\%$\\\hline
$\widetilde{u}_{R}\rightarrow\widetilde{g}u$ & $69\%$ & $64\%$\\\hline
\end{tabular}
.}%
\end{equation}
Similarly for the left-handed sdown and sup, the dominant decay is to gluino
plus quark, followed by the decay to chargino plus quark. These decays
are all sensitive to the PQ deformation only at the level of a
few percent change in the branching fraction.

Next consider the decay of the chargino $\widetilde{\chi}_{1}^{\pm}$ and the second neutralino $\widetilde{\chi}_{2}^{0}$.
These channels are especially interesting because their production is not directly accompanied by colored objects. Moreover,
in contrast to the other channels considered so far, the branching fractions of $\widetilde{\chi}_{1}^{\pm}$ and
$\widetilde{\chi}_{2}^{0}$ both have strong dependence on the PQ deformation.
The branching fractions for the decay of $\widetilde{\chi}_{1}^{\pm}$
are:%
\begin{equation}
\text{%
\begin{tabular}
[c]{|l|c|c|}\hline
Channel & $\Delta_{PQ}=0$ & $\Delta_{PQ}=250$ GeV\\\hline
$\widetilde{\chi}_{1}^{\pm}\rightarrow\widetilde{\tau}_{1}^{\pm}\nu_{\tau}$ &
$51\%$ & $85\%$\\\hline
$\widetilde{\chi}_{1}^{\pm}\rightarrow\widetilde{\chi}_{1}^{0}W^{\pm}$ & $49\%$ &
$15\%$\\\hline
\end{tabular}
.}%
\end{equation}
Because increasing the size of the PQ deformation parameter decreases the
mass of the stau, more phase space is available
for a decay into a stau and tau-neutrino. This large effect may mean
it is possible to measure non-zero $\Delta_{PQ}$ by triggering on
events with $W^{\pm}$ and $\tau$'s. In certain cases, this is possible
because the $\tilde{\chi}_{1}^{\pm}$ production cross section can
be $\geq0.1$~pb.

Next consider the decay of $\widetilde{\chi}_{2}^{0}$. In this case,
the relevant branching fractions are:
\begin{equation}
\text{%
\begin{tabular}
[c]{|l|c|c|}\hline
Channel & $\Delta_{PQ}=0$ & $\Delta_{PQ}=250$\\\hline
$\widetilde{\chi}_{2}^{0}\rightarrow\widetilde{\tau}_{1}^{\pm}\tau^{\mp}$ & $51\%$ & $84\%$\\\hline
$\widetilde{\chi}_{2}^{0}\rightarrow\widetilde{\chi}_{1}^{0}h$ & $43\%$ & $13\%$\\\hline
$\widetilde{\chi}_{2}^{0}\rightarrow\widetilde{\chi}_{1}^{0}Z$ & $5\%$ &
$2\%$\\\hline
\end{tabular}
.}%
\end{equation}
Much as in the decay of $\widetilde{\chi}_{1}^{+}$, the increase in the branching
fraction to staus as $\Delta_{PQ}$ increases follows from the fact that as
the mass of the stau decreases, more phase space becomes available
for a decay into a stau and tau.

\subsubsection{Multiple Messenger Case} \label{MultiMess}

In contrast to the single messenger case, in the case of multiple
messengers, the masses of the scalars are typically lower in comparison
to the gauginos. One consequence of this change is that the gluino
can now decay to two on-shell products. In addition, the effects of
the PQ\ deformation are typically weaker in the range of values where
the bino is the NLSP. This is due to the fact that the stau is already
closer in mass to the bino prior to turning on any PQ\ deformation.
For these reasons, we shall focus on decay channels present at zero
PQ\ deformation.

Restricting now to the case of zero PQ\ deformation, it is of interest
to compare multiple messenger models with similar spectra. Letting
$\Lambda_{i}$ denote the value of $\Lambda$ in the case of $i$
messengers, we expect a rough degeneracy in the mass spectrum when
either the gauginos or scalars have similar masses. A similar gaugino
mass spectrum leads to the condition:
\begin{equation}
i\cdot\Lambda_{i}=j\cdot\Lambda_{j}%
\end{equation}
while a similar scalar mass spectrum requires:%
\begin{equation}
\sqrt{i}\cdot\Lambda_{i}=\sqrt{j}\cdot\Lambda_{j}\text{.}%
\end{equation}
A rough similarity in the mass spectra can therefore be expected in the range
of values:%
\begin{equation}
\sqrt{\frac{i}{j}}\leq\frac{\Lambda_{i}}{\Lambda_{j}}\leq\frac{i}{j}\text{,}%
\end{equation}
where without loss of generality, $i\geq j$. As an example, we compare the
decay channels for a two and three messenger model such that:%
\begin{align}
\Lambda_{2}  &  =8\times10^{4}\text{ GeV}\\
\Lambda_{3}  &  =5.7\times10^{4}\text{ GeV.}%
\end{align}
As in the case of the single messenger models, we now determine the primary
decay channels for gluinos, right-handed first and second generation squarks,
and charginos. The distinction between the various branching fractions then
provide a means to extract signals which can distinguish between these cases.
In addition, we find that in all cases, there is little shift in the branching
fractions at non-zero PQ deformation in the bino NLSP\ regime.

The dominant decay channel for gluinos is given by decays to squarks.
Comparing the two and three messenger models presented above, we find:
\begin{equation}\label{gtt}
\text{%
\begin{tabular}
[c]{|l|c|c|}\hline
Channel & $N_{5}=2$ & $N_{5}=3$\\\hline
$\widetilde{g}\rightarrow\widetilde{t}^{\pm}t^{\mp}$ & $82\%$ & $45\%$\\\hline
$\widetilde{g}\rightarrow\widetilde{b}^{\pm}b^{\mp}$ & $17\%$ & $30\%$\\\hline
$\widetilde{g}\rightarrow\widetilde{q}_{R}^{\pm}q^{\mp}$ & $0.3\%$ &
$25\%$\\\hline
\end{tabular}
.}%
\end{equation}
By inspection, gluinos decay to stops and tops with large branching fractions,
thus giving rise to a spectacular four top signature at the LHC \cite{Acharya:2009gb}.
It is interesting to note that similar signature can also arise
from the $G_{2}$-MSSM \cite{Acharya:2008zi}. Further, note that this decay
mode is more favored in the two messenger case.

Next consider the decay of the right-handed sdown and sup. We find that in
both cases, the dominant decay of the squark is to the lightest neutralino and
quark with all other decay channels entirely negligible. This is to be contrasted
with the single messenger case, where the dominant decay mode is to a gluino
and quark due to the fact that in the single messenger model, the squark is
typically heavier than the gluino, whereas in the multiple messenger models,
the situation is reversed. The change in the decay of left-handed sdown and sup is
similar. Both of these decay dominantly to $\tilde \chi_{1}^{\pm}+q'$ and
$\tilde \chi_{2}^{0} + q$. The decay to gluino is now suppressed kinematically.

The decay of $\widetilde{\chi}_{1}^{\pm}$ is roughly similar in the two
and three messenger cases, with branching fractions:
\begin{equation}
\text{%
\begin{tabular}
[c]{|l|c|c|}\hline
Channel & $N_{5}=2$ & $N_{5}=3$\\\hline
$\widetilde{\chi}_{1}^{\pm}\rightarrow\widetilde{\tau}_{1}^{\pm}\nu_{\tau}$ &
$79\%$ & $82\%$\\\hline
$\widetilde{\chi}_{1}^{\pm}\rightarrow\widetilde{\chi}_{1}^{0}W^{\pm}$ & $21\%$ &
$18\%$\\\hline
\end{tabular}
,}%
\end{equation}
which is again a small effect.

Finally, we also consider decays of $\widetilde{\chi}_{2}^{0}$. The branching
fractions are only mildly sensitive to a change in the number of messengers:
\begin{equation}
\text{%
\begin{tabular}
[c]{|l|c|c|}\hline
Channel & $N_{5}=2$ & $N_{5}=3$\\\hline
$\widetilde{\chi}_{2}^{0}\rightarrow\widetilde{\tau}_{1}^{\pm}\tau^{\mp}$ & $78\%$ & $80\%$\\\hline
$\widetilde{\chi}_{2}^{0}\rightarrow\widetilde{\chi}_{1}^{0}h$ & $19\%$ & $16\%$\\\hline
$\widetilde{\chi}_{2}^{0}\rightarrow\widetilde{\chi}_{1}^{0}Z$ & $2\%$ &
$2\%$\\\hline
\end{tabular}
.}%
\end{equation}
To summarize, we therefore see that the predominant
difference between the branching fractions present in the two and three
messenger models are dictated by the decays of the gluino. Such signals then
provide a means to distinguish between two and three messenger models. Moreover,
as indicated earlier, in the regime of parameters where the bino is the NLSP,
the PQ deformation in the multiple messenger cases does not
lead to a significant change in the sparticle mass spectrum. Therefore, it is expected
to be more difficult to distinguish them from mGMSB models.

\subsection{F-theory GUTs with a Stau NLSP}

\label{STAUDISCUSS}

While the primary focus of this paper is F-theory GUT scenarios with
a bino NLSP, the stau NLSP\ scenario is also a viable option, and
is especially likely for F-theory GUTs with multiple messenger fields.
In this subsection we briefly sketch some features of the expected
signals in this case, and discuss how to distinguish such models from
high messenger scale mGMSB\ models with a quasi-stable stau NLSP.

When the lightest stau is the NLSP of an F-theory model, it will either
leave a highly ionizing track in the tracking chamber or \textquotedblleft fake
muons \textquotedblright \ in the muon chamber of a detector at the LHC.
The mass of the lightest stau can be determined by the energy-loss
($dE/dt$) and Time-of-Flight measurement. The other particles further
up the decay chain can be constructed as well in principle \cite{Ellis:2006vu}. \ For example,
by determining the invariant mass resulting from the on shell
decays $\widetilde{\chi}_{1,2}^{(0)}\rightarrow\tau^{\pm}\widetilde{\tau}_{1}^{\mp}$,
it should then be possible to reconstruct the mass of $\widetilde{\chi}_{1,2}^{0}$.
\ Kinematic considerations require $m_{\widetilde{\chi}_{1,2}^{0}}>m_{\widetilde{\tau}_{1}}$. In addition, the decay of a first or second-generation right-handed squark via the process $\widetilde{q}_{R}\rightarrow \widetilde{\chi}_{1}^{0} q \rightarrow \widetilde{\tau}_{1}^{\pm} \tau^{\mp} q$ can also be used to extract detailed properties of the spectrum. For example, by observing the trajectory of the lightest stau, it is then possible to reconstruct the four-momentum of $\widetilde{\chi}_{1}^{0}$. Thus, once the mass of $\widetilde{\chi}_{1}^{0}$ has been extracted, the corresponding squark mass can also be specified.
While a completely accurate reconstruction may require about $10-30$~
fb$^{-1}$ of integrated luminosity, this can in principle be accomplished
with data from the first three years of the LHC, and therefore provides one reliable
method for determining detailed features of the spectrum. In addition
to this type of direct mass reconstruction, many of the methods based
on a footprint analysis will carry over to the stau NLSP\ case as
well.

Once the masses of the $\widetilde{\tau}_{1}$, $\widetilde{\chi}_{1,2}^{0}$
and $\widetilde{q}_{R}$ have been determined, it will be
possible to compare these values with the spectrum expected from
mGMSB$\ $with a high messenger scale. Since the masses of the
sparticles in mGMSB\ are determined by $N_{5}$, $\Lambda$ and to a far
weaker extent $M_{\text{mess}}$, the masses of the remaining sparticles
are fixed by the values of $m_{\text{bino}}$ (or $m_{\text{wino}}$)
and $m_{\widetilde{q}_{R}}$. In particular, it is possible to then
determine the mass of the lightest stau in a minimal GMSB scenario. Assuming
that this determination has been performed, measuring the mass of
the lightest stau will then provide a direct way to distinguish between
the mGMSB\ prediction, and the F-theory prediction with non-zero PQ\ deformation.

\section{Distinguishing F-theory GUTs From Other Models \label{DISTINGUISH}}

One of the primary aims of this paper is to determine whether the
LHC\ will be able to distinguish F-theory GUTs from other potential
extensions of the Standard Model. At the first level of analysis,
it is important to establish whether the signatures from the LHC\ are
compatible with the MSSM. This is a topic which has been discussed
extensively in the literature, for some recent studies of this kind see
\cite{Wang:2008sw,Kane:2008kw,Hubisz:2008gg,Burns:2008cp,Gedalia:2009ym}
and references therein. We shall therefore assume that evidence
compatible with the MSSM\ has been found.

At the next stage of analysis, we would like to establish whether
F-theory GUTs can be distinguished from other models with an MSSM
spectrum such as mSUGRA, as well as mGMSB scenarios with a low messenger
scale. Once we have ruled out these possibilities as potential candidates
which can mimic the effects of F-theory GUTs, it is then important
to establish that F-theory GUTs can be distinguished from high messenger
scale mGMSB\ scenarios. Because of the similarities between F-theory GUTs and high scale
gauge mediation scenarios, we shall postpone this analysis to section
\ref{DETFth}.

The extent to which we can distinguish a given class of models depends
on the integrated luminosity of LHC\ data. For the most part, we
shall simulate $5$~fb$^{-1}$ of integrated luminosity. Interestingly,
we find that even with just $5$~fb$^{-1}$ of simulated LHC\ data, it is possible
to distinguish F-theory GUTs from small A-term mSUGRA\ models. The
primary limitation in this determination is that the squark and gluino
masses must be light enough to be produced by the LHC. We also find that
it is possible to distinguish large A-term mSUGRA models
from single messenger F-theory GUTs. On the other hand, large A-term
mSUGRA\ models can more effectively mimic
some of the signatures of multiple messenger F-theory GUTs. Finally,
we also show that F-theory GUTs can be distinguished from
minimal gauge mediation models with a lower messenger scale.

To perform this analysis, we have adapted the general footprint method
of \cite{Kane:2006yi,KaneFootprint} to the case of F-theory GUTs. This consists
of developing a set of signatures which will allow us to distinguish
between an F-theory GUT and other models, such as mSUGRA models, or
even between distinct F-theory GUT models. Given a model, one would
like to determine a set of $N$ signatures (the exact value of $N$
depending on the details of the footprint) which are likely to be
sensitive to the input parameters of the model. Thus a single model
generates an $N$-dimensional vector. The proximity or lack thereof
between the vectors of two such theories can then be used to distinguish between different models.

The methodology of the footprint is very general, and consists of
scanning over various signatures in a class of models and varying
the allowed parameters within a given class. Comparing with other
classes of models it is then possible to single out the signatures
which are most effective in distinguishing them. These signatures
can in principle consist of just counting events, or some combination
of counting events with more refined observables. In the footprint
method only actual measurable signatures are used, not quantities
difficult to measure such as sparticle masses, soft-breaking parameters,
or $\tan\beta$. Further, parameters not explicitly fixed by prior
considerations are scanned. In this way, we can avoid deriving misleading
conclusions based on specific features of a single model or simulation.

Once we have isolated a class of signals which can potentially serve
to distinguish an F-theory GUT from other models, the next step is
to determine whether at a more quantitative level it is possible to
distinguish various models. To this end, we introduce a chi-square
like measure of distinguishability, and show that with respect to
this measure, it is indeed possible to distinguish between many models.

The rest of this section is organized as follows. We first describe
the simulation of signals and Standard Model background. After this,
we define in more precise terms the class of mSUGRA\ models and low
scale mGMSB\ models which we shall compare to F-theory GUTs, paying
special attention to signatures which can potentially be used to distinguish
F-theory GUTs from these models. After introducing a more quantitative notion of distinguishability
in both theory and signature space, we develop a set of signatures which comprise our footprint.
Using this set of signatures, we show that in most cases, it is possible
to distinguish between F-theory GUTs and mSUGRA\ models, as well
as low messenger scale mGMSB\ models. We defer a full comparison
of F-theory GUTs to high messenger scale mGMSB\ models to section
\ref{DETFth}.

\subsection{Simulation of Signals and Background\label{SIMULATION}}

Before proceeding to potential signals of interest, we first describe
our simulation of LHC signatures. To simulate the signals of F-theory
based models, we use the spectrum and decay table obtained (in SLHA format)
by linking \texttt{SOFTSUSY} and \texttt{SDECAY} to generate events
using \texttt{PYTHIA 6.4}. These events are then passed
to \texttt{PGS4} \cite{PGS} for detector simulation. To reduce the
number of background events, we have used the \texttt{PGS} trigger with high
thresholds, the details of which can be found in Appendix D. In principle,
one should include the next-to-leading order correction to the tree-level
cross sections used in the \texttt{PYTHIA}. However, since our main
purpose is to compare different classes of models, it is not particularly
important to include such corrections. In addition, although \texttt{PGS}
is a simplified detector simulator and is not tuned to match either
ATLAS or CMS, it is still useful as a way to obtain characteristic
LHC signatures.

We have simulated $5$~fb$^{-1}$ of Standard Model background by
including the most important channels $t\overline{t}$, $W/Z$ + jets
and $WW/WZ/ZZ$ + jets. In many cases, a significant reduction in
QCD background can be achieved by means of sophisticated event selection
cuts, which we shall therefore typically neglect. These events are
generated using the \texttt{PYTHIA}-\texttt{PGS} package in the same
way as the signals. Of course, even though the use of parton-shower
Monte Carlo to estimate multi-jet backgrounds is known to underestimate
the backgrounds, for our present purposes where $S\gg\sqrt{B}$, it
suffices to include just the effects of the \texttt{PYTHIA} simulation.

To obtain LHC signatures, we selected subsets of events with different
final state objects: jets, leptons, missing energy. Before the event
selection, we impose the following object-level cut:
\begin{itemize}
\item photon, lepton, tau: $P_{T}>10$ GeV and $\eta<2.5$
\item jets: $\eta<3.0$
\end{itemize}
Here no jet $P_{T}$ cut is imposed until the event selection.

In order to reduce SM background, we first consider two typical SUSY search
channels based on inclusive two jets plus $\ds{\not}E_{T}$, and inclusive
four jets plus $\ds{\not}E_{T}$, with selection cuts given in table \ref{selection-1}.
Both channels are well studied in ATLAS \cite{Aad:2009wy} and CMS \cite{Ball:2007zza}, and the background
(mainly from $t\bar{t}$ and W/Z + jets) is known to be controllable
in the large effective mass region.

\begin{table}[pth]
\begin{center}%
\begin{tabular}
[c]{|c|c|c|}\hline
Selection & $2$ jets + $\ds{\not}E_{T}$ & $4$ jets + $\ds{\not}E_{T}$\\\hline
$P_{T}$ (GeV) & $> 150$, $100$ & $> 100$, $50$, $50$, $50$\\\hline
$\ds{\not}E_{T}$ (GeV) & $>150$& $>100$\\\hline
$S_{T}$ & $>0.2$ & $>0.2$\\\hline
$M_{eff}$ (GeV) & $>1200$ & $>1200$\\\hline
\end{tabular}
\end{center}
\caption{Inclusive SUSY search channel selection cuts used in order to reduce SM background.}
\par
\label{selection-1}
\end{table}

Here, the Effective Mass $M_{eff}$ is defined as the sum of $P_{T}$
of all objects in an event including the missing energy:
\begin{equation}
M_{eff}\equiv\sum_{i=all}P_{T}^{i}+\ds{\not}E_{T}\text{.}
\end{equation}
In addition, $0<S_{T}<1$ denotes the transverse sphericity:
\begin{equation}
S_{T}=\frac{2\lambda_{\text{large}}}{\lambda_{\text{small}}+\lambda_{\text{large}}}
\end{equation}
where $\lambda_{\text{large}}$ and $\lambda_{\text{small}}$ denote
the large and small eigenvalues of the $2\times2$ transverse sphericity
tensor:
\begin{equation}
\mathbf{S=}\left(\begin{array}{cc}
\Sigma p_{x}^{2} & \Sigma p_{x}p_{y}\\
\Sigma p_{x}p_{y} & \Sigma p_{y}^{2}
\end{array}\right)\text{.}
\end{equation}
In the case of SUSY events, the transverse sphericity is typically
closer to $S_{T}=1$, indicating a more isotropic event.

From these two inclusive selections, we further divide the events
according to the number of leptons, taus and b-jets in the final states.
Other signatures include multijets plus $\ds{\not}E_{T}$ with hard jet
$P_{T}$ thresholds, and inclusive one tau signatures with various
different $P_{T}$ thresholds. For completeness, we also include
the signatures which have been found to be effective for testing
gaugino mass unification in \cite{Altunkaynak:2009tg}. In total,
we have studied $103$ signatures, which are listed in Appendix C.

Although we have included a large number of candidate signatures, we will
see later that only a small subset of these are really effective in distinguishing
models. See table \ref{siglist1} for the $10$ signatures selected
using the footprint analysis. The details of this footprint analysis will
be applied later in the context of specific models. These signatures
will also be used later to distinguish F-theory GUTs from other
candidate models using a chi-square like measure $\Delta S^{2}$ which shall be
introduced in subsection \ref{DSTWO}.

\begin{table}[pth]
\begin{center}
\begin{tabular}[c]{|c|c|}
\hline
 & Signature List A\\
\hline
\hline
$1$ & $0\tau$($P_{T}>40$),$\ge2$jets($P_{T}>150,100$)\\
\hline
$2$ & $\ge1\tau$($P_{T}>40$),$\ge2$jets($P_{T}>150,100$)\\
\hline
$3$ & $\ge2\tau$($P_{T}>40$),$\ge2$jets($P_{T}>150,100$)\\
\hline
$4$ & $0b$($P_{T}>50$),$\ge2$jets($P_{T}>150,100$)\\
\hline
$5$ & $\ge2b$($P_{T}>50$),$\ge2$jets($P_{T}>150,100$)\\
\hline
$6$ & $\ge6$jets($P_{T}>100,100,20,20,20,20$)\\
\hline
$7$ & $\ge4$jets($P_{T}>250,250,150,150$)\\
\hline
$8$ & $0l$($P_{T}>20$),$\le4$jets($P_{T}>50$), $\ds{\not}E_{T}>500$ \\
\hline
$9$ & $0l$($P_{T}>20$),$\ge5$jets($P_{T}>50$), $0.5<r_{\rm jet}<1.0$\\
\hline
$10$ & $\ge1l$($P_{T}>20$),$\ge5$jets($P_{T}>50$), $0.05<\ds{\not}E_{T}/M_{eff}<0.35$\\
\hline
\end{tabular}
\end{center}
\caption{List of signatures obtained from a footprint analysis which are effective in distinguishing F-theory
GUTs from mSUGRA and low scale mGMSB. These signatures will also be used to distinguish models using
a chi-square like measure $\Delta S^{2}$, which shall be introduced in
subsection \ref{DSTWO}. All the $P_T$ and $\ds{\not}E_{T}$ thresholds are in
units of GeV. The variable $r_{\rm jet}$ is defined as
$r_{\rm jet}\equiv (P_T^{\rm jet3}+P_T^{\rm jet4})/(P_T^{\rm jet1}+P_T^{\rm jet2})$,
where $P_T^{{\rm jet}i}$ is the transverse momentum of the $i$-th hardest jet in the event.}
\par
\label{siglist1}
\end{table}

Before comparing signals for different classes of new physics models,
it is important to check that the candidate signal $S$ is sufficiently
large compared to the SM background $B$ so that $S>5\sqrt{B}$. As
reviewed for example in \cite{Altunkaynak:2009tg}, the discovery
condition can also be written as a condition on luminosity
\begin{equation}
L>\gamma^{2}\frac{\sigma_{SM}}{\sigma^{2}}
\end{equation}
where $\sigma$ and $\sigma_{SM}$ are respectively the cross sections of a specific
channel, and the Standard Model background, and $\gamma$ characterizes the number of sigma for a given
confidence level, so that for example a five sigma level of confidence corresponds to $\gamma = 5$. For a given luminosity, only those
channels which could lead to discovery are important, and these are the only ones we shall consider.

To distinguish models, this criterion must be supplemented by the
condition that the difference of two signals is also above the Standard
Model background. For example, consider a signal $S_{1}$ and $S_{2}$
with corresponding cross sections $\sigma_{1}$ and $\sigma_{2}$
for two models obtained for given luminosity $L$. When $S<B$,
the condition for two models to be distinguishable is:
\begin{equation}
\vert\sigma_{1}L-\sigma_{2}L\vert>\gamma\sqrt{\sigma_{SM}L}
\end{equation}
The minimal luminosity required for distinguishability is then given by:
\begin{equation}\label{LMIN}
L_{min}^{DT}=\gamma^{2}\frac{\sigma_{SM}}{(\sigma_{1}-\sigma_{2})^{2}}
\end{equation}
By inspection, $L_{min}^{DT}$ increases as the difference in signatures
decreases, and could be much larger than the discovery limit. Conversely,
as the difference in signatures increases, the minimal luminosity
required for distinguishability decreases.

\subsection{Mimicking F-theory GUTs} \label{MIMIC}

In this section we analyze some models with an MSSM spectrum which
can potentially mimic the effects of F-theory GUTs. While it is in
principle possible to perform a complete scan over all of the soft
supersymmetry breaking terms of the MSSM, in practice, this is computationally
not feasible. For this reason, we shall focus on some well known examples
with similar spectra to those of F-theory GUTs. To this end, we consider
mSUGRA\ models with small and large universal A-terms, and minimal
gauge mediation models with low and high messenger scales. We
defer a full comparison between high scale mGMSB models and F-theory
GUTs to section \ref{DETFth}.

Even when the spectra of two models are similar, there will typically
be a difference in the branching fractions which can translate into
experimental observables. In this section we discuss from a theoretical
standpoint signatures which can potentially distinguish between these
models and F-theory GUTs. As throughout this paper, we focus on the
case of F-theory GUTs with a bino NLSP.

We begin by discussing the case of mSUGRA\ models. See Appendix A
for further details on the scan of mSUGRA models performed. There
are two regimes of interest corresponding to mSUGRA\ models with
small universal A-terms, and large universal A-terms. By the {}``small
A-term\textquotedblright\ regime we shall mean the region in mSUGRA\ parameter
space where the universal A-term is small, and such that
the slepton masses are large compared to the electro-weakinos:
\begin{equation}
m_{\tilde{\tau}}\approx m_{\tilde{e},\tilde{\mu}}>m_{\tilde{\chi}_{1,2}^{0}}\text{,}\label{slepchibound}
\end{equation}
which roughly mimics the case of F-theory GUTs, although there is
some difference in that $m_{\tilde{\tau}}>m_{\tilde{\chi}_{2}^{0}}$.
For these mSUGRA\ models, $\chi_{2}^{0}$ and $\chi_{1}^{\pm}$ will
typically decay into an LSP plus a gauge boson or a Higgs boson. This
is in contrast to the case in F-theory based models where decay through
the lightest stau $\tilde{\tau}_{1}$ is always important, giving
$2\;\tau$ plus missing $E_{T}$ or $4\;\tau$ plus missing $E_{T}$
signatures. Therefore, to distinguish this subset of mSUGRA models,
one can look for lepton flavor universality, or more explicitly, whether
there are more taus than electrons and muons in the events.

Large A-term mSUGRA\ models can also potentially mimic the spectrum
of F-theory GUTs. By the \textquotedblleft large A-term\textquotedblright\ regime
we shall mean the region in mSUGRA\ parameter space where the universal
A-term is large, and where the mass of the lightest stau
$\tilde{\tau}_{1}$ is suppressed due to either a large Yukawa, or
an A-term such that:
\begin{equation}
m_{\tilde{\chi}_{1}^{0}}<m_{\tilde{\tau}_{1}}<m_{\tilde{\chi}_{2}^{0}}\text{.}\label{chistauchi}
\end{equation}
This is the same ordering of masses present in F-theory GUTs with a bino NLSP.
Therefore, large A-term models can mimic signals based on the decay
of $\chi_{2}^{0}$ to an on-shell stau. Further, although the A-terms of F-theory
GUTs are zero at the messenger scale, the high messenger scale implies that
radiative corrections can still generate contributions to the A-terms
of F-theory GUTs at lower energy scales. In fact, as we will see in
section \ref{F-theory-large-A}, this subset of mSUGRA models can for low enough luminosity
indeed sometimes be difficult to distinguish from the signatures of F-theory based models.

Minimal GMSB scenarios constitute another class of models which
could mimic the signatures of F-theory GUTs. Here, there are
two qualitative regimes of interest corresponding to models
with a high, or low messenger scale. Since we shall consider
the case of high messenger scale models in greater detail in section
\ref{DETFth}, in this section we focus on the comparison between
F-theory GUTs and low messenger scale mGMSB\ models. In contrast
to F-theory GUTs where the NLSP\ always decays outside the detector,
in the case of a low scale GMSB\ model, the NLSP decays inside the
detector. In principle, the NLSP\ can correspond either to a bino-like
neutralino, or the lightest stau. To determine whether we can distinguish
F-theory GUTs from these possibilities, we compare the two cases of
a bino or stau NLSP for F-theory GUTs, respectively denoted
by $F_{\text{bino}}$ and $F_{\text{stau}}$ with the analogous options for low scale mGMSB, which we denote by
$\text{mGMSB(LO)}_{\text{bino}}$ and $\text{mGMSB(LO)}_{\text{stau}}$:
\begin{equation}
\text{%
\begin{tabular}
[c]{|l|l|}\hline
$F_{\text{bino}}$ vs $\text{mGMSB(LO)}_{\text{bino}}$ & $F_{\text{bino}}$ vs $\text{mGMSB(LO)}_{\text{stau}}$\\\hline
$F_{\text{stau}}$ vs $\text{mGMSB(LO)}_{\text{bino}}$ & $F_{\text{stau}}$ vs $\text{mGMSB(LO)}_{\text{stau}}$\\\hline
\end{tabular}
.} \label{Square}%
\end{equation}
In the case of a low scale model with bino NLSP, the decay of a pair
of binos emits two hard photons, providing a relatively clean signature
which is absent in the case of F-theory GUT models. Next consider
the two remaining possibilities denoted by the second column of (\ref{Square}).
When the stau is the NLSP\ of an F-theory GUT, it will leave a charged
track in the detector, which should again provide a relatively straightforward
means by which to distinguish this class of models from low scale models.

The final case of comparison, corresponding to a high scale model
with a bino NLSP versus a low scale model with a stau NLSP\ is more
delicate. Note, however, that when the stau NLSP decays promptly inside
the detector, the lightest neutralino can decay to opposite sign taus
and a gravitino, as in $\tilde{\chi}_{1}^{0}\rightarrow\tau^{\pm}\tau^{\mp}\tilde{G}$,
leading to a $4\;\tau$ plus missing $E_{T}$ signature. Although
such signatures can also arise from F-theory GUTs with a bino NLSP
via the decay channel $\tilde{\chi}_{2}^{0}\rightarrow\tau^{\pm}\tau^{\mp}\tilde{\chi}_{1}^{0}$,
they are less common since the corresponding branching ratio is
usually smaller than that associated with channels involving gauge bosons and the Higgs.
In addition, a large fraction of sparticles decay into chargino $\tilde{\chi}_{1}^{\pm}$
which only gives one $\tau$ for each chargino. Therefore, the counting
signatures related to multiple taus in the final state should be helpful
in distinguishing F-theory models from low scale mGMSB models with
a stau NLSP.

Having specified some examples of models with very different theoretical
motivations, we next proceed to determine to what extent we can distinguish
these possibilities from F-theory GUTs, both at the level of soft
terms, as well as more importantly, from the perspective of potential
experimental signatures.

\subsection{Theory Space Measure: $\Delta P^{2}$}

Although our primary focus will be on the extent to which we can distinguish
collider signatures of F-theory GUTs from other models, it is also
of interest to see whether there are any degeneracies at the level
of the effective field theory. In such cases, the resulting experimental
signatures may also be quite similar. However, even differences between
soft masses on the order of $50$~GeV are typically sufficient to
produce some differences in the branching fractions of a given decay.
To get a rough sense of the level of the theoretical degeneracy, we
have used a similar measure of distinguishability to that given in
\cite{LHCInverse}. Summing over the soft parameters of a given model,
we define a notion of theoretical distinguishability between two models
$M_{1}$ and $M_{2}$ as:
\begin{equation}
\Delta P^{2}(M_{1},M_{2})=\frac{1}{N_{\text{param}}}\underset{i=1}{\overset{N_{\text{param}}}{\sum}}\left(
\frac{\vert p_{i}^{(1)}\vert-\vert p_{i}^{(2)}\vert}{\overline{p}_{i}^{(12)}}\right)^{2}\text{,}
\label{DPSQU}
\end{equation}
where the sum $i$ runs over parameters of the
model, such as the soft masses or trilinear couplings, $N_{\text{param}}$
denotes the total number of such parameters, and $\overline{p}_{i}^{(12)}=(\vert p_{i}^{(1)}\vert+\vert p_{i}^{(2)}\vert)/2$
denotes the average between the two values. Here, we have taken the
norm of all parameters because we shall typically be interested in
differences based on branching fractions and masses.

The quantity $\Delta P^{2}$ roughly measures the quadrature average
of the percentage difference between two models. Note that since we
shall always consider the absolute value of a parameter, $\Delta P^{2}$
is bounded by the inequalities:
\begin{equation}
0\leq\Delta P^{2}(M_{1},M_{2})\leq4\text{.}
\end{equation}
Typically, we shall compare models up to the level of about $\Delta P^{2}>0.01$,
although in principle this can be refined further.

In all cases, we shall compare models in theory space using the dominant
soft supersymmetry breaking parameters, because the corresponding
soft breaking terms serve to determine the physical masses and branching
fractions of a given model. The soft breaking parameters include the
twenty soft mass parameters of the MSSM (three gaugino mass parameters,
two Higgs soft mass parameters and $3\times5=15$ soft scalar mass
parameters for the scalar partners of Standard Model fermions), as
well as the $(3,3)$ component of all the trilinear A-term couplings given
by $A_{\tau}$, $A_{b}$ and $A_{t}$, in the self-explanatory notation.
The Higgs sector of the theory is determined by the parameters $m_{H_{u}}^{2}$,
$m_{H_{d}}^{2}$, $\mu$ and $B\mu$. Proper electroweak symmetry
breaking imposes one real condition on these parameters, so that in
any viable model, there are in fact only three independent parameters
to scan over. Since we have already included the soft masses squared
of the Higgs fields, either $\mu$ or $B\mu$ can be used as the third
parameter. In fact, we shall typically specify the third parameter
of the Higgs sector by including just the value of $\tan\beta=\langle H_{u}\rangle/\langle H_{d}\rangle$.

To summarize, we shall compare distinct models in theory space using
the $24$ soft parameters just specified. Thus, in this
case $N_{\text{param}}=24$, and the sum of equation (\ref{DPSQU})
will range over these $24$ parameters.

\subsection{Signature Space Measure:\ $\Delta S^{2}$} \label{DSTWO}

To quantify the extent to which a given set of signals is able to
distinguish two models, we must have some notion of {}``distance\textquotedblright\ between
various models. Given two models $M_{1}$ and $M_{2}$, we are interested
in knowing whether a given set of $N$ observables $\mathcal{O}_{1}^{(1)},...,\mathcal{O}_{N}^{(1)}$
and $\mathcal{O}_{1}^{(2)},...,\mathcal{O}_{N}^{(2)}$ effectively
distinguish these models. To this end, we can introduce a chi-square
like variable which defines a notion of distinguishability:
\begin{equation}
\Delta S^{2}\left(M_{1},M_{2}\right)=\frac{1}{N}\underset{i=1}{\overset{N}{\sum}}\frac{
\left(\mathcal{O}_{i}^{(1)}-\mathcal{O}_{i}^{(2)}\right)^{2}}{(\sigma_{i}^{(1)})^{2}+
(\sigma_{i}^{(2)})^{2}+(\sigma_{i}^{(SM)})^{2}}\text{,}
\end{equation}
where in the above, $(\sigma_{i}^{(1)})^{2}$ denotes the variance
in the $i^{\text{th}}$ signature of $\mathcal{O}^{(1)}$, with similar
notation for $(\sigma_{i}^{(2)})^{2}$. In the above, we have also
included the background from Standard Model processes, in the form
of $(\sigma_{i}^{(SM)})^{2}$. Although quite similar to a chi-square
measure, this definition is typically reserved for global fits to
the data, and so to avoid any confusion, we shall instead refer to
the relevant measure as $\Delta S^{2}$. In the actual signatures
considered in this paper, we will primarily focus on counting events
so that for $n$ events of a particular signature, we shall assign
a variance of $(\sigma(n))^{2} = n + 1$.

We shall say that two models are distinguishable provided:
\begin{equation}
\Delta S^{2}\left(M_{1},M_{2}\right)>\gamma_{N}(p)\text{,}
\end{equation}
where $\gamma_{N}(p)$ is some cutoff greater than one which depends
on the number of signals, and the confidence level $0\leq p\leq1$.
In the context of an actual chi-square, $\gamma_{N}(p)$ is implicitly
defined by the relation:
\begin{equation}
\Gamma\left(\frac{N}{2},\frac{N}{2}\gamma_{N}(p)\right)=\Gamma\left(\frac{N}{2}\right)(1-p)\text{.}
\end{equation}
In the above, $\Gamma(a,b)$ denotes the incomplete upper gamma function. In this paper
we shall use this value of $\gamma_{N}(p)$ as a rough measure of distinguishability,
and will compare models at the $p =99\%$ level.

Depending on which signatures turn out to be useful in discriminating between different
models, the actual sum over signals may in principle depend on the type of
comparison performed. In comparing mSUGRA models and low scale mGMSB models with F-theory GUTs,
we will exclusively use the $10$ signatures detailed in table \ref{siglist1} of subsection \ref{SIMULATION}. This defines
a particular measure which we refer to as $\Delta S_{(A)}^{2}$. Later in
section \ref{DETFth} when we compare F-theory GUTs with high scale mGMSB scenarios
as well as other F-theory GUT models, we will find that there are in fact two classes of signatures,
one set of $10$ signatures which produce relatively distinguishable footprints, and another set of
$13$ signatures (also determined by the footprint method) detailed in table \ref{siglist3} of subsection \ref{SIGLIST} which are especially
helpful in a $\Delta S^{2}$ analysis. These signatures define another measure, which we shall refer to
as $\Delta S_{(D)}^{2}$. Indeed, one can envision first using the $\Delta S_{(A)}^{2}$ measure to broadly distinguish between models,
with $\Delta S_{(D)}^{2}$ serving as a more refined measure of distinguishability. We shall say that
we can distinguish between models for these $\Delta S^{2}$ measures provided:
\begin{align}
\Delta S_{(A)}^{2}& >\gamma _{10}(0.99)=2.3 \\
\Delta S_{(D)}^{2}& >\gamma _{13}(0.99)=2.1\text{.}
\end{align}

\subsection{F-theory Versus Small A-term mSUGRA}

We now analyze the extent to which LHC\ observables are capable of
distinguishing between F-theory GUTs and other well-motivated extensions
of the MSSM. We begin by considering the case of mSUGRA\ models with
small A-terms. Further details on the precise scan over small A-term
mSUGRA (mSUGRA(SA)) models considered may be found in Appendix A.

\begin{figure}[ptb]
\begin{center}
\includegraphics[
height=5.5625in,
width=5.7017in
]{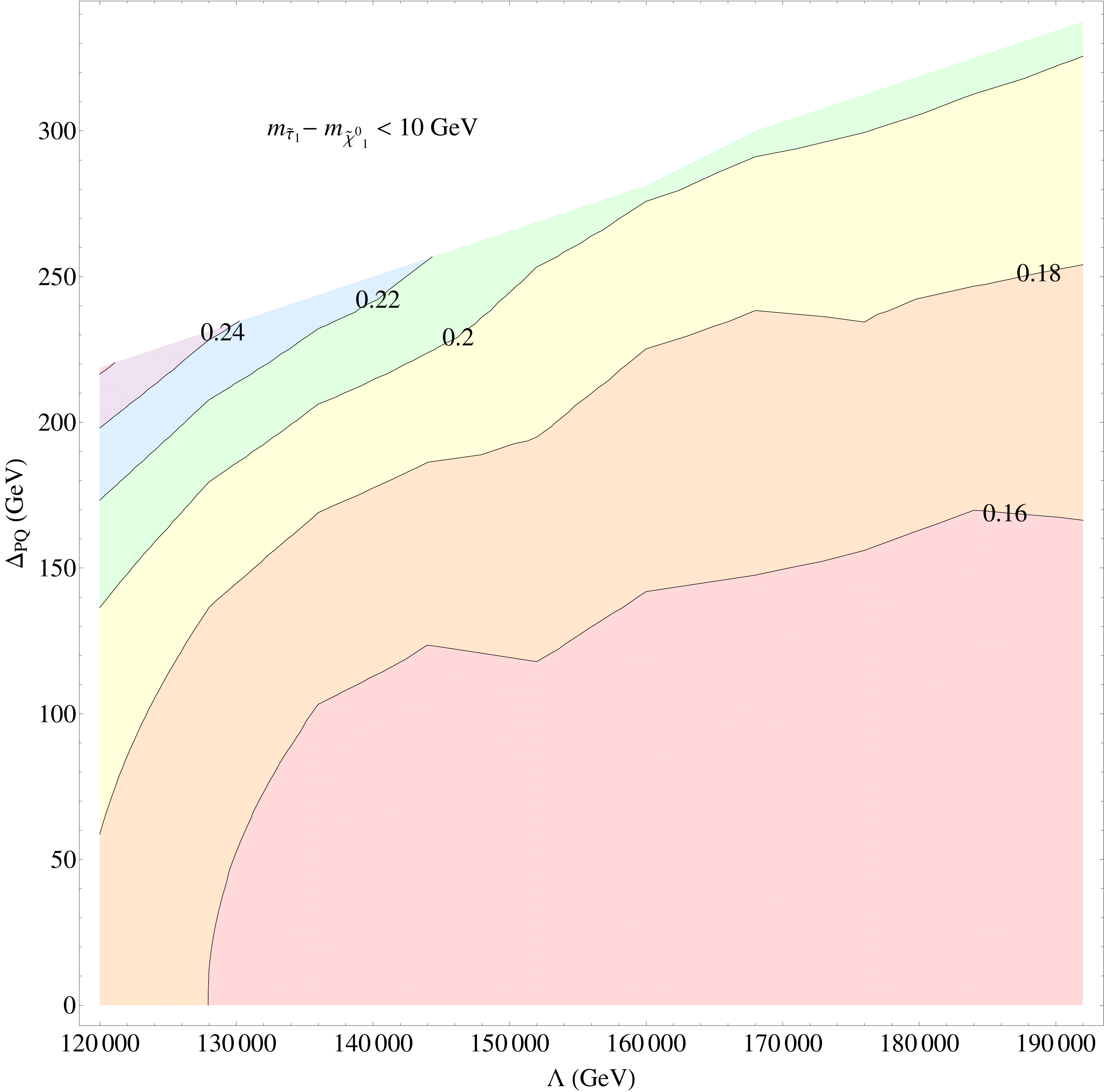}
\end{center}
\caption{Contour plot of the value of $\Delta P^{2}$ obtained by fixing a
particular value of $\Lambda$ and $\Delta_{PQ}$ of an F-theory GUT model with
$N_{5}=1$, and minimizing with respect to all small A-term mSUGRA\ models. We adopt
a rough criterion for theoretical distinguishability specified by the requirement that $\Delta P^2 > 0.01$. By
inspection, $\Delta P^2$ is greater than $0.15$, indicating that
such models are distinguishable at the theoretical level from F-theory GUTs. See
figure \ref{FIG:softspecthrsugrasa} in Appendix
F for a similar plot for $N_{5} = 3$ F-theory GUTs.}%
\label{FIG:softspeconesugrasa}%
\end{figure}

As a first check, we first show to what extent these models are distinguishable
as distinct particle physics models. To this end, we have computed
the value of $\Delta P^{2}$ between F-theory GUT\ models and mSUGRA(SA)
models. To gauge the level of distinguishability in
the {}``worst case scenario\textquotedblright, we next minimized
$\Delta P^{2}$ over all such mSUGRA(SA) models:
\begin{equation}
P_{SA}\left(N_{5},\Lambda,\Delta_{PQ}\right)=\underset{m_{\text{mSUGRA(SA)}}}{\min}
\Delta P^{2}(m_{F},m_{\text{mSUGRA(SA)}})\text{,}
\end{equation}
scanning over the parameters $M_{1/2}$, $m_{0}$, $A$ and $\tan\beta$
over the range of small A-terms specified in Appendix A. For a
generic F-theory GUT\ model, the value of $P_{SA}$ satisfies
the inequality:
\begin{equation}
P_{SA}\left(N_{5},\Lambda,\Delta_{PQ}\right)\gtrsim0.15\text{.}
\end{equation}
See figure \ref{FIG:softspeconesugrasa}, and figure \ref{FIG:softspecthrsugrasa} in Appendix F
for contour plots of the one and three messenger F-theory GUT models.
In some sense, this level of distinguishability is to be expected,
because mSUGRA models typically exhibit a more compressed spectrum
where the masses of the squarks and sleptons are more degenerate.
The fact that in gauge mediation models, there is more separation
in the mass of the squarks, sleptons and gauginos reflects this difference.

Having established that at a theoretical level it is possible to distinguish
between F-theory GUTs and small A-term mSUGRA\ models, we now determine
experimental observables which discriminate between these two classes
of models.

\subsubsection{Footprint Analysis \label{smallFOOT}}

\begin{figure}[ptb]
\begin{center}
\includegraphics[
height=3.8in,
width=5.91131in
]{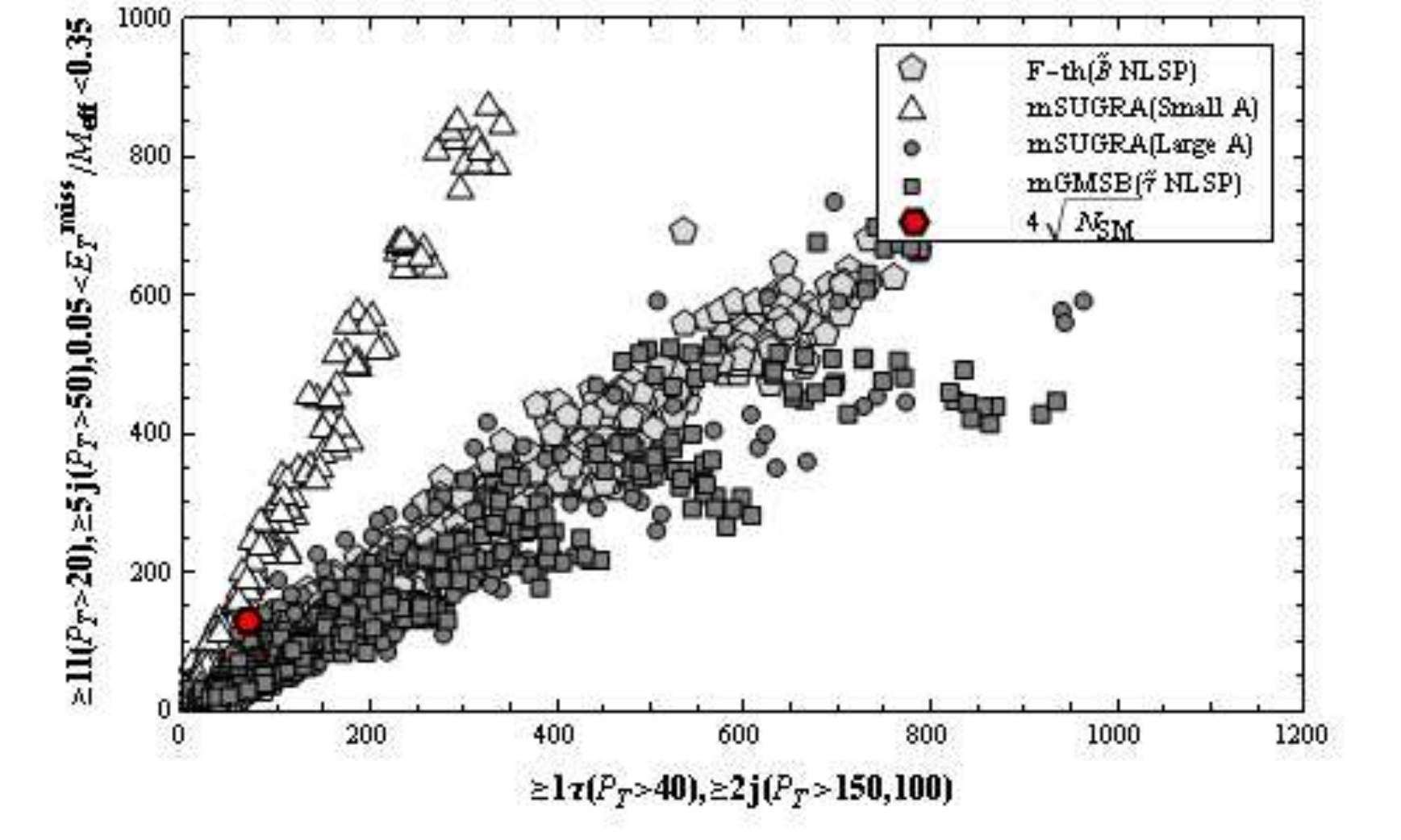}
\includegraphics[
height=3.8in,
width=5.91131in
]{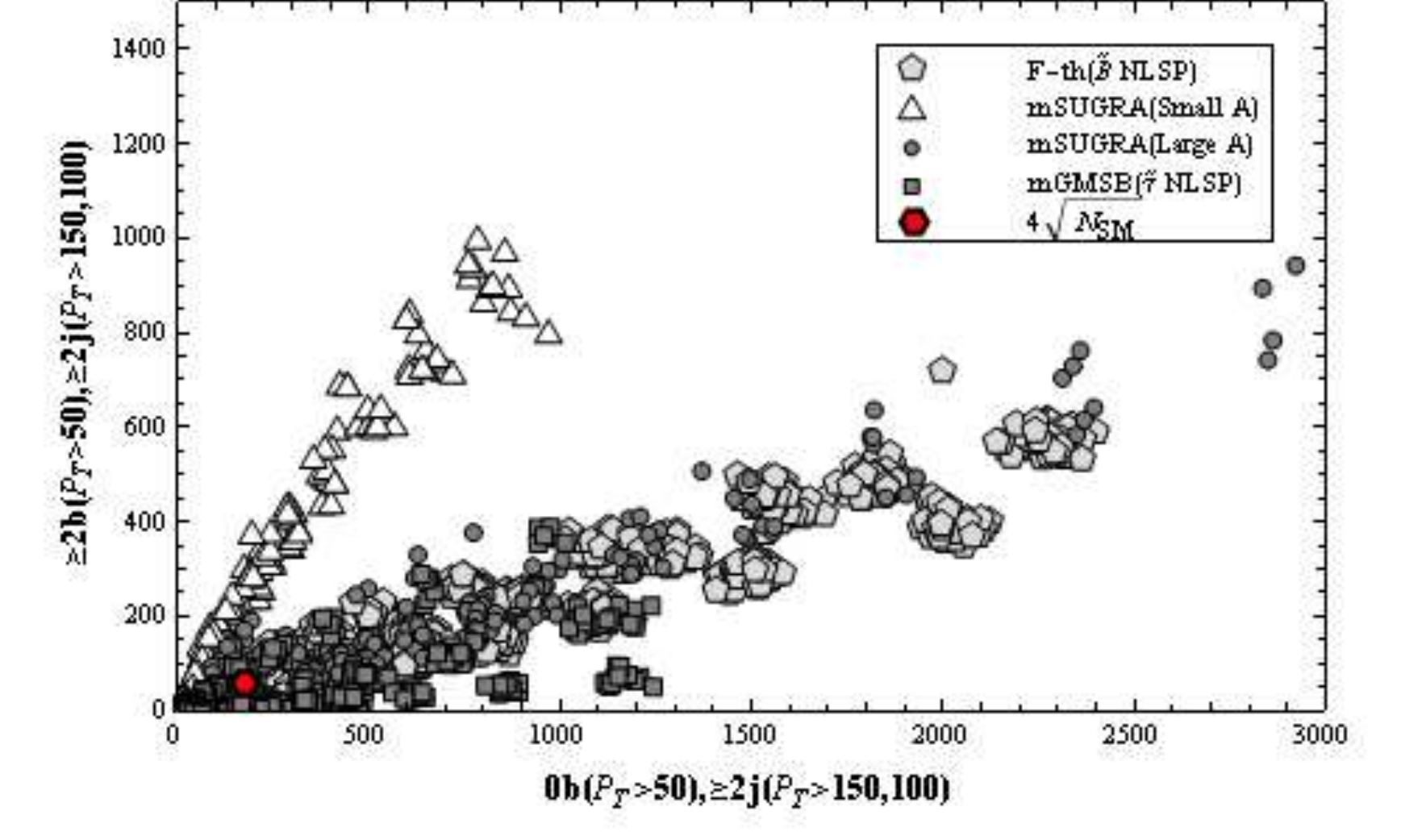}
\end{center}
\caption{Footprint of LHC signatures (without SM background) for
distinguishing F-theory GUTs and small A-term mSUGRA models with
$5$~fb$^{-1}$ integrated luminosity. The red circle denotes the
$4\sigma$ deviation from the SM background.}%
\label{sig-1}%
\end{figure}

In this subsection we use the general footprint method described earlier to compare the signatures of
F-theory GUTs with small A-term mSUGRA models. To this end,
we have computed LHC signatures from the models simulated using
\texttt{PGS}. Upon constructing a footprint for these models, we
next compare correlations of different pairs of signatures by
making two dimensional plots. From these plots, we can distinguish
at a qualitative level between F-theory models and small A-term
mSUGRA models. In particular, for certain pairs of signatures,
there is little to no overlap between the two classes of models.
See the plots in figure \ref{sig-1} for examples of such signature plots. We
find that signatures which distinguish between the two classes of
models typically include b-jets and $\tau$ leptons.

The differences in LHC\ signatures between these classes of models
can be understood based on their different spectra. As mentioned
before, F-theory GUTs have a light $\tilde{\tau}_{1}$, and
so in models with a bino NLSP, the subsequent decay of the $\tilde{\tau}_{1}$
will produce more events with taus compared to electrons and muons.
Additional leptons will be generated through the decays of the $Z$,
$h$ and $t$.

This is to be contrasted with the case of small A-term mSUGRA\ models where leptons typically
do not originate from the decay of the $\tilde{\tau}_{1}$, but rather,
from the decays of the $Z$, $h$ and $t$. Since these decays are
typically accompanied by b-jets, it follows that by vetoing on events
with b-jets, we can expect a greater number of lepton events in F-theory
GUTs in comparison with small A-term mSUGRA\ models. This distinction
in the number of leptons to hadrons in various events explains why
it is possible to distinguish between such models using signatures
with taus and signatures with lepton + $0$ b-jets.

\subsubsection{$\Delta S^{2}$ Analysis}

We now show that with only $5$~fb$^{-1}$
of simulated LHC data, it is also possible to distinguish
at a more quantitative level between F-theory GUTs and small A-term
mSUGRA models. In comparing F-theory GUTs and small A-term mSUGRA\ models, we included all $10$ of the
signatures detailed in table \ref{siglist1} of subsection \ref{SIMULATION}. These signatures are not identical
to the ones presented for the footprint plots, but instead appear
to provide a cleaner way to minimize over $\Delta S^{2}$. We shall refer to the corresponding measure defined by signature list A as
$\Delta S_{(A)}^{2}$. Much as in our discussion of $\Delta P^{2}$,
we have computed the value of $\Delta S_{(A)}^{2}$ between F-theory
GUTs and small A-term mSUGRA\ models. The value of $\Delta S_{(A)}^{2}$
between an F-theory GUT point and a typical mSUGRA\ point is typically
far greater than $10$, and so such models are easily distinguished. However,
it is still possible that certain models could share similar characteristics
with F-theory GUTs. Fixing a given F-theory model, we next scanned
over all of the relevant mSUGRA\ models, minimizing the corresponding
value of $\Delta S_{(A)}^{2}$:

\begin{equation}
S_{SA}\left(  N_{5},\Lambda,\Delta_{PQ}\right)  =\underset
{m_{\text{mSUGRA(SA)}}}{\min}\Delta S_{(A)}^{2}(m_{F},m_{\text{mSUGRA(SA)}%
})\text{,}%
\end{equation}
where here, the minimization is performed over all small A-term mSUGRA\ models
of potential interest. See figure \ref{FIG:OneSugra} and figure \ref{FIG:ThrSugra} in Appendix F
for density plots of $S_{SA}$ for $N_{5}=1$ and $3$ as a function
of $\Lambda$ and $\Delta_{PQ}$ at $5$~fb$^{-1}$ of LHC data. A
similar result holds in the case of two messenger F-theory GUT models.
\begin{figure}[ptb]
\begin{center}
\includegraphics[
height=5.0929in,
width=6.2128in
]{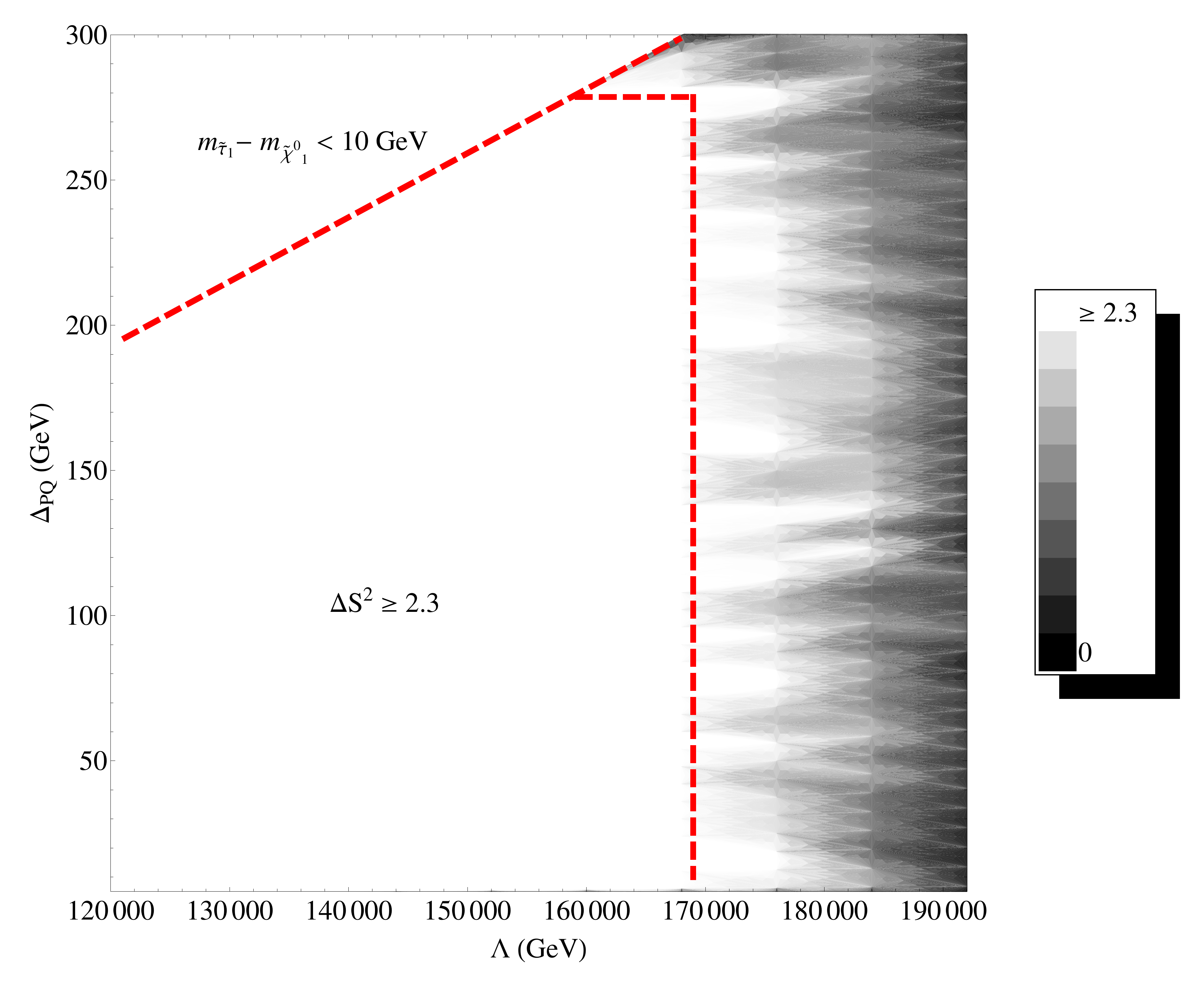}
\end{center}
\caption{Density plot of $\Delta S_{(A)}^{2}$ defined by the signature list
in table \ref{siglist1} of subsection \ref{SIMULATION} comparing the minimal value of a given
$N_{5}=1$ F-theory GUT model with a scan over small A-term mSUGRA models. The signals used are
obtained with $5$ fb$^{-1}$ of simulated LHC data. Here, we have used a rough notion of distinguishability based
on $99\%$ confidence and $10$ signals so that at $\Delta S_{(A)}^{2}>2.3$ we shall
say that two models are distinguishable. By inspection,
for most of the plot, it is possible to distinguish between F-theory GUTs and
mSUGRA models. See figure \ref{FIG:ThrSugra} in Appendix F for a similar
plot in the case of $N_{5} = 3$ F-theory GUTs.}%
\label{FIG:OneSugra}%
\end{figure}

These plots illustrate that it is typically possible to distinguish
between F-theory GUTs and mSUGRA models. It is only by minimizing
over all such points that we even come close to saturating the cutoff
for distinguishability. In a certain sense, this is to be expected
because the spectra and branching fractions are typically different
in such models, so that in principle, there should exist signals which
distinguish between these two possibilities.

\subsection{F-theory Versus Large A-term mSUGRA\label{F-theory-large-A}}

In the previous subsection we found that it is indeed possible to distinguish
between small A-term mSUGRA\ models and
F-theory GUTs. Large A-term mSUGRA\ models (mSUGRA(LA)) can also
potentially mimic F-theory GUTs. Indeed, since the
universal trilinear term in mSUGRA\ models couples the
scalars of the theory, when the Higgs develops a vev, this
will induce a shift in the soft masses. This by itself is important,
because it can allow mSUGRA models to mimic the effects of the PQ\ deformation.
On the other hand, introducing a large A-term changes the branching fractions
in the decay of scalars. Thus, one might hope that even if the
the mass spectra are somewhat similar, the decay channels in large
A-term mSUGRA\ models could still be distinguished from F-theory GUTs.

\begin{figure}[ptb]
\begin{center}
\includegraphics[
height=5.5625in,
width=5.7017in
]{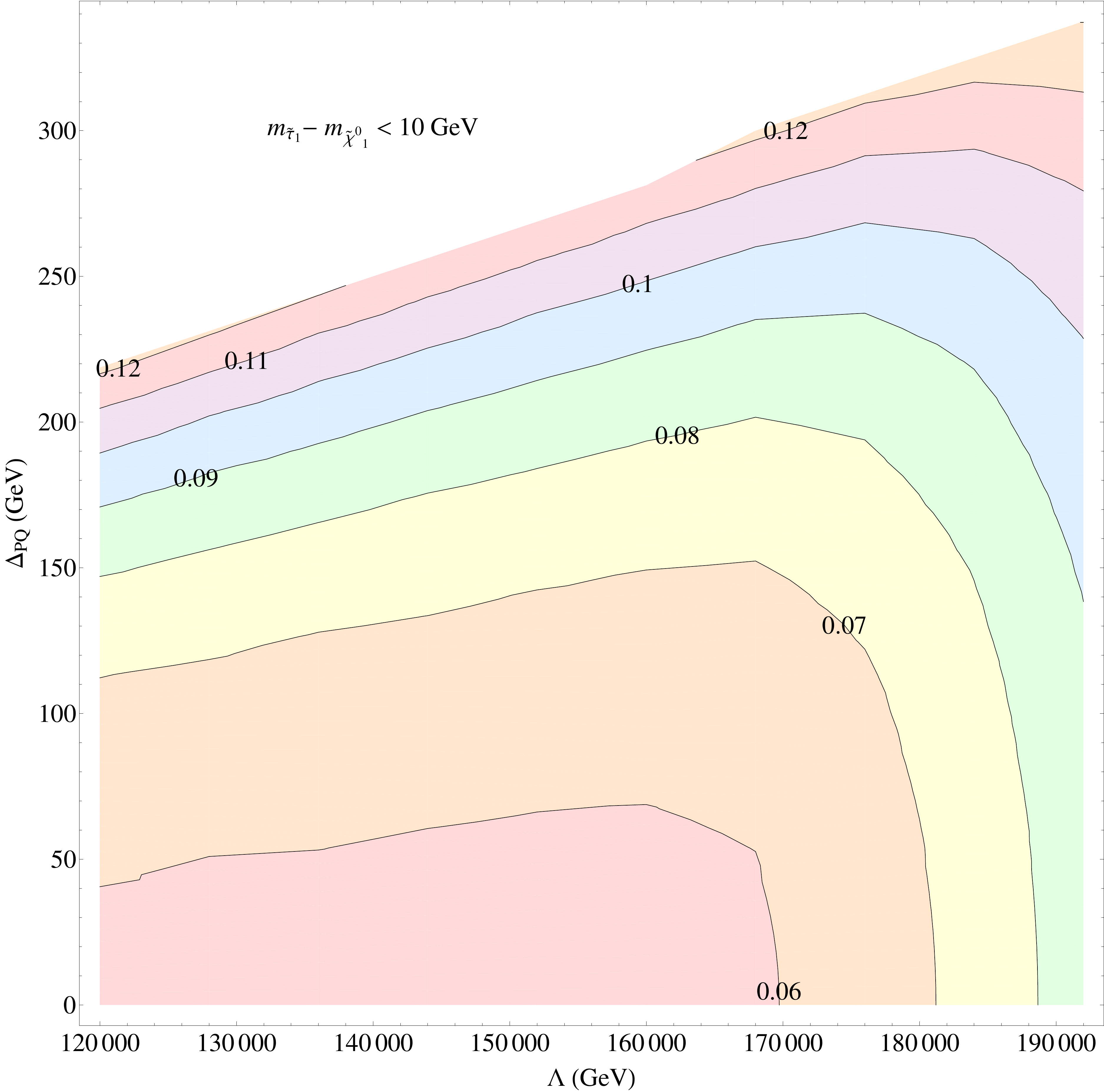}
\end{center}
\caption{Contour plot of the value of $\Delta P^{2}$ obtained by fixing a
particular value of $\Lambda$ and $\Delta_{PQ}$ of an F-theory GUT model with
$N_{5}=1$, and minimizing with respect to all large A-term mSUGRA\ models. We adopt
a rough criterion for theoretical distinguishability specified by the requirement that $\Delta P^2 > 0.01$. By
inspection, $\Delta P^2$ is greater than $0.05$. This is to be
contrasted with the level of distinguishability found for small A-term mSUGRA
models where $\Delta P^{2}$ is greater than $0.15$. See figure
\ref{softspecthrsugrala} in Appendix F for a similar plot for F-theory GUTs with $N_{5} = 3$.}%
\label{softspeconesugrala}%
\end{figure}

To a certain extent, it is indeed possible to distinguish at a theoretical
level to distinguish mSUGRA(LA) and F-theory GUT scenarios, although
in comparison with small A-term models, the soft terms are closer
to those of F-theory GUTs. Fixing a given F-theory GUT\ model,
and minimizing over all mSUGRA(LA) models, we define:
\begin{equation}
P_{LA}\left(  N_{5},\Lambda,\Delta_{PQ}\right)  =\underset
{m_{\text{mSUGRA(LA)}}}{\min}\Delta P^{2}(m_{F},m_{\text{mSUGRA(LA)}})\text{.}%
\end{equation}
we have computed the value of $P_{LA}$ for one, two and three messenger
models.

In comparison to the case of small A-term mSUGRA\ models, the value
of $P_{LA}$ is typically somewhat smaller, but is still bounded below
by:
\begin{equation}
P_{LA}\left(N_{5},\Lambda,\Delta_{PQ}\right)\gtrsim0.05\text{.}
\end{equation}
Figure \ref{softspeconesugrala} and figure \ref{softspecthrsugrala} in Appendix F
show contour plots of $P_{LA}$ as a function
of $\Lambda$ and $\Delta_{PQ}$. Note that in comparison to the small
A-term mSUGRA\ model contour plots depicted in figure \ref{FIG:softspeconesugrasa}
and figure \ref{FIG:softspecthrsugrasa} in Appendix F, the soft parameters of this class
of models can more effectively mimic F-theory GUTs.

\subsubsection{Footprint Analysis}

\begin{figure}[ptb]
\begin{center}
\includegraphics[
height=3.8in,
width=5.91131in
]{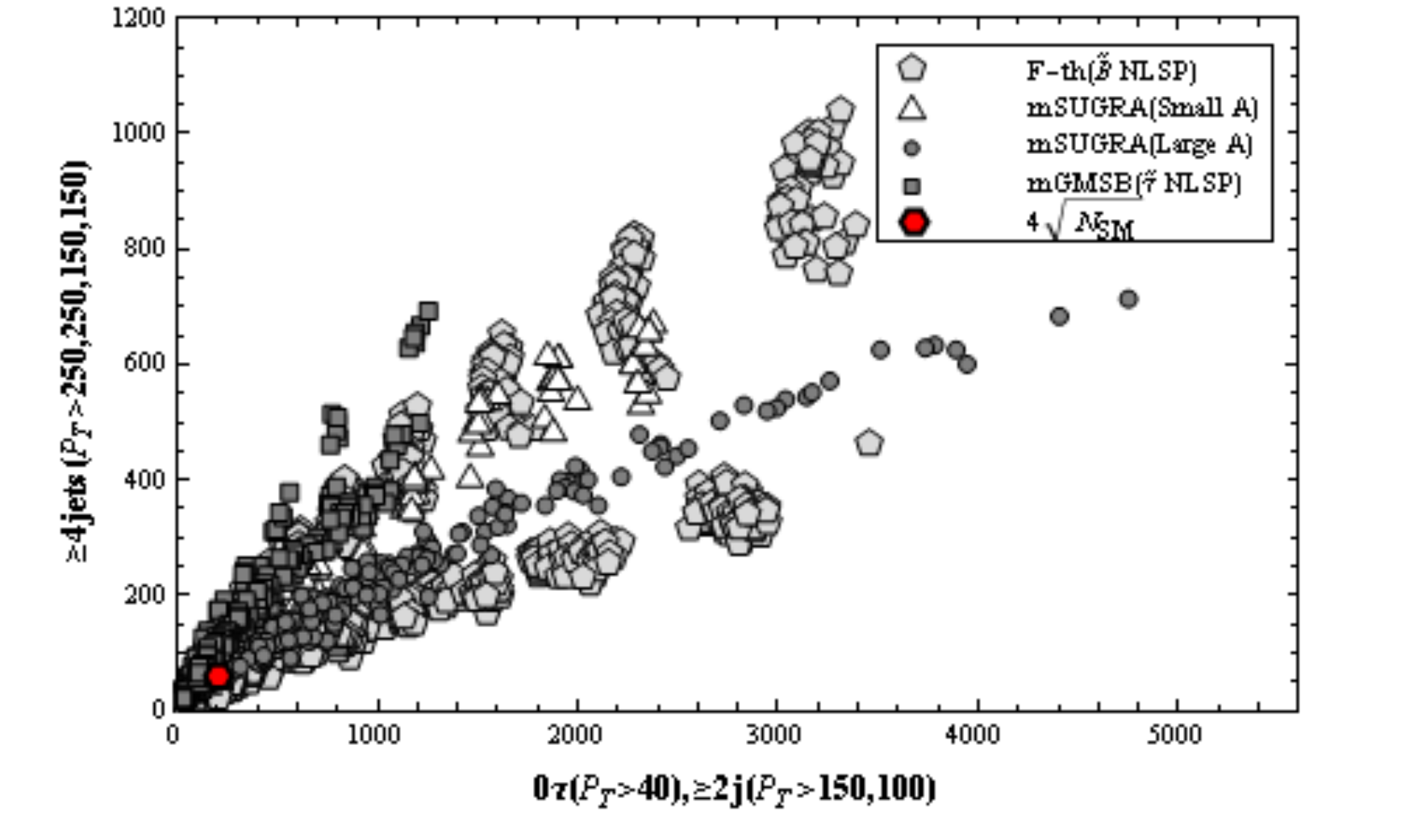} \includegraphics[
height=3.8in,
width=5.91131in
]{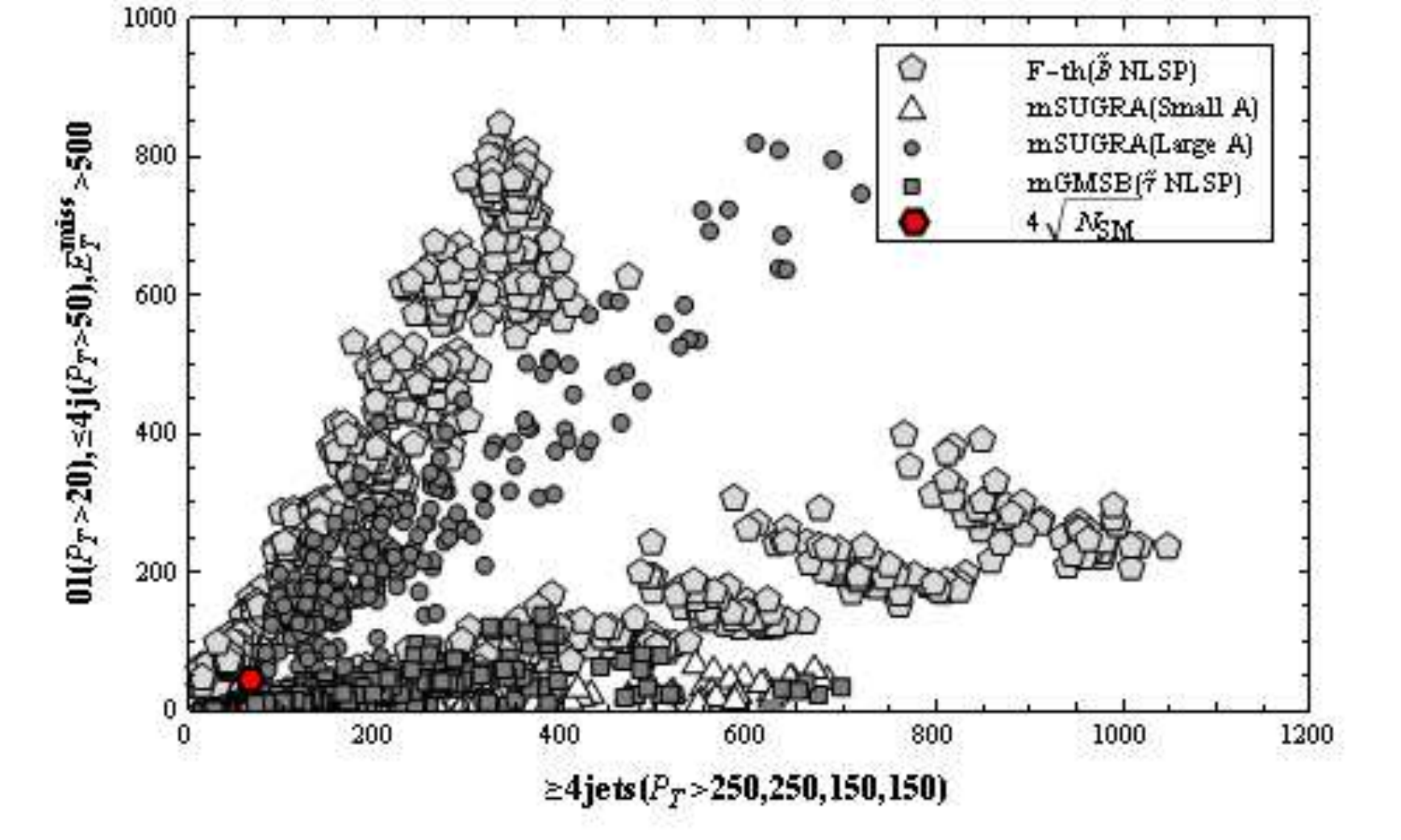}
\end{center}
\caption{Footprint of LHC signatures (without SM background) for
distinguishing F-theory GUTs and large A-term mSUGRA models with
$5$~fb$^{-1}$ integrated luminosity. The red circle denotes the
$4\sigma$ deviation from the SM background.}%
\label{sigs-3}%
\end{figure}

In this section, we discuss whether the footprints of large A-term
mSUGRA and F-theory GUTs can be used to distinguish these two classes
of models. As before, we generated two dimensional footprints using
the $103$ signatures listed in Appendix C. By inspecting many footprint
plots, we have found a few signatures which can be used
to distinguish F-theory GUTs and large A-term mSUGRA models as in
figure \ref{sigs-3}. These plots also show that there is still a large overlap
in the region corresponding to models with relatively small cross section.
The primary overlap region is mainly comprised of two and three messenger
F-theory GUT models. For this reason, we find that it is not possible to
cleanly distinguish F-theory GUTs with multiple messengers in the bino NLSP
regime from large A-term mSUGRA models with our implementation of signatures and
limited integrated luminosity. We also notice that the simple signatures
involving lepton as well as tau and b-jet cannot
distinguish these two classes of models. As we shall explain in greater
detail in subsection \ref{MOREON}, this can be roughly understood
from the fact that the two and three messenger F-theory GUT models
share the same decay topology with large A-term mSUGRA
models. For the two messenger case, there are differences in the lightest
stau and Higgsino mass in the region of small $\Lambda$. Nevertheless,
these distinctions do not seem to be sufficient for producing discriminating
LHC signatures with $5$~fb$^{-1}$ of LHC data. On the other hand, figure \ref{sigs-3} illustrates that
multi-jet signatures partially resolve this \textquotedblleft degeneracy\textquotedblright. This could be due to the difference
in the branching ratio of gluino decay into tops and stops.

\subsubsection{$\Delta S^{2}$ Analysis}

In this section we discuss the extent to which F-theory GUTs can be
distinguished from large A-term mSUGRA\ models. We find that although
single messenger models can typically be distinguished from such large
A-term scenarios, in the case of two and three messengers, the value
of $\Delta S^{2}$ does not effectively distinguish between all such
F-theory GUTs and large A-term mSUGRA (mSUGRA(LA)) models. As in the
case of the small A-term mSUGRA models, we have used the $10$ signatures
detailed in table \ref{siglist1} of subsection \ref{SIMULATION}. We find
that there is typically a small region at small and large $\Lambda$ where
it is possible to distinguish multiple messenger F-theory GUT models from mSUGRA(LA) models.
Although in many cases the value of $\Delta S_{(A)}^{2}$ between F-theory GUTs and mSUGRA(LA)
models is on the order of $10$ or more, other mSUGRA(LA) models can effectively
mimic the signatures of F-theory GUTs.
\begin{figure}[ptb]
\begin{center}
\includegraphics[
height=5.0929in,
width=6.2128in
]{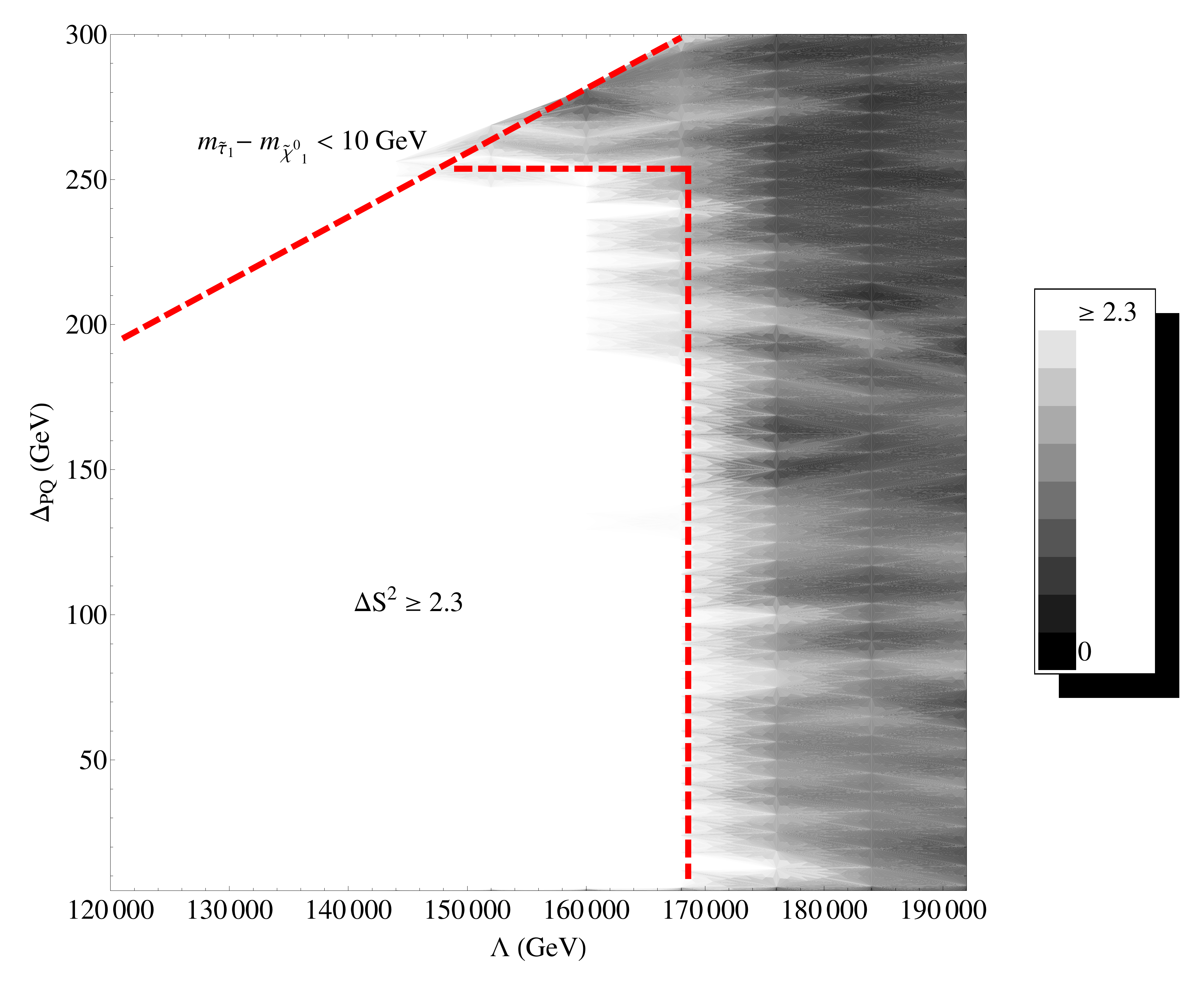}
\end{center}
\caption{Density plot of $\Delta S_{(A)}^{2}$ defined by the signature list
in table \ref{siglist1} of subsection \ref{SIMULATION} comparing the minimal value of a given
$N_{5}=1$ F-theory GUT model with a scan over large A-term mSUGRA models. The signals used are
obtained with $5$ fb$^{-1}$ of simulated LHC data. Here, we have used a rough notion of distinguishability based
on $99\%$ confidence and $10$ signals so that at $\Delta S_{(A)}^{2}>2.3$ we shall
say that two models are distinguishable. The plot shows the
minimal value of $\Delta S^{2}$ for a fixed F-theory GUT\ point. By
inspection, we see that just as for the small A-term case, single messenger
F-theory GUTs can be distinguished from mSUGRA\ models over much of the range
of interest. This is in contrast to the case of multiple messenger models
where the same signatures do not cleanly distinguish between F-theory and mSUGRA(LA) models. See
figures \ref{twosugrala} and \ref{thrsugrala} of Appendix F for the analogous plots in the multiple messenger case.}
\label{onesugrala}
\end{figure}

To quantify the extent to which F-theory GUTs can be distinguished
from such models, we have computed the value of the function:
\begin{equation}
S_{LA}\left(N_{5},\Lambda,\Delta_{PQ}\right)=\underset
{m_{\text{mSUGRA(LA)}}}{\min}\Delta S_{(A)}^{2}(m_{F},m_{\text{mSUGRA(LA)}})
\end{equation}
obtained by minimizing over all large A-term mSUGRA\ models of interest.
Figure \ref{onesugrala}\ shows that for $N_{5}=1$, typically $S_{LA}>2.3$.
On the other hand, figures \ref{twosugrala} and \ref{thrsugrala}
in Appendix F show that for $N_{5}=2$ and $N_{5}=3$, it is more
difficult to distinguish F-theory GUTs from large A-term scenarios.

\subsubsection{More on the Multiple Messenger Case}

\label{MOREON}

In the previous subsections we found that single messenger F-theory
GUT models are distinguishable from large A-term mSUGRA models, both
at the level of footprint plots and at a semi-quantitative level using
the measures $\Delta S^{2}$. On the other hand, the large A-term mSUGRA models
appear to more effectively mimic the multiple messenger F-theory GUTs
with a bino NLSP. In this subsection we discuss some further features
of such models which partially explain the observed overlaps in signatures.

\begin{figure}[ptb]
\begin{center}
\includegraphics[
height=3.98in,
width=6.4in
]{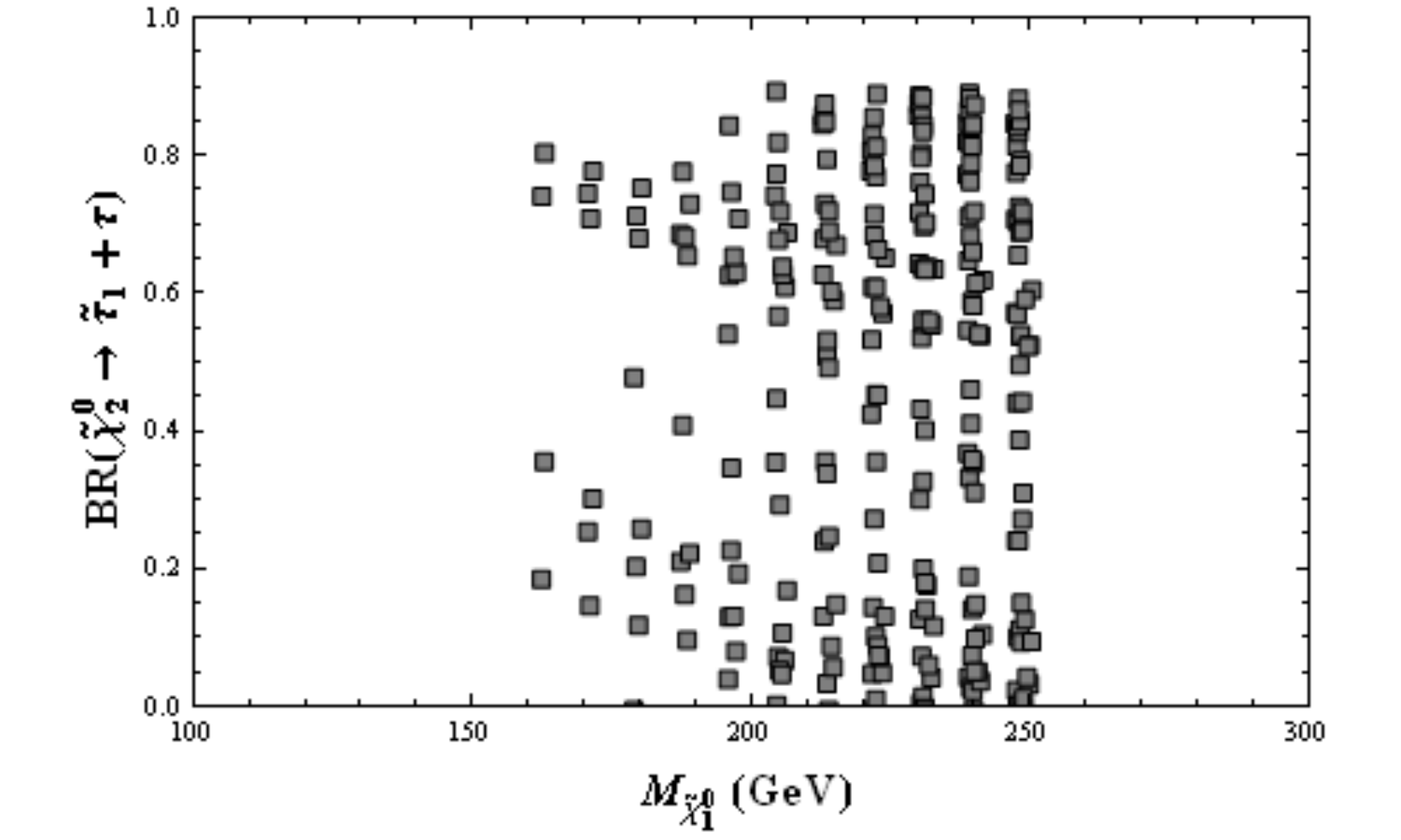}
\end{center}
\caption{Plot of second neutralino branching fraction $\widetilde{\chi}_{2}^{0}\rightarrow\widetilde{\tau}_{1}^{\pm}+\tau^{\mp}$
in large A-term mSUGRA models versus the LSP mass. The large
distribution of values indicates that such models can effectively
mimic the branching fraction present in other models, such as F-theory
GUTs. Note, however, that there is a gap in the lower left part of the distribution.
The relation between the mass of the stau and the two lightest neutralinos in this
region is similar to that of two and three messenger F-theory GUTs with a bino NLSP.}
\label{MSUGRA-LA-STAU-BR}%
\end{figure}

\begin{figure}[ptb]
\begin{center}
\includegraphics[
height=3.4272in,
width=5.719in
]{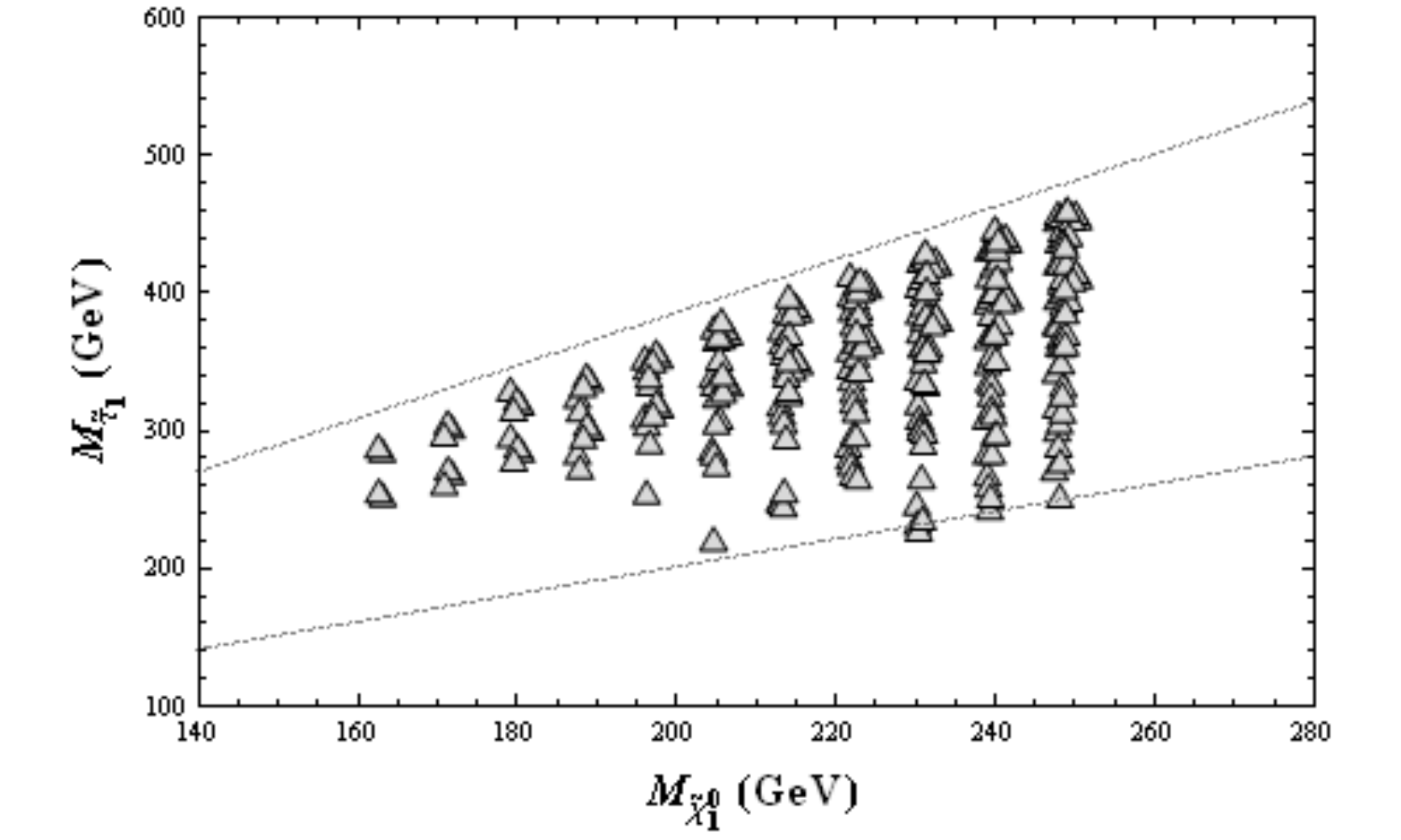}
\end{center}
\caption{Plot of the mass of the lightest stau in large A-term mSUGRA models versus
the LSP mass. The upper line in the plot corresponds to $m_{\tilde{\chi}_{2}^{0}}$, while
the lower one corresponds to $m_{\tilde{\chi}_{1}^{0}}$. By inspection,
the mass of the $\widetilde \tau_1$ effectively covers the entire range between the mass of
$\widetilde \chi_{1}^{0}$ and $\widetilde \chi_{2}^{0}$, so that such models can
effectively mimic the mass hierarchy present in F-theory GUTs.}
\label{MSUGRA-LA-MASS}
\end{figure}

Because of the similarity in the stau and neutralino masses, the decay
of $\widetilde{\chi}_{2}^{0}$ via an on-shell $\tilde{\tau}_{1}$ can enable
large A-term models to mimic the leptonic signals of F-theory GUTs.
Figure \ref{MSUGRA-LA-STAU-BR} shows the value of the LSP mass versus
the branching ratio $\tilde{\chi}_{2}^{0}\rightarrow\tilde{\tau}_{1}+\tau$ for a range of mSUGRA(LA) models.
The large variation in values implies that such models can effectively mimic the value of the
branching fraction present in F-theory GUTs. Further, although we have seen that
single messenger models can indeed be distinguished from mSUGRA(LA) models, some of the
parameters of such models such as the masses of the lightest neutralinos $\tilde{\chi}_{1}^{0}$
and $\tilde{\chi}_{2}^{0}$ are quite similar. Figure \ref{MSUGRA-LA-MASS} shows
that the mass of the lightest stau can essentially cover the entire range of values between
$m_{\tilde{\chi}_{1}^{0}}$ and $m_{\tilde{\chi}_{2}^{0}}$. Note, however, that at
smaller values of the neutralino mass, a gap exists in the range of values for the lightest stau mass.
Since multiple messenger F-theory GUTs with a bino NLSP typically have a $\tilde{\tau}_{1}$ which
is close in mass to $\tilde{\chi}_{1}^{0}$, this difference may lead to observable differences
between multiple messenger F-theory GUTs and mSUGRA(LA) models.

\begin{figure}[ptb]
\begin{center}
\includegraphics[
height=3.4272in,
width=5.719in
]{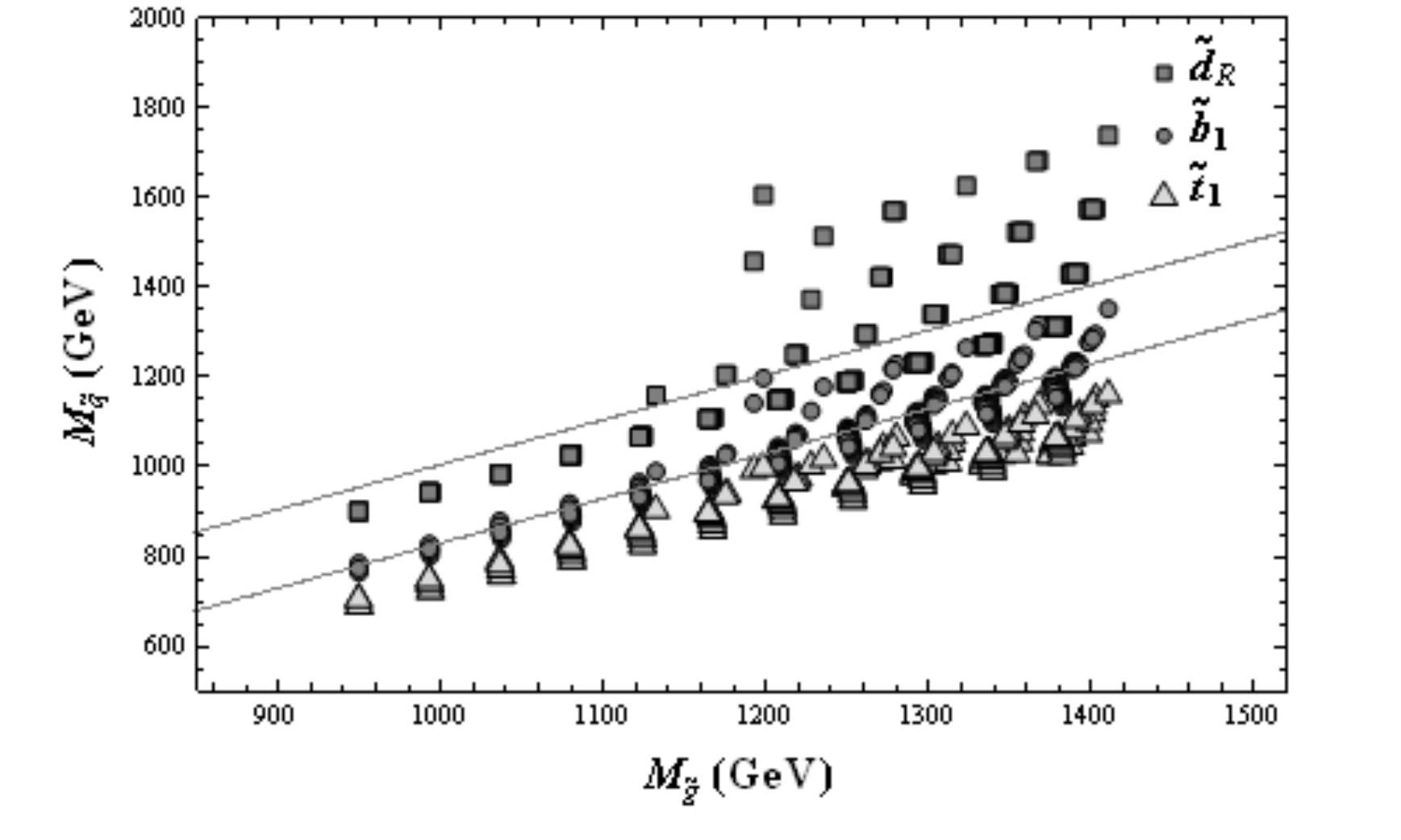}
\includegraphics[
height=3.4272in,
width=5.719in
]{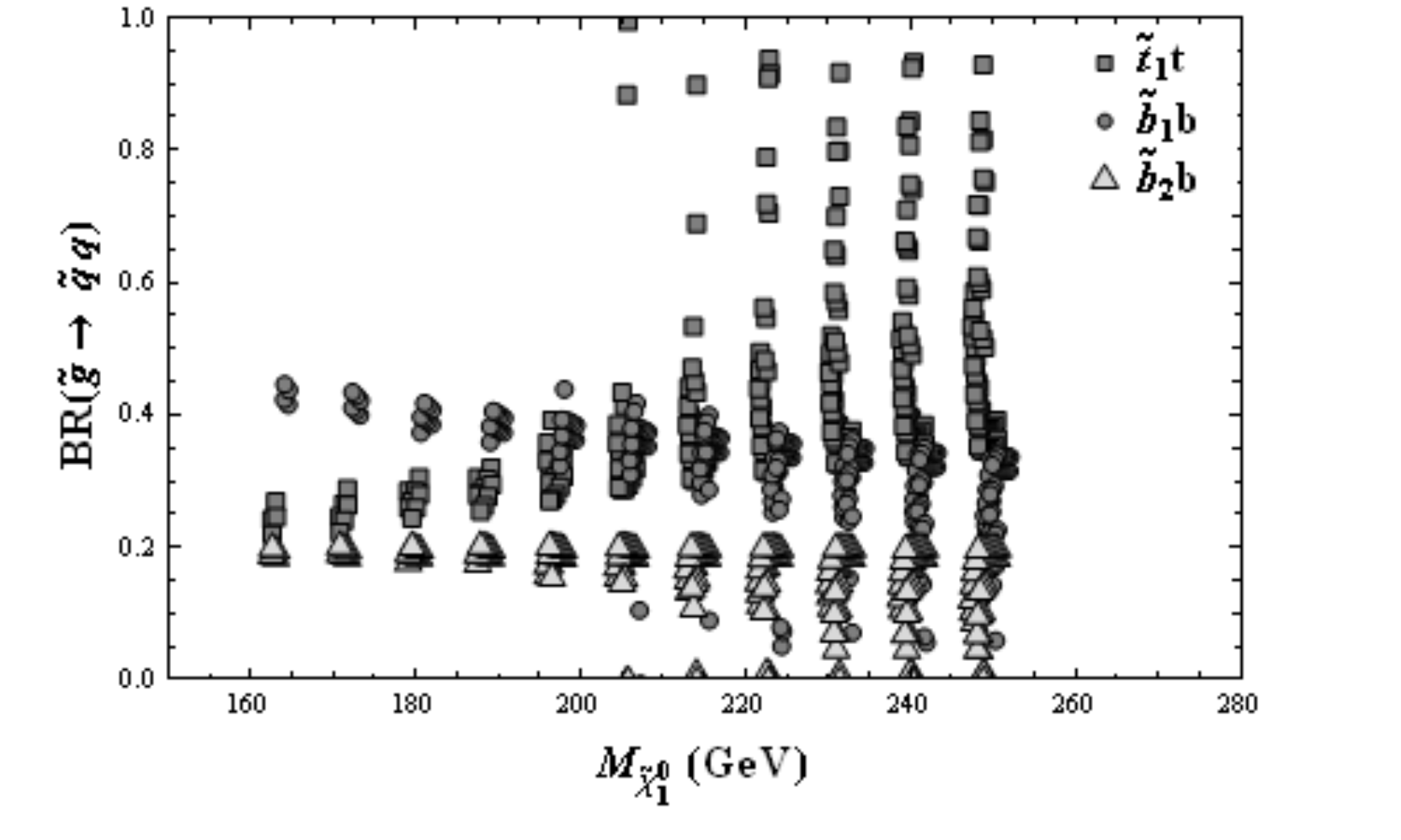}
\end{center}
\caption{Top figure: the squark mass distribution $(\tilde{t}_{1},\tilde{b}_{1},\tilde{d}_{R})$
in large A-term mSUGRA models. The upper line corresponds to $m_{\tilde{g}}$
and the lower line to $m_{\tilde{g}}-m_{t}$. Bottom figure: the
gluino branching fraction in large A-term mSUGRA models versus
the LSP mass in the decay channels $\widetilde{g}\rightarrow\widetilde{t}_{1}+t$,
$\widetilde{g}\rightarrow\widetilde{b}_{1}+b$, and $\widetilde{g}\rightarrow\widetilde{b}_{2}+b$.}
\label{MSUGRA-LA-GLUINO-SQUARK}
\end{figure}

Large A-term models can also mimic F-theory GUT signals controlled by
the decay of colored particles. In large mSUGRA(LA) models, the main decay channels for the gluino are
$\tilde{g}\rightarrow\tilde{t}_{1}t$, $\tilde{g}\rightarrow\tilde{b}_{1}b$ and $\tilde{g}\rightarrow\tilde{b}_{2}b$. See
figure \ref{MSUGRA-LA-GLUINO-SQUARK} for a plot of each of these branching ratios.
The total branching ratio into third generation quarks is around $90\%$.
Since the gluino decay is a two-body process, multi-jets plus missing $E_T$
signature should distinguish these mSUGRA models with single messenger
F-theory GUTs, as we have already seen. Indeed, this is consistent with the discussion in section \ref{FATLHC} that the gluino
decays via a three-body process in single messenger models. Note, however, that in two and three messenger F-theory GUTs,
the gluino decay is also a two body process and is quite similar to large A-term mSUGRA models.
In fact, the total branching ratio to third generation squarks and quarks is
respectively $100\%$ and $80\%$. This explains why simple b-jet counting
does not help in distinguishing between F-theory models and large A-term mSUGRA models.
Further note that for two messenger F-theory GUTs, there is a larger branching ratio to
$\tilde{t}_{1}t$, which is, however, not present in the three messenger case. At first, this
would seem to provide a promising class of processes to focus on since the increase in the
number of top quarks from gluino decay will leads to an increase in the number of $W$ bosons,
which would seemingly produce a sizable difference in the lepton signatures. Unfortunately,
the heavy neutralinos $\widetilde{\chi}_{3,4}^{0}$ are generally lighter
in F-theory models, and thus give rise to softer gauge bosons and subsequently produced leptons.
This leads to opposite effects on the lepton signatures compared to the large
$\tilde t_{1} t$ branching ratio. Therefore, at least in the case of low integrated luminosity,
distinguishing F-theory models with two and three messenger from large A-term mSUGRA models
appears to be much more challenging at the LHC.

\begin{figure}[ptb]
\begin{center}
\includegraphics[
height=3.4272in,
width=5.719in
]{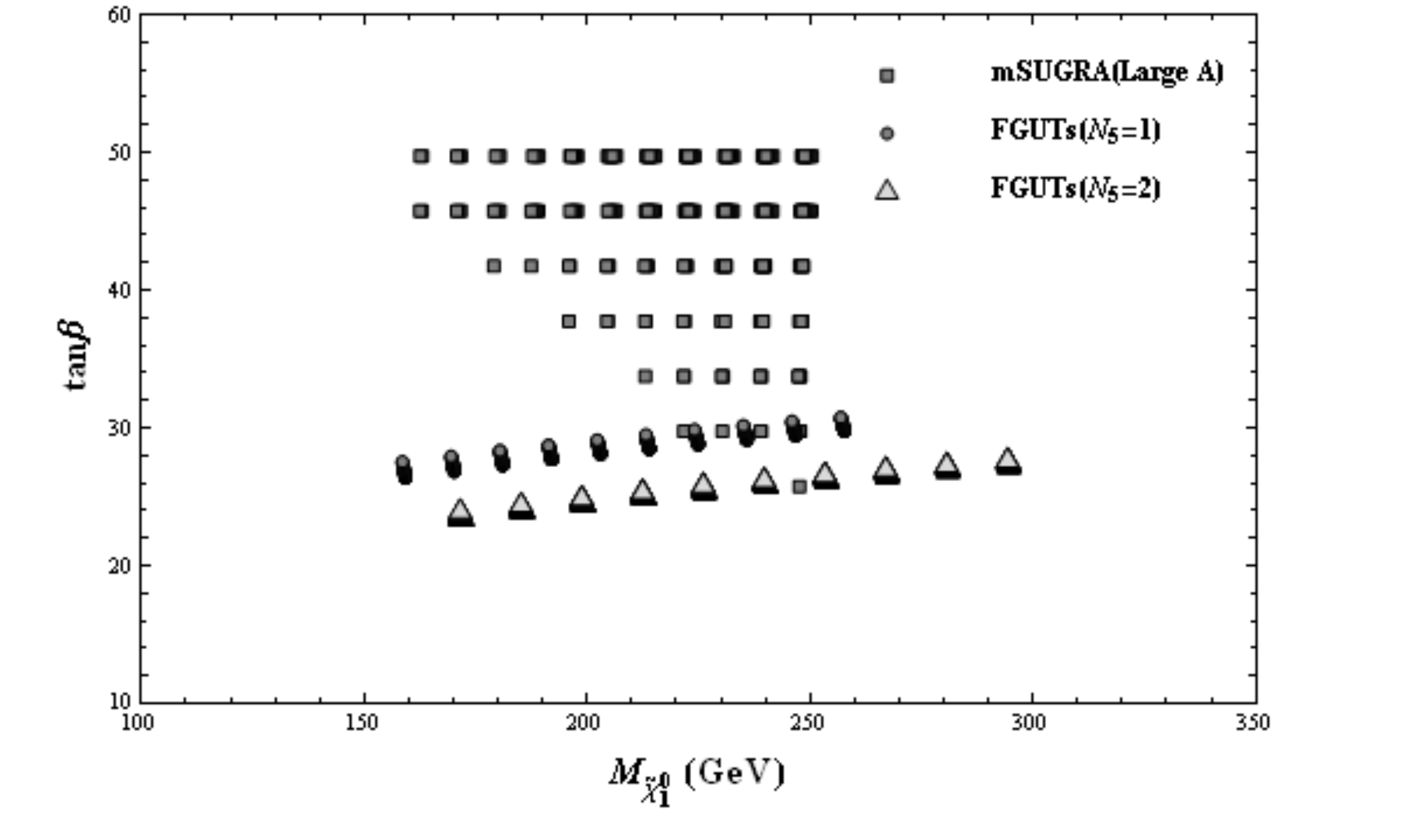}
\end{center}
\caption{Plot of $\tan\beta$ in large A-term mSUGRA models versus the mass of the LSP. This
is to be compared with the values of $\tan\beta$ in F-theory GUTs which have been
overlaid on the same plot as a function of $\Lambda$ (axis not shown). The $N_5=3$
case is omitted because it is similar to the $N_5=2$ case.
This plot shows that the value of $\tan\beta$ in large A-term mSUGRA models is typically larger
than that in F-theory GUTs. In particular, $\tan\beta$ enhanced processes could therefore potentially distinguish
between F-theory GUTs and such mSUGRA models.}
\label{MSUGRA-LA-TANB}
\end{figure}

To further distinguish between large A-term scenarios and F-theory
GUTs, increased luminosity will certainly help (see for example equation (\ref{LMIN}) of subsection \ref{SIMULATION}).
At the same time, one could look for signatures which are
sensitive to the $\tilde{t}_{1}t$ channel of gluino decay.
In addition, $\tan\beta$ enhanced signatures could also be useful,
since, as figure \ref{MSUGRA-LA-TANB} shows, large A-term mSUGRA
models typically have larger $\tan\beta$ than F-theory GUTs. An effective observable may come from
rare B-decays such as $B_s \rightarrow \mu^{+}\mu^{-}$ where the additional
gluino contribution to the decay rate can be significantly enhanced
by a factor proportional to $\tan^{6}\beta$ in the large $\tan\beta$ regime
(see \cite{Carena:2008ue} for a recent study).
However, a full analysis of these possibilities is beyond the scope of this paper, and
we leave this issue for future investigations.

Beyond collider signatures, astrophysical probes could provide another
means by which to distinguish large A-term mSUGRA models from multiple messenger F-theory GUTs.
In mSUGRA models, the lightest neutralino provides a natural dark matter candidate.
This is in contrast to the situation in F-theory GUTs
where the neutralino is unstable, eventually decaying
to a gravitino LSP. The difference in the nature of the LSP will affect
the dark matter relic density. For example, in the bino LSP case,
generating an acceptable dark matter relic density from binos requires a
light stau which is almost degenerate in mass with the LSP. This further
criterion would exclude most of the large A-term mSUGRA models in our study, thus providing
better separation between F-theory GUTs and large A-term models. In
addition, both direct and indirect dark matter detection experiments can
potentially see evidence of a neutralino LSP, but not gravitinos.
Therefore, non-collider signature could provide an additional means
by which to distinguish F-theory GUTs and mSUGRA models.

\subsection{F-theory Versus Low Scale mGMSB}

In this subsection we compare F-theory GUTs with minimal GMSB
scenarios with a low messenger scale, deferring a comparison
with high messenger scale scenarios to section \ref{DETFth}.
Returning to the discussion near (\ref{Square}), recall that a stau NLSP\ low scale mGMSB\ model
can potentially mimic some of the signatures of F-theory GUTs with
a bino NLSP. Indeed, all other scenarios have qualitatively distinct
behavior, and so we shall focus on this one remaining case.

\begin{figure}[ptb]
\begin{center}
\includegraphics[
height=5.5625in,
width=5.7017in
]{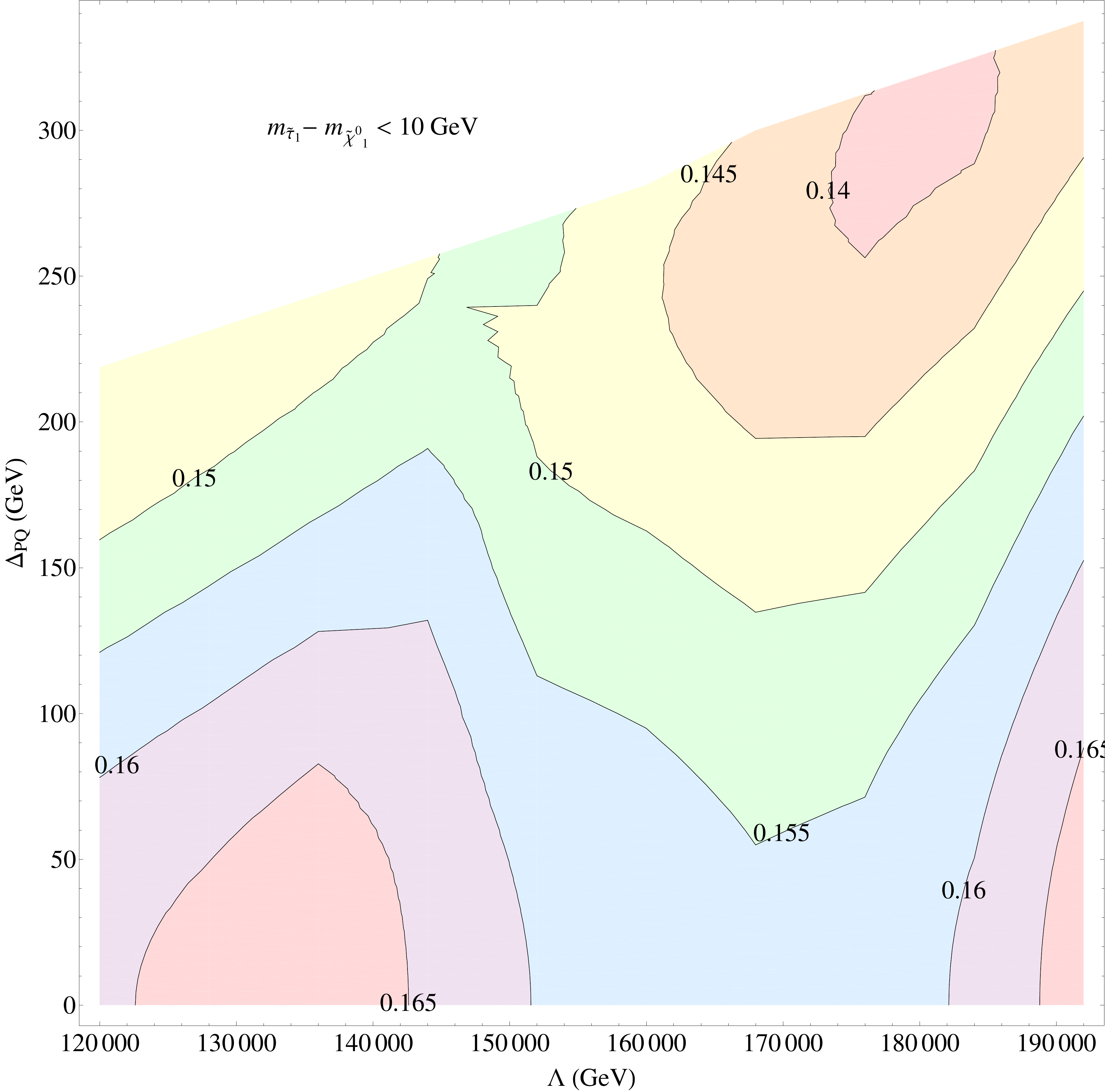}
\end{center}
\caption{Contour plot of the value of $\Delta P^{2}$ obtained by fixing a
particular value of $\Lambda$ and $\Delta_{PQ}$ of an F-theory GUT model with
$N_{5}=1$, and minimizing with respect to all low scale minimal GMSB\ models
with a stau\ NLSP. We adopt a rough criterion for theoretical
distinguishability specified by the requirement that $\Delta P^2 > 0.01$. By inspection, $\Delta P^2$ is greater than
$0.1$, indicating that such models are distinguishable at the theoretical
level from F-theory GUTs. See figure \ref{softspecthrgmsbstau} in Appendix F
for a similar plot for $N_{5}$ = 3 F-theory GUTs.}%
\label{softspeconegmsbstau}%
\end{figure}

Even though the presence of signatures with missing $E_{T}$ can potentially
be mimicked, there are still important differences in both the soft
parameters, and collider signatures. As in our comparison with mSUGRA\ models, we have scanned over low
scale mGMSB\ models with a stau NLSP (see Appendix B for a description of this scan),
and computed the value of $\Delta P^{2}$ between these models and F-theory GUTs. Minimizing over all such
models, we have determined the value of
\begin{equation}
P_{LO}\left(  N_{5},\Lambda,\Delta_{PQ}\right)  =\underset{m_{\text{mGMSB(LO)}%
}}{\min}\Delta P^{2}(m_{F},m_{\text{mGMSB(LO)}})\text{.}%
\end{equation}
Figure \ref{softspeconegmsbstau} and figure \ref{softspecthrgmsbstau}
in Appendix F show contour plots of $P_{LO}$ as a function of $\Lambda$
and $\Delta_{PQ}$. These plots illustrate that
\begin{equation}
P_{LO}\gtrsim0.1\text{.}
\end{equation}

It is interesting to note that although both classes of models share
the same mechanism for the mediation of supersymmetry breaking, the
soft terms of large A-term mSUGRA\ models can more effectively mimic
the soft terms of F-theory GUTs. Having established this distinction,
we now discuss footprints which distinguish between the signatures
of these two classes of models.

\subsubsection{Footprint Analysis}

\begin{figure}[ptb]
\begin{center}
\includegraphics[
height=4.12in,
width=6.4091in
]{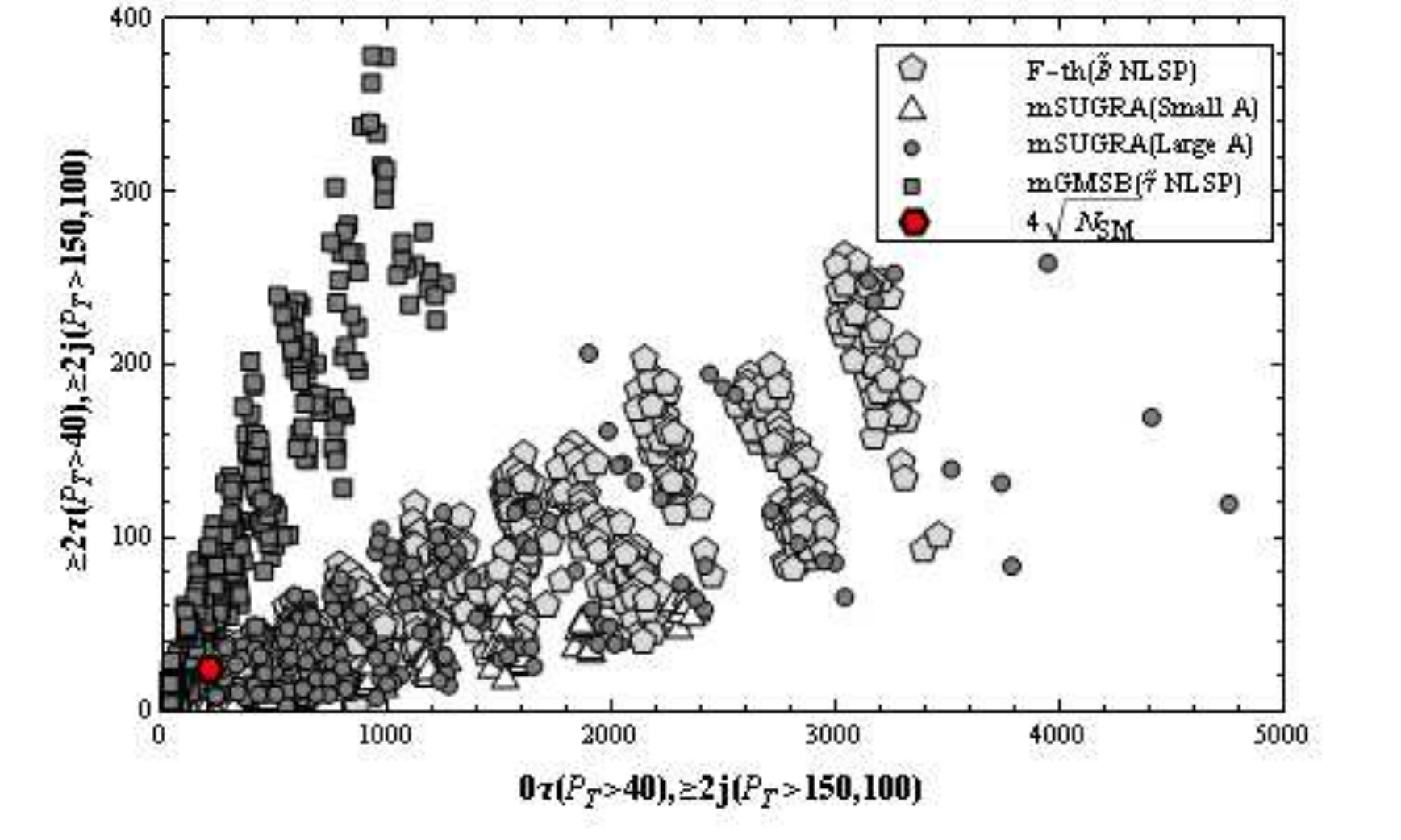}
\end{center}
\caption{Footprint of LHC signatures (without SM background) for
distinguishing F-theory GUTs and low scale mGMSB models with a stau NLSP with
$5$~fb$^{-1}$ integrated luminosity. The red circle denotes the $4\sigma$
deviation from the SM background.}
\label{sig-2}
\end{figure}

The footprint of low scale mGMSB models with a stau NLSP can be easily
separated from that of F-theory GUTs by using $0\tau$ and $\ge2\tau$
signatures as seen in figure \ref{sig-2}. As explained in section
\ref{MIMIC}, this is because there exist long decay chains where the lightest
neutralinos decay into a stau and then to a gravitino, leading to
events with many taus.

\subsubsection{$\Delta S^{2}$ Analysis}

\begin{figure}[ptb]
\begin{center}
\includegraphics[
height=5.0929in,
width=6.2128in
]{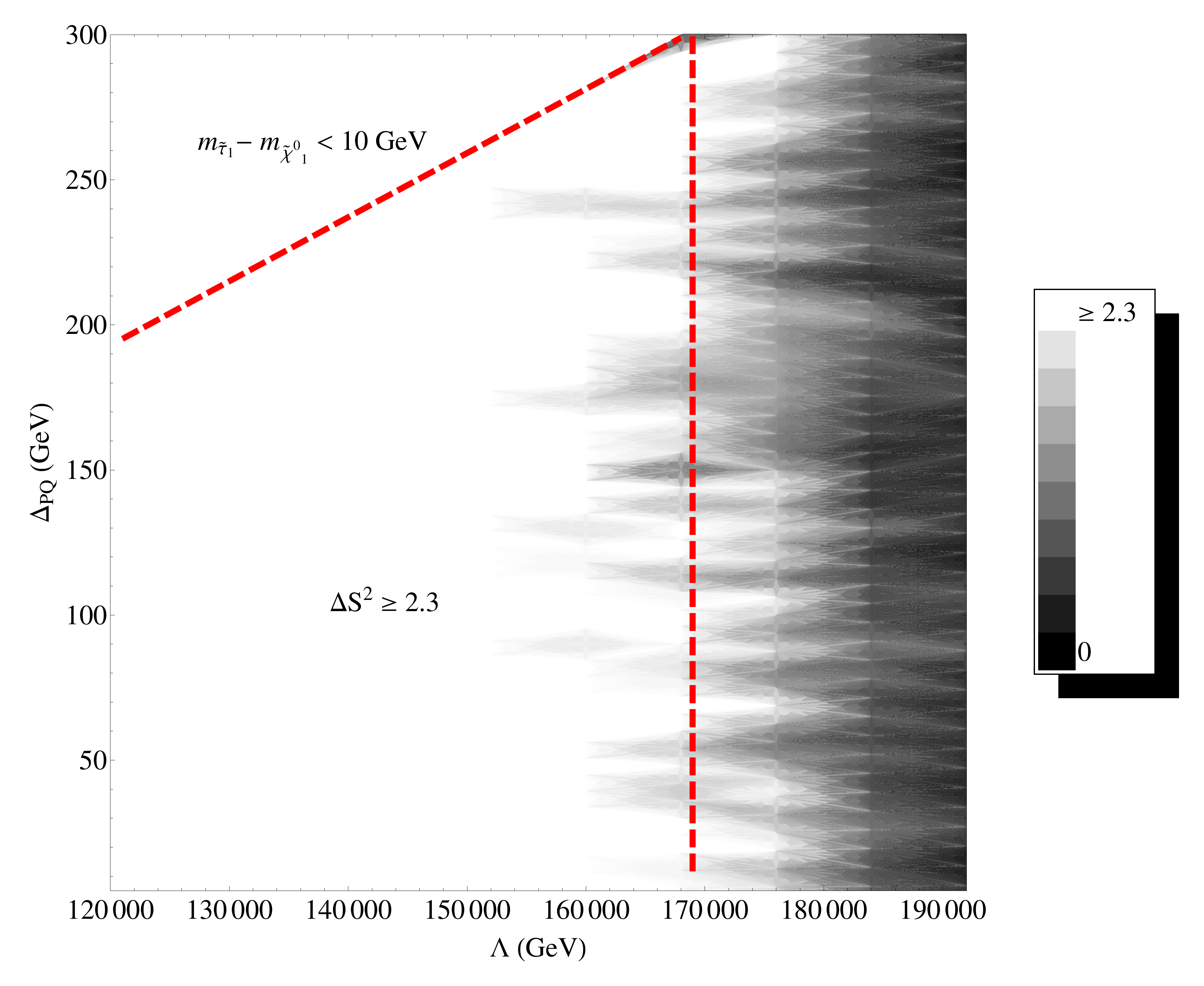}
\end{center}
\caption{Density plot of $\Delta S_{(A)}^{2}$ defined by the signature list
in table \ref{siglist1} of subsection \ref{SIMULATION} comparing the minimal value of a given
$N_{5}=1$ F-theory GUT model with a scan over low scale mGMSB models with a stau NLSP. The signals used are
obtained with $5$ fb$^{-1}$ of simulated LHC data. Here, we have used a rough notion of distinguishability based
on $99\%$ confidence and $10$ signals so that at $\Delta S_{(A)}^{2}>2.3$ we shall
say that two models are distinguishable. By inspection, much of the scanned
region of F-theory GUTs is distinguishable from such models.}%
\label{onegmsbstau}%
\end{figure}

Having seen that the footprints of F-theory GUTs are quite distinct from low scale mGMSB models,
we next compute using the same $10$ signatures detailed in table \ref{siglist1} of subsection \ref{SIMULATION} the value of
$\Delta S_{(A)}^{2}$ between these two classes of models. Minimizing over all such low scale models, we define
the function:
\begin{equation}
S_{LO}\left(  N_{5},\Lambda,\Delta_{PQ}\right)  =\underset{m_{\text{mGMSB(LO)}%
}}{\min}\Delta S_{(A)}^{2}(m_{F},m_{\text{mGMSB}})\text{.}%
\end{equation}
In this case, we find that over the entire range of parameters scanned for
$\Lambda$ and $\Delta_{PQ}$,
\begin{equation}
S_{LO}\left(N_{5},\Lambda,\Delta_{PQ}\right)\geq2.3
\end{equation}
when $N_{5}=2,3$. This is the threshold for distinguishability at $99\%$ confidence,
and so we find that it is relatively easy to distinguish between these models.
Moreover, figure \ref{onegmsbstau} shows that even in
the case of single messenger F-theory GUT models, much of
the range of F-theory GUTs is distinguishable using these same signatures.
This illustrates that in general, we can expect to distinguish between
F-theory GUTs and low scale mGMSB\ models.

%EndExpansion

\section{Determination of F-theory Parameters} \label{DETFth}

In this section we determine the extent to which two distinct
F-theory GUT\ models can be distinguished from one another. Note
that since high messenger scale mGMSB\ models correspond to a subclass
of F-theory GUTs with $\Delta_{PQ}=0$, performing this analysis will
also address whether F-theory GUTs can be distinguished from a minimal
GMSB\ model with a high messenger scale.

\begin{figure}[ptb]
\begin{center}
\includegraphics[
height=5.5625in,
width=5.7017in
]{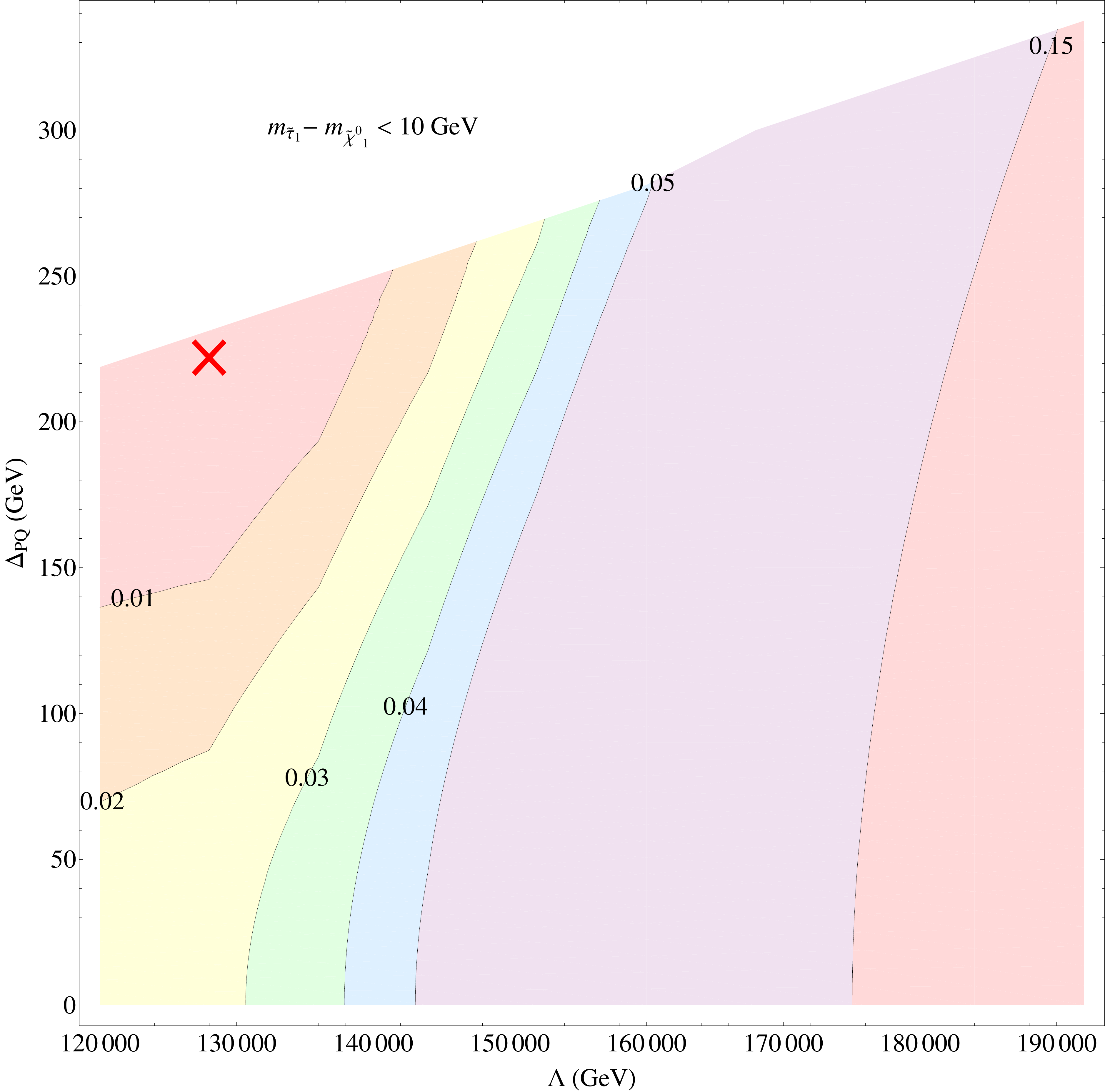}
\end{center}
\caption{Contour plot of $\Delta P^{2}$ between a fixed F-theory GUT\ model with
$N_{5}=1$, $\Lambda=1.28\times10^{5}$ GeV and $\Delta_{PQ}=222$ GeV
(indicated by red X), and single messenger F-theory GUT models. We adopt
a rough criterion for theoretical distinguishability specified by the requirement that $\Delta P^2 > 0.01$. This figure
shows that $\Delta P^{2}$ minimizes in the vicinity of the representative
point, and that in particular, there is no degeneracy in the soft terms in a
scan over single messenger models. However, figure \ref{softspecexampthrfth} of Appendix F shows that in plots of
$\Delta P^{2}$ between this point and three messenger models, there are still regions
where a local minimum is potentially possible, although this value is greater than $0.02$.}%
\label{softspecexamponefth}%
\end{figure}

Using the same $\Delta P^{2}$  measure of theoretical distinguishability based on soft terms utilized earlier, we have computed the value of
$\Delta P^{2}$ between a fixed F-theory GUT specified by the parameters
$N_{5}^{(0)},\Lambda^{(0)},\Delta_{PQ}^{(0)}$, and the class of all
F-theory GUTs:
\begin{equation}
P_{FTH}\left(N_{5},\Lambda,\Delta_{PQ}\right)=\Delta P^{2}(m_{F}
(N_{5}^{(0)},\Lambda^{(0)},\Delta_{PQ}^{(0)}),m_{F}(N_{5},\Lambda,\Delta_{PQ}))\text{.}
\end{equation}

We find that for each $N_{5}$, $P_{FTH}$ achieves a local minimum
only for a small strip of values of $\Lambda$, which typically extends
over a range of values for $\Delta_{PQ}$. However, the absolute minimum
of $P_{FTH}$ is always smaller when $N_{5}=N_{5}^{(0)}$. Typically,
single messenger models are more easily distinguished from multiple
messenger models since the spectrum and soft terms of single messenger F-theory GUTs are distinct.
For example, in single messenger models, some of the squarks are
typically heavier than the gluino, whereas in multiple messenger
models, the gluino is always heavier. Due to the similarities in the
sparticle spectrum, two and three messenger F-theory GUTs can exhibit similar behavior.

\begin{figure}[ptb]
\begin{center}
\includegraphics[
height=5.5625in,
width=5.7017in
]{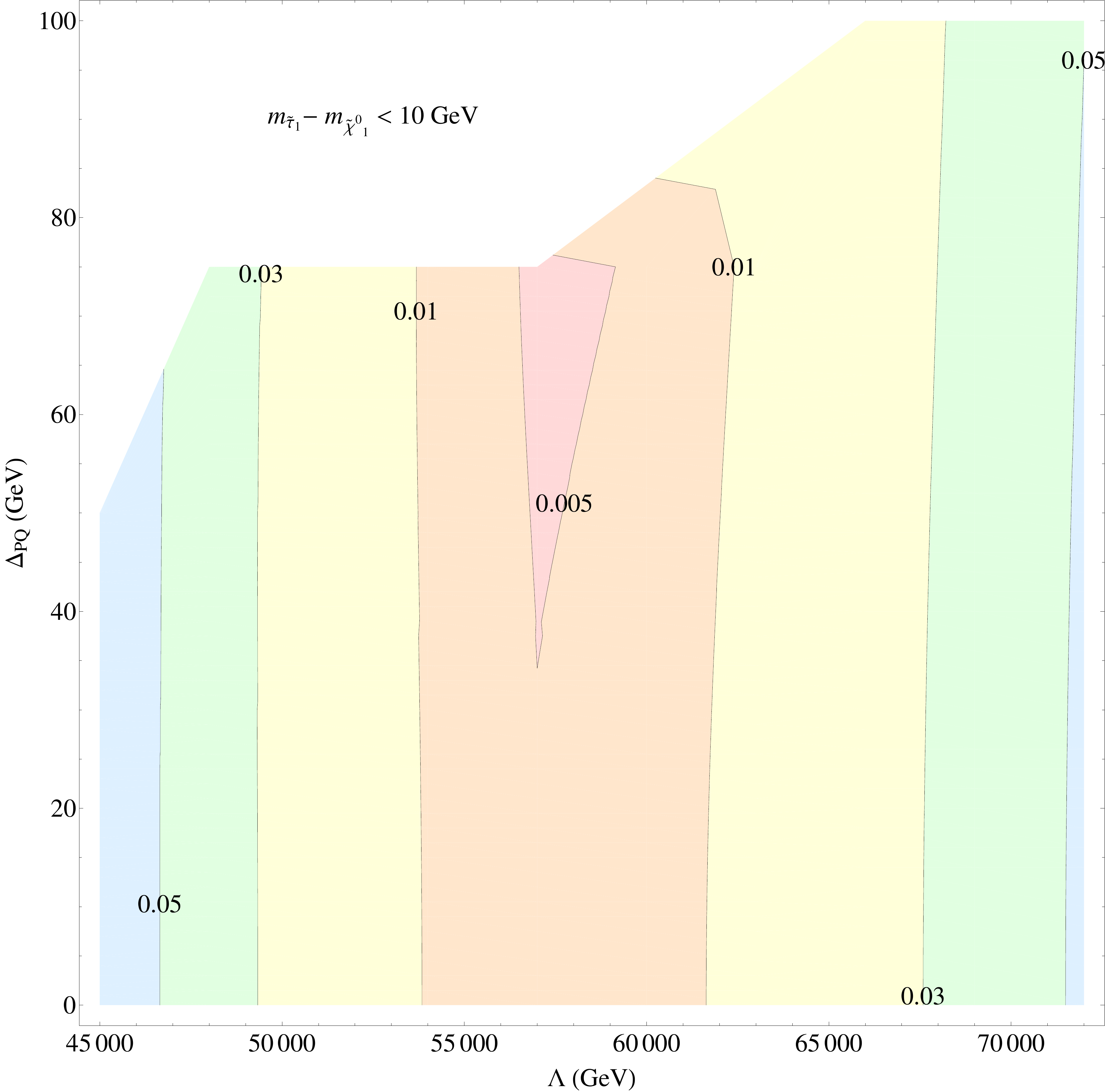}
\end{center}
\caption{Contour plot of $\Delta P^{2}$ between a fixed F-theory GUT\ model with
$N_{5}=2$, $\Lambda=8\times10^{4}$ GeV and $\Delta_{PQ}=104$ GeV and
three messenger F-theory GUT models. This figure
shows that even though the number of messengers are different, the value of
$\Delta P^{2}$ can minimize to a value below our rough criterion for
distinguishability given by $0.01$. Note, however, that although
$\Delta P^2 < 0.01$, the range of values for which $\Delta P^2 < 0.005$ is far smaller, indicating that
at least in principle, two and three messenger models are distinguishable at a theoretical level.}%
\label{softspecexampcthrsec4}%
\end{figure}

Fixing a representative single messenger model with
$\Lambda^{(0)} = 1.28 \times 10^5$ GeV and $\Delta^{(0)}_{PQ} = 222$ GeV as the \textquotedblleft LHC point\textquotedblright,
figure \ref{softspecexamponefth} shows that in a scan over all single
messenger models, there is a single local minimum for $\Delta P^{2}$.
In addition, while there is certainly a similar local minimum present
in the scan over three messenger model shown in figure \ref{softspecexampthrfth}
of Appendix F, the minimal value of $\Delta P^{2}$ in such cases is bounded
below by roughly $0.03$, so that with our rough criterion, such models are distinguishable. Similar considerations
apply for other representative \textquotedblleft LHC point\textquotedblright\ with a single messenger.
By contrast, figure \ref{softspecexampcthrsec4} illustrates that when
the \textquotedblleft LHC point\textquotedblright\ corresponds to a two messenger model
with $\Lambda^{(0)} = 8 \times 10^4$ GeV and $\Delta^{(0)}_{PQ} = 104$ GeV, there exists a strip of three messenger models with $\Delta P^2 < 0.01$, which falls below the rough bound we have adopted for distinguishability. Even so, it is important to note that at no point does $\Delta P^2$ appear to vanish. With this in mind, the fact that $\Delta P^2$ can be small should be viewed as a rough guide for what to expect when searching for discriminating signatures.

In the remainder of this section we investigate the extent to which
different F-theory GUTs generate distinguishable experimental signatures. To
this end, in subsection \ref{SIGLIST} we first list candidate signatures
of interest which have strong dependence on the parameters of F-theory
GUTs. Using these signatures, we compute the corresponding value of
$\Delta S^{2}$ between a fixed \textquotedblleft LHC\ point\textquotedblright\ and
various F-theory GUTs. In particular, we show that this type of analysis
is capable of determining both $N_{5}$ and $\Lambda$, and moreover,
can distinguish between scenarios with small and large PQ\ deformation.
Restricting to the case of single messenger F-theory GUTs and a
particular value of $\Lambda$, we next compute the value
of $\Delta S^{2}$ between models with different values
of $\Delta_{PQ}$. We find that at $5$~fb$^{-1}$ of simulated data,
$\Delta_{PQ}$ can be determined up to an uncertainty of $\sim \pm 80$~GeV, while at $50$
fb$^{-1}$, this improves to an uncertainty of $\sim \pm 10$~GeV. In particular,
this shows that the LHC is indeed sensitive, albeit indirectly, to
string scale physics! Finally, although outside the main focus of this paper,
in Appendix E we consider the sensitivity of the endpoint of the ditau
invariant mass distribution as a function of $\Delta_{PQ}$.

\subsection{Signature List \label{SIGLIST}}

\begin{figure}[ptb]
\begin{center}
\includegraphics[
height=3.8in,
width=5.91131in
]{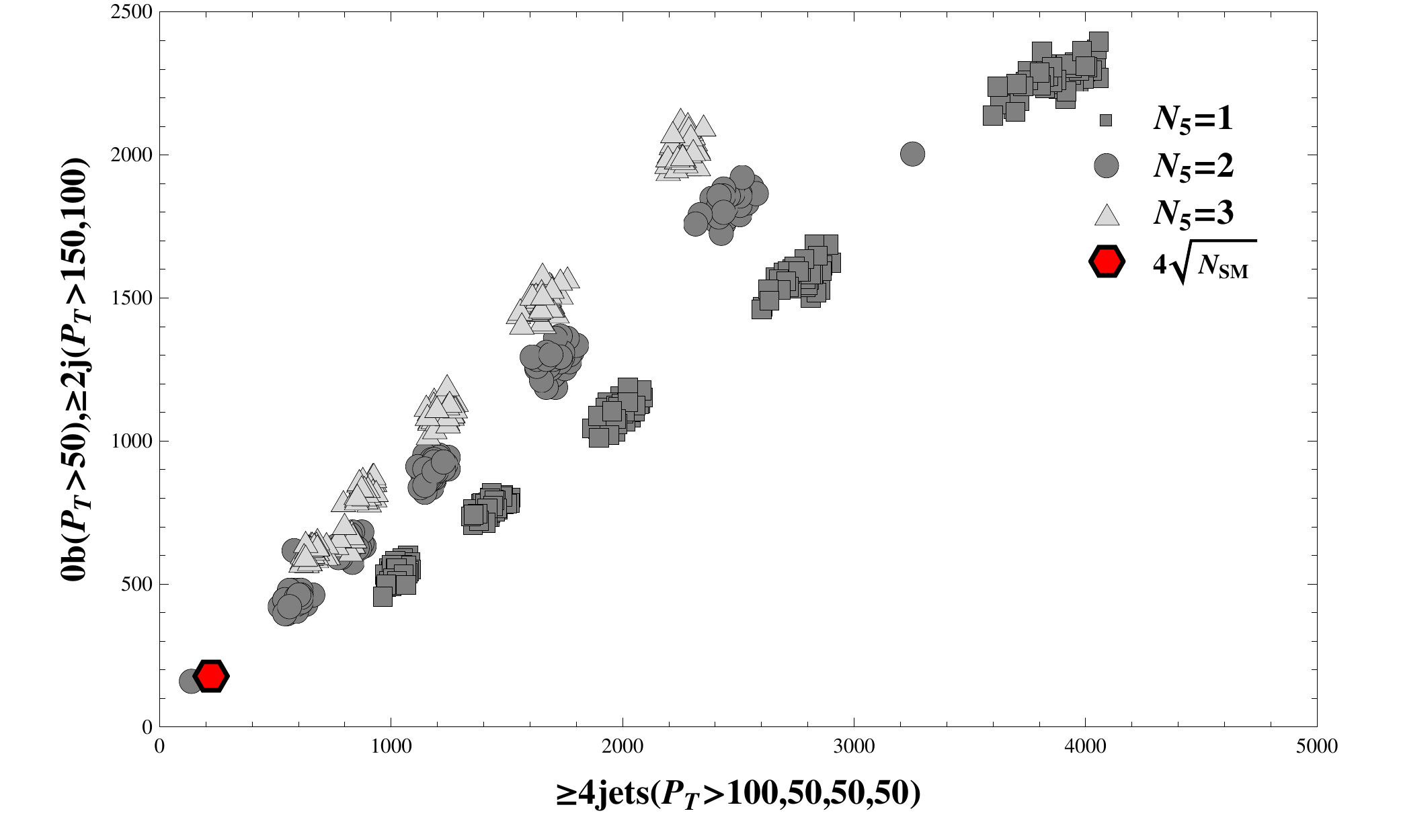}
\includegraphics[
height=3.8in,
width=5.91131in
]{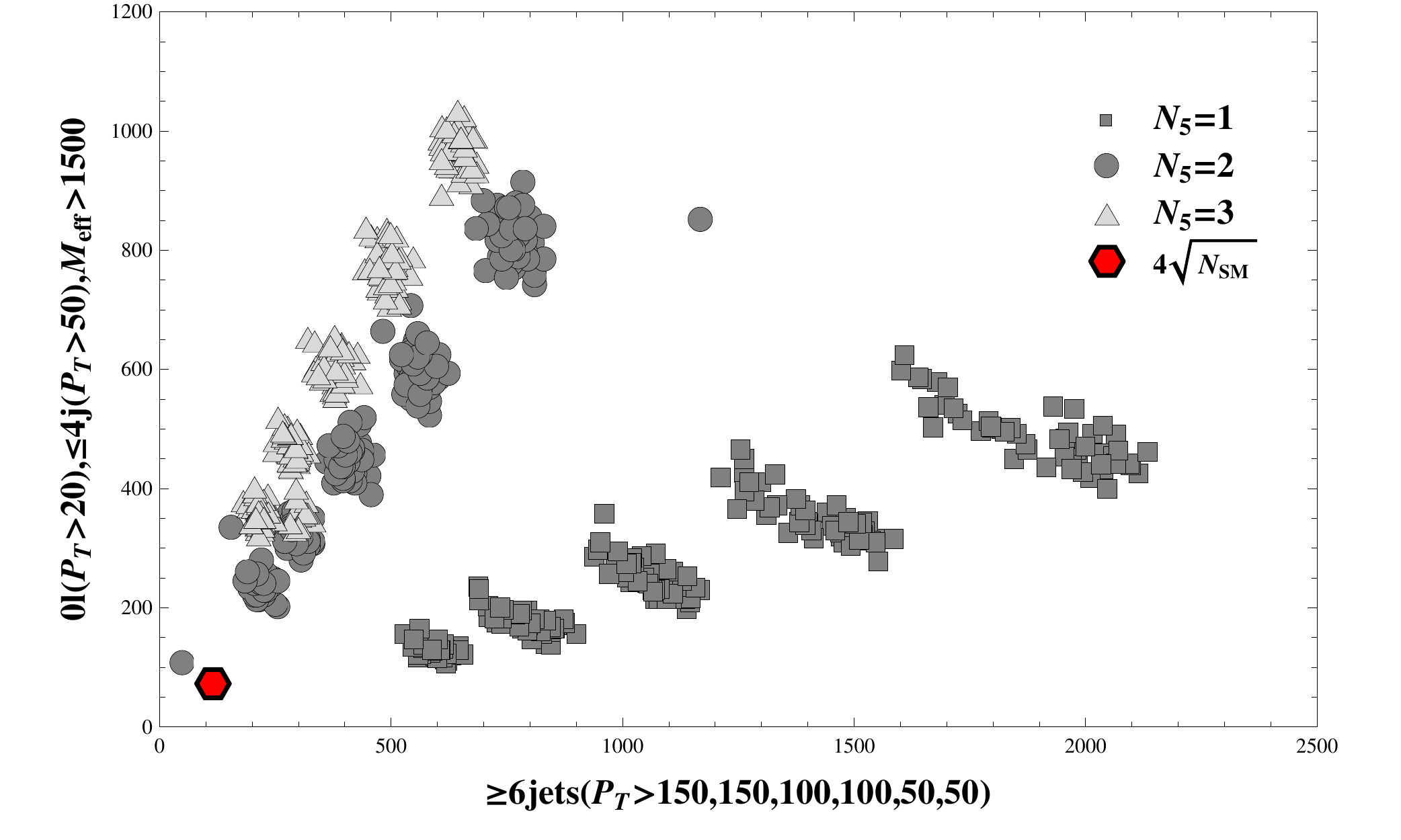}
\end{center}
\caption{Footprint of F-theory models with different $N_{5}$ with $5$~fb$^{-1}$ of integrated
luminosity. These plots shows the footprint can distinguish
F-theory based models with different numbers of messengers $N_{5}$.}
\label{sig-N5}
\end{figure}

\begin{figure}[ptb]
\begin{center}
\includegraphics[
height=3.8in,
width=5.91131in
]{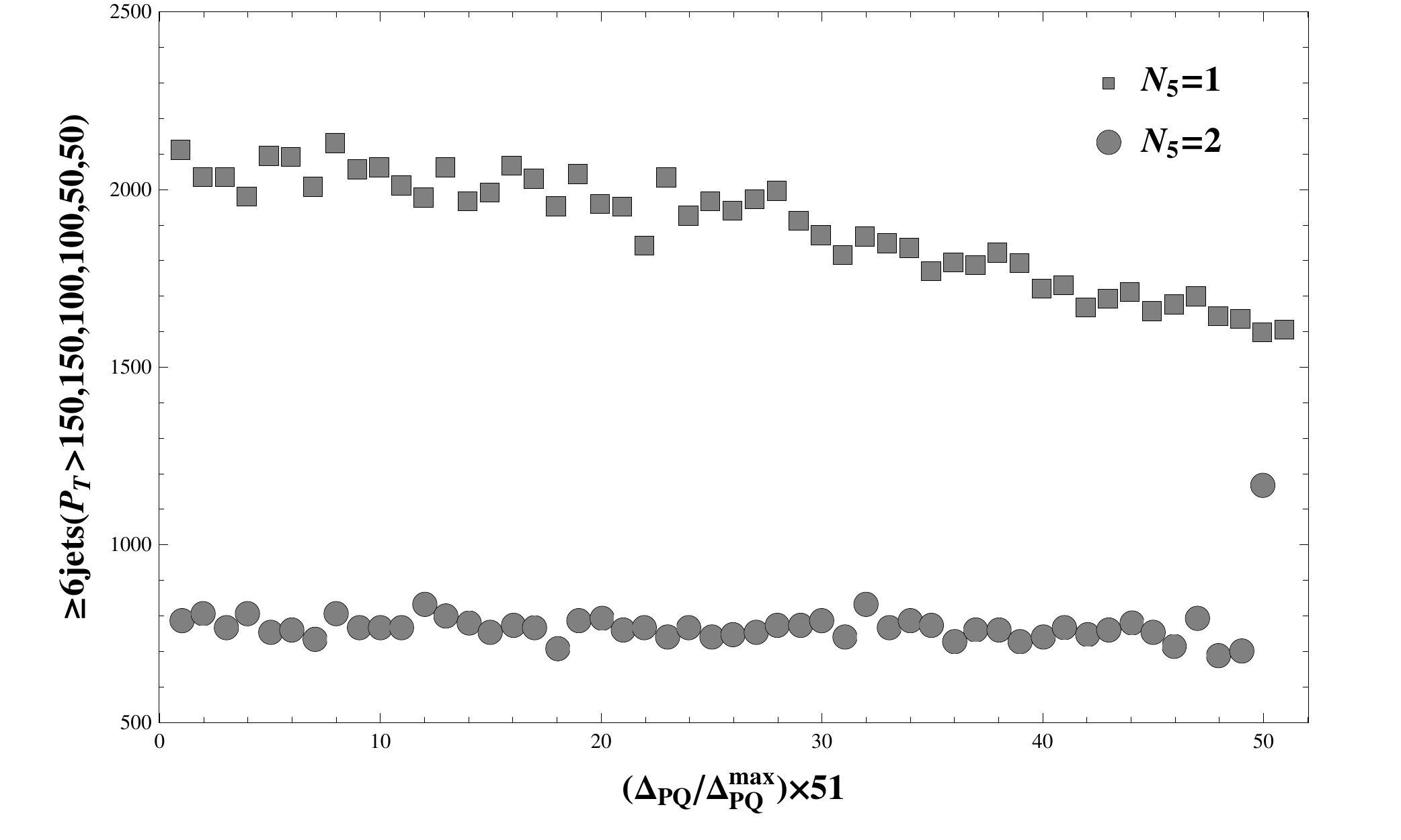} \includegraphics[
height=3.8in,
width=5.91131in
]{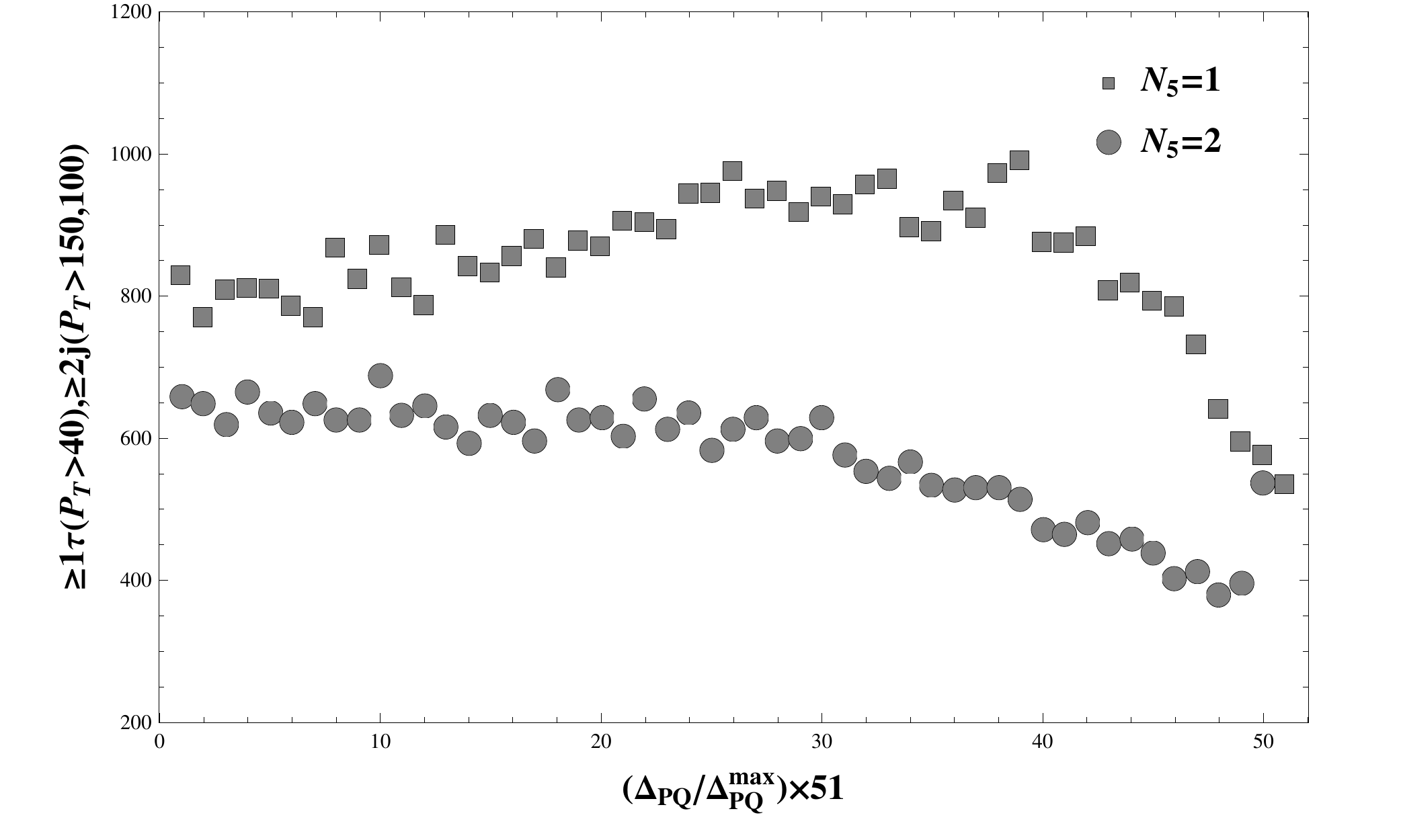}
\end{center}
\caption{LHC signatures as functions of $\Delta_{PQ}$ for F-theory GUTs with
($N_{5}=1$,$\Lambda=1.2\times10^{5}$~GeV) and ($N_{5}=2$,$\Lambda=6.5\times10^{4}$~GeV).
The signatures are simulated with $5$~fb$^{-1}$ of integrated luminosity.
These plots show that the LHC signatures are sensitive to the ``stringy''
parameter $\Delta_{PQ}$.}
\label{sig-deltaPQ-1}
\end{figure}

We now proceed to discuss signatures which are especially sensitive
to $N_{5}$, $\Lambda$ and $\Delta_{PQ}$. As explained in
subsection \ref{SIMULATION}, we have generated a total of $103$ LHC signatures
for all $1530$ F-theory GUT models. However, not all of these signatures
are sensitive to changes in the parameters of F-theory GUTs. In some cases, either poor statistics,
or a correlation in seemingly different signatures can obstruct the use of these observables. A general discussion
of selecting signatures can be found in reference \cite{Altunkaynak:2009tg}. As explained there,
there exists some optimal choice and number of signatures which
minimize the luminosity needed for distinguishing between two models. Although
a systematic optimization is certainly useful, it goes beyond the scope
of this paper. For our study, we will instead consider a manual selection
of signatures by examining the footprint of F-theory GUTs using all
possible pairs of signatures. For example, to find the signatures
sensitive to $N_{5}$, we examined all possible two-dimensional plots constructed
out of $103$ signatures for models with different $N_{5}$. We then
selected by examining such plots a class of signatures which do not
appear to share any significant correlations. One such
set of signatures which are especially helpful in a footprint analysis are
shown as signature list B in table \ref{siglist2}. Figure \ref{sig-N5}
illustrates an example of some signatures which distinguish between
single and multiple messenger models. Notice that the one messenger
models are more easily separated from the two and three messenger models.
This is consistent with the fact explained in section \ref{FATLHC} that these models have different
gluino and squark decay topologies. Similarly, we have also find a set of signatures
referred to as signature list C which are sensitive to the PQ deformation
parameter $\Delta_{PQ}$. Figure \ref{sig-deltaPQ-1} shows two signatures
as functions of $\Delta_{PQ}$. Finally, in our $\Delta S^2$ analysis,
we have used signature list D given in table \ref{siglist3}.

\begin{table}
\begin{center}
\begin{tabular}{|c|c|}
\hline
 & Signature List B \\
\hline
\hline
1  & $\ge4$ jets($P_{T}>100,50,50,50$)\\
\hline
2 & $\ge6$ jets($P_{T}>150,150,100,100,50,50$)\\
\hline
3 & $0b(P_T>50),\ge 2$jets($P_T>150,100$)\\
\hline
4 & $\ge 2b(P_T>50),\ge 2$jets($P_T>150,100$)\\
\hline
5 & $0l(P_T>20),\le4$jets($P_T>50$), $M_{eff}>1500$\\
\hline
\hline
& Signature List C \\
\hline
1 & $\ge6$jets($P_T > 150,150,100,100,50,50$)\\
\hline
2 &  $\ge1\tau$($P_T > 40$),$\ge2$jets($P_T>150,100$)\\
\hline
3 & $\ge1l$($P_T > 40$),$\ge4$jets($P_T>100,50,50,50$)\\
\hline
4 & $0\tau$($P_T > 10$),$\ge 4$jets($P_T>100,50,50,50$)\\
\hline
5 & $\ge1\tau$($P_T <40$),$\ge2$jets($P_T > 70$)\\
\hline
6  & $0l$($P_T > 20$),$\le4$jets($P_T>50$), $\ds{\not}E_T>500$\\
\hline
\end{tabular}
\end{center}
\caption{The collection of signatures which are effective for distinguishing models with distinct
values of $N_{5}$ (List B) and $\Delta_{PQ}$ (List C) using footprint plots.
All cuts are in units of GeV. Table \ref{siglist3} contains an
alternate set of signatures which are especially helpful in
distinguishing models using the $\Delta S^2$ measure.}
\par
\label{siglist2}
\end{table}

\begin{table}
\begin{center}
\begin{tabular}{|c|c|}
\hline
 & Signature List D\\
\hline\hline
 $1$ & $\ge 4$jets($P_T>100,50,50,50$)\\
\hline
$2$ & $\ge 4$jets($P_T>250,250,100,100$)\\
\hline
$3$ & $\ge 2$jets($P_T>350,350$)\\
\hline
$4$ & $\ge 6$jets($P_T>150,150,100,100,50,50$)\\
\hline
$5$ & $0\tau$($P_T>20$),$\ge 4$jets($P_T>100,50,50,50$)\\
\hline
$6$ & $1\tau$($P_T>40$),$\ge 4$jets($P_T>100,50,50,50$)\\
\hline
$7$ & $1b$($P_T>50$),$\ge 4$jets($P_T>100,50,50,50$)\\
\hline
$8$ & $0l$($P_T>10$),$\le 4$jets($P_T>50$), $M_{eff}>1400$\\
\hline
$9$ & $0l$($P_T>10$), $\le 4$jets($P_T >50$), $M_{eff}>1400$, $M_{inv}({\rm jets})> 800$\\
\hline
$10$ & $0l$($P_T>10$), $\ge 5$jets($P_T >50$), $M_{eff}>1400$\\
\hline
$11$ & $0l$($P_T>10$), $\ge 5$jets($P_T>50$), $M_{eff}>800$, $P_T(\textrm {4th hardest jet})>140$\\
\hline
$12$ & $\ge 1l$($P_T>10$), $\ge 5$jets($P_T >50$), $M_{eff}>1400$\\
\hline
$13$ & $\ge 1l$($P_T>10$), $\ge 5$jets($P_T>50$), $0.1< \ds{\not}E_{T}/M_{eff}< 0.3$\\
\hline
\end{tabular}
\end{center}
\caption{The collection of LHC signatures used in our $\Delta S^2$
analysis. All cuts are in units of GeV. In all signatures where
a cut is not explicitly stated, we have imposed
a base cut of $\ds{\not}E_{T}>100$~GeV and $M_{eff}>1200$~GeV}
\par
\label{siglist3}
\end{table}

In tables \ref{siglist2} and \ref{siglist3}, the corresponding signatures solely consist of
inclusive multijet (with particular $P_{T}$ cuts) plus missing
$E_{T}$ signatures, or signatures with one more lepton or tau (with
particular $P_{T}$ cuts). This is not so surprising as most other
more exclusive event selections such as those based on two or three lepton
signatures have low statistics at $5$~fb$^{-1}$. Even assuming sufficient statistics, we find
that in comparison with one lepton signatures, such signatures typically do not distinguish
as cleanly between various models.

We now discuss in greater detail the signatures in table \ref{siglist2}.
The first two signatures in signature list B are multijet + missing $E_T$
signatures. We indicate $P_{T}$ thresholds of jets in the parentheses.
The detailed selection criteria are provided in table \ref{selection-1}
of subsection \ref{SIMULATION}. These multijet signatures with different choices of
thresholds are chosen to capture the differences in the jet multiplicity
and $P_{T}$ distributions, and will thus be sensitive to the gluino and squark decays.
The signatures $0b(P_T > 50),\ge 2$ jets and $2b(P_T >50),\ge 2$ jets are
selected by veto b-jets or requiring at least
two b-jets with $P_{T}$ threshold given in the parentheses from the
inclusive selections $\ge2$jets($P_T >150,100$). The remaining signatures
in list B are $0l(P_T >20),\le4$jets($P_T >50$) with $M_{eff}>1500$.
For signature list C, the signatures are basically inclusive signatures
with one lepton, one tau and multijets. The reason why these signatures
are sensitive to $\Delta_{PQ}$ can be traced back to the shift in
the masses of $\tilde{t}_{1}$ and $\tilde{\tau}_{1}$, though the
detail reason could be complicated due to the mixture of different
production and decay channels. Similar considerations apply for signature
list D in table \ref{siglist3}.\footnote{We note that the selection
cuts used in signature list D are slightly different from those
which appear in Appendix C.} Here, it is important to stress that different analyses are better
suited to different sets of signatures. For example, in our footprint analysis,
we find that signatures from table \ref{siglist2} can typically distinguish
between models with different values of $N_5$ and $\Delta_{PQ}$. On the other
hand, we find that with more quantitative measures such as $\Delta S^{2}$,
the class of signatures provided by table \ref{siglist3} are instead better
suited for distinguishing between distinct F-theory GUTs. For this
reason, we shall often refer to this measure as $\Delta S_{(D)}^{2}$.

Given the prominent role of the PQ deformation in F-theory GUTs, it is important
to find signatures sensitive to the value of $\Delta{PQ}$. Because the effects of $\Delta_{PQ}$ are
most prominent for bino NLSP F-theory GUTs with a single messenger, we confine our discussion to this case.
As explained in section \ref{FATLHC}, the PQ deformation alters both the masses and branching fractions. A
non-zero PQ deformation leads to two major observable effects.
First, the charginos produced either directly or from
gluino and squark decay will undergo cascade decays of the form
$\tilde{\chi}_{1}^{\pm}\rightarrow\tilde{\tau}^{\pm}_{1}%
\nu_{\tau}\rightarrow\tilde{\chi}_{1}^{0}\tau^{\pm}\nu_{\tau}$.
The branching ratio of this channel will increase as $\Delta_{PQ}$
increases, and at the same time, the taus produced from the decay of the $\tilde{\tau}_{1}$
will become less energetic. Based on these changes, our expectation
is that signatures related to $\tau$'s or soft jets in the final
state will be helpful in distinguishing these models. Second, in comparison
to the other squarks, the mass of lightest stop $\tilde{t}_{1}$ is
more sensitive to $\Delta_{PQ}$. This leads to changes in the branching
ratio of the gluino decay into third generation quarks, which is particularly
important in the case of single messenger models. This can be probed
by using signatures related to third generation quarks in the final state.

\subsection{Determining $N_{5}$ and $\Lambda$}

\begin{figure}
[ptb]
\begin{center}
\includegraphics[
height=5.0929in,
width=6.2128in
]%
{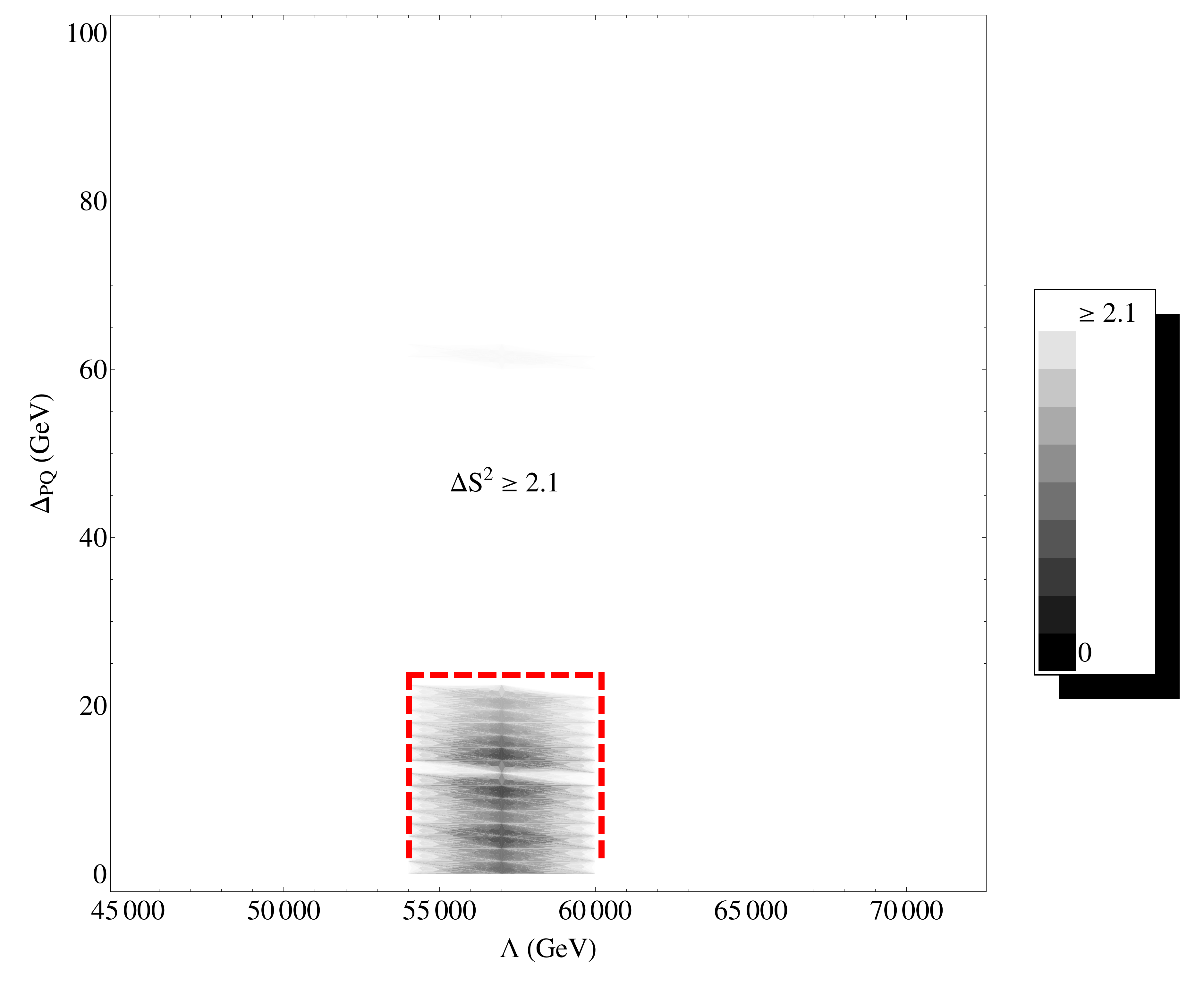}%
\caption{Density plot of $\Delta S_{(D)}^{2}$ defined by the signature list
in table \ref{siglist3} of subsection \ref{SIGLIST} between a two messenger F-theory
GUT\ model with $\Lambda=8\times10^{4}$ GeV and $\Delta_{PQ}=104$ GeV and
three messenger F-theory GUTs. Using a rough notion of
distinguishability compatible with a chi-square measure, models with $\Delta
S_{(D)}^{2}>2.1$ are distinguishable from this \textquotedblleft LHC
point\textquotedblright\ at the $99\%$ level for $13$ signals. Note that
$\Delta S_{(D)}^{2}$ possesses a local minimum, even though the number of
messengers is incorrect. This illustrates that distinguishing $N_5 = 2$
and $N_5 = 3$ is more subtle with limited luminosity.}%
\label{thrbexamp}%
\end{center}
\end{figure}

Having established a set of signatures with strong dependence on the parameters of F-theory GUTs, in this section we compute
the value of $\Delta S_{(D)}^2$ specified by the signatures of table \ref{siglist3} between a given F-theory GUT \textquotedblleft LHC point\textquotedblright and all other F-theory GUTs. Minimizing $\Delta S_{(D)}^2$ over the class of all F-theory GUTs with a bino NLSP, we can then in principle extract the corresponding values of $N_{5}$ and $\Lambda$. In computing the value of $\Delta S_{(D)}^{2}$ between models with a single
messenger, and models with multiple messengers, we find that $\Delta S_{(D)}^{2}$
is typically greater (sometimes far greater) than $10$. Based on this,
we conclude that it is typically possible to distinguish single messenger
models from multiple messenger models. On the other hand, distinguishing between
multiple messenger models appears more challenging. As an explicit example,
figure \ref{thrbexamp} shows the value of $\Delta S_{(D)}^{2}$
between three messenger F-theory models and a two messenger
F-theory GUT model with $\Lambda=8\times10^{4}$~GeV and $\Delta_{PQ}=104$~GeV,
which contains a region of models with $\Delta S_{(D)}^2 < 2.3$. Indeed, as explained earlier,
the mass spectra are much closer in such cases. Even so, the decay of the gluino is
different depending on the number of messengers (for example, see the branching fractions of (\ref{gtt})
in subsection \ref{MultiMess}). Although beyond the scope of this paper, it would be interesting
to study whether signatures sensitive to the decay $\widetilde{g}\rightarrow\widetilde{t}t$ could
provide a means to distinguish between two and three messenger F-theory GUTs.

\begin{figure}
[ptb]
\begin{center}
\includegraphics[
height=5.0929in,
width=6.2128in
]%
{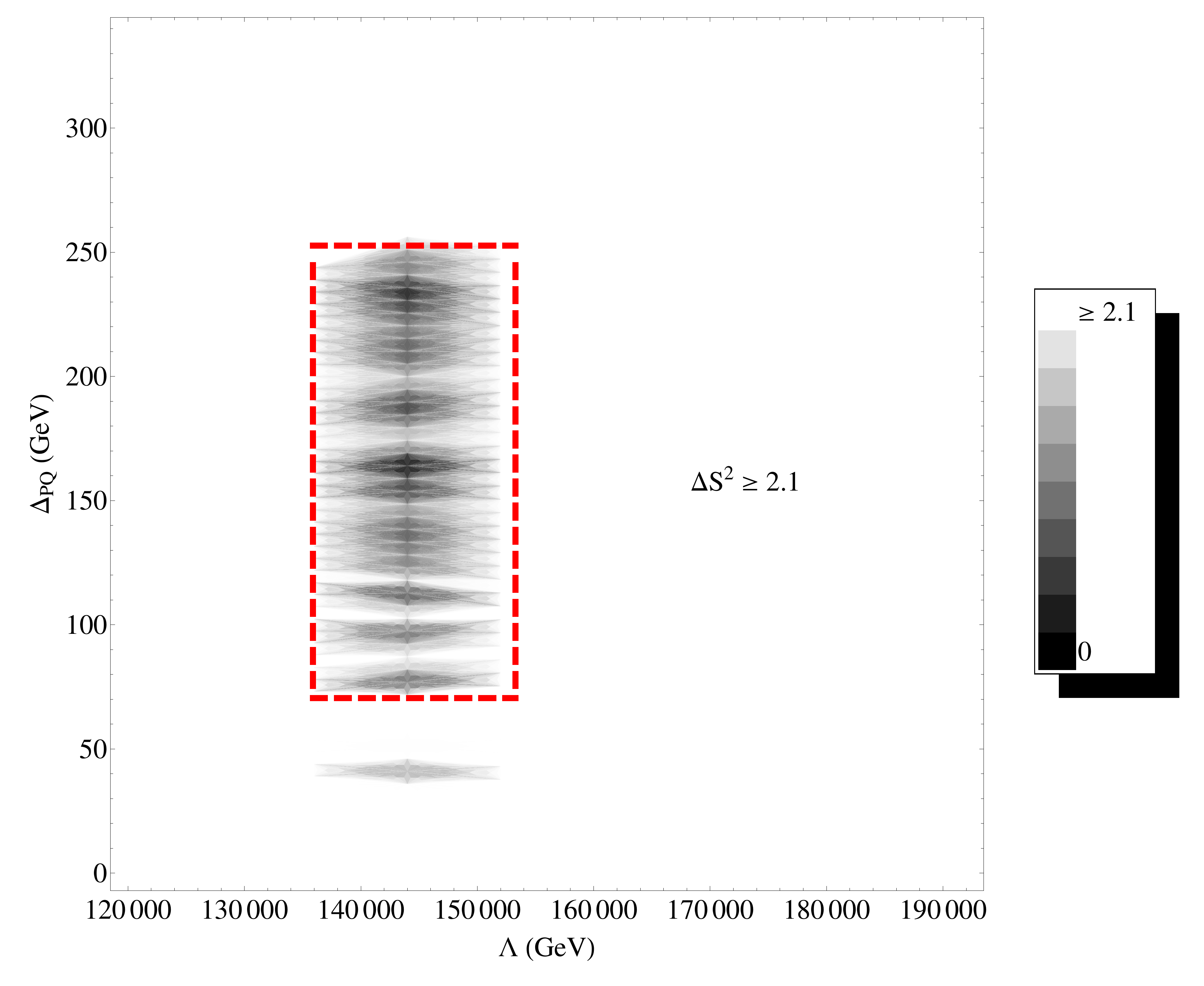}%
\caption{Density plot of $\Delta S_{(D)}^{2}$ defined by the signature list
in table \ref{siglist3} of subsection \ref{SIGLIST} between a single messenger F-theory
GUT\ model with $\Lambda=1.44\times10^{5}$ GeV and $\Delta_{PQ}=164$ GeV and
other single messenger F-theory GUTs. Here, we have used a rough notion of distinguishability based
on $99\%$ confidence and $13$ signals so that at $\Delta S_{(D)}^{2}>2.3$ we shall
say that two models are distinguishable. Note that
$\Delta S_{(D)}^{2}$ minimizes near the correct value of $\Lambda$, and that
moreover, the value of $\Delta_{PQ}$ can be distinguished as a non-zero value
up to $\sim \pm100$ GeV. The small shaded area just outside the dashed box is absent
when SM background is turned off.}%
\label{twoexamp}%
\end{center}
\end{figure}

Nevertheless, this analysis does establish that it is indeed possible to distinguish between single and multiple messenger models.
Assuming that the discrete parameter $N_{5}$ has been determined, the largest difference
between various models is the value of $\Lambda$. Focussing for simplicity
on the case of a single messenger model, we find that the value of
$\Delta S_{(D)}^{2}$ minimizes at a single value of $\Lambda$. As an explicit
example, figure \ref{twoexamp} shows the value of $\Delta S_{(D)}^{2}$
between all single messenger F-theory models and a single messenger
F-theory GUT model with $\Lambda=1.44\times10^{5}$~GeV and $\Delta_{PQ}=164$~GeV.
Note in particular that the density plot minimizes at a particular
value of $\Lambda$, and that furthermore, there is a lower bound to the value
of $\Delta_{PQ}$ in the region where $\Delta S_{(D)}^{2} < 2.3$. We therefore conclude that even at $5$~fb$^{-1}$,
it is possible to distinguish single messenger models from multiple
messenger models, and moreover, once the number of messengers has
been determined, to extract the value of $\Lambda$, and a crude bound on $\Delta_{PQ}$.

\subsection{Determining $\Delta_{PQ}$ at $5$ fb$^{-1}$ and $50$ fb$^{-1}$}

To completely distinguish between F-theory GUTs and a generic mGMSB\ model,
we next turn to the determination of $\Delta_{PQ}$. As opposed to
the discrete parameter $N_{5}$ and the gaugino mass unification scale
$\Lambda$, the dependence of the mass spectrum on the parameter $\Delta_{PQ}$
is somewhat smaller. In particular, in the case of multiple messenger
models, the most pronounced effect of the PQ\ deformation only occurs
in a regime of parameter space where the stau is already the NLSP.
Since the primary focus of our current analysis is on the more difficult
case of a bino NLSP, we shall therefore restrict attention to the
single messenger case.

\begin{figure}[ptb]
\begin{center}
\includegraphics[
height=5.0929in,
width=6.2128in
]{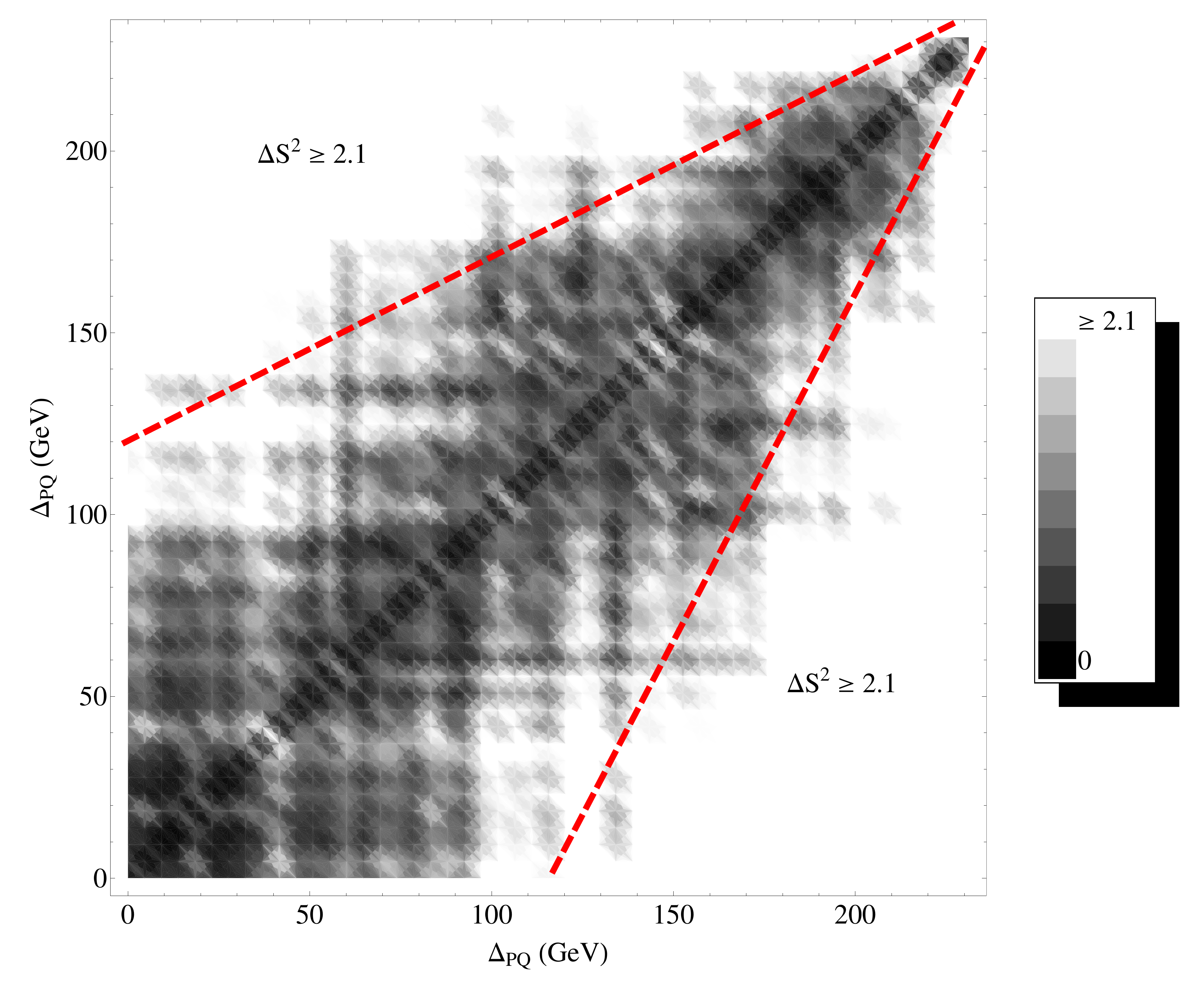}
\end{center}
\caption{Density plot of $\Delta S_{(D)}^{2}$ defined by the signature list
in table \ref{siglist3} of subsection \ref{SIGLIST} at 5 fb$^{-1}$ of simulated data
comparing F-theory GUTs with $N_{5} = 1$ and $\Lambda= 1.28 \times10^{5}$ GeV,
but with varying values of the PQ deformation parameter. Here, we have used a rough notion of distinguishability based
on $99\%$ confidence and $13$ signals so that at $\Delta S_{(D)}^{2}>2.3$ we shall
say that two models are distinguishable. This plot illustrates
that $\Delta S_{(D)}^{2}$ minimizes in a neighborhood of the region $\Delta
^{(1)}_{PQ} = \Delta^{(2)}_{PQ}$. Note in addition that the
determination of $\Delta_{PQ}$ becomes more accurate at larger values of the
PQ deformation.}%
\label{FIG:5fbPQ}%
\end{figure}

\begin{figure}[ptb]
\begin{center}
\includegraphics[
height=5.0929in,
width=6.2128in
]{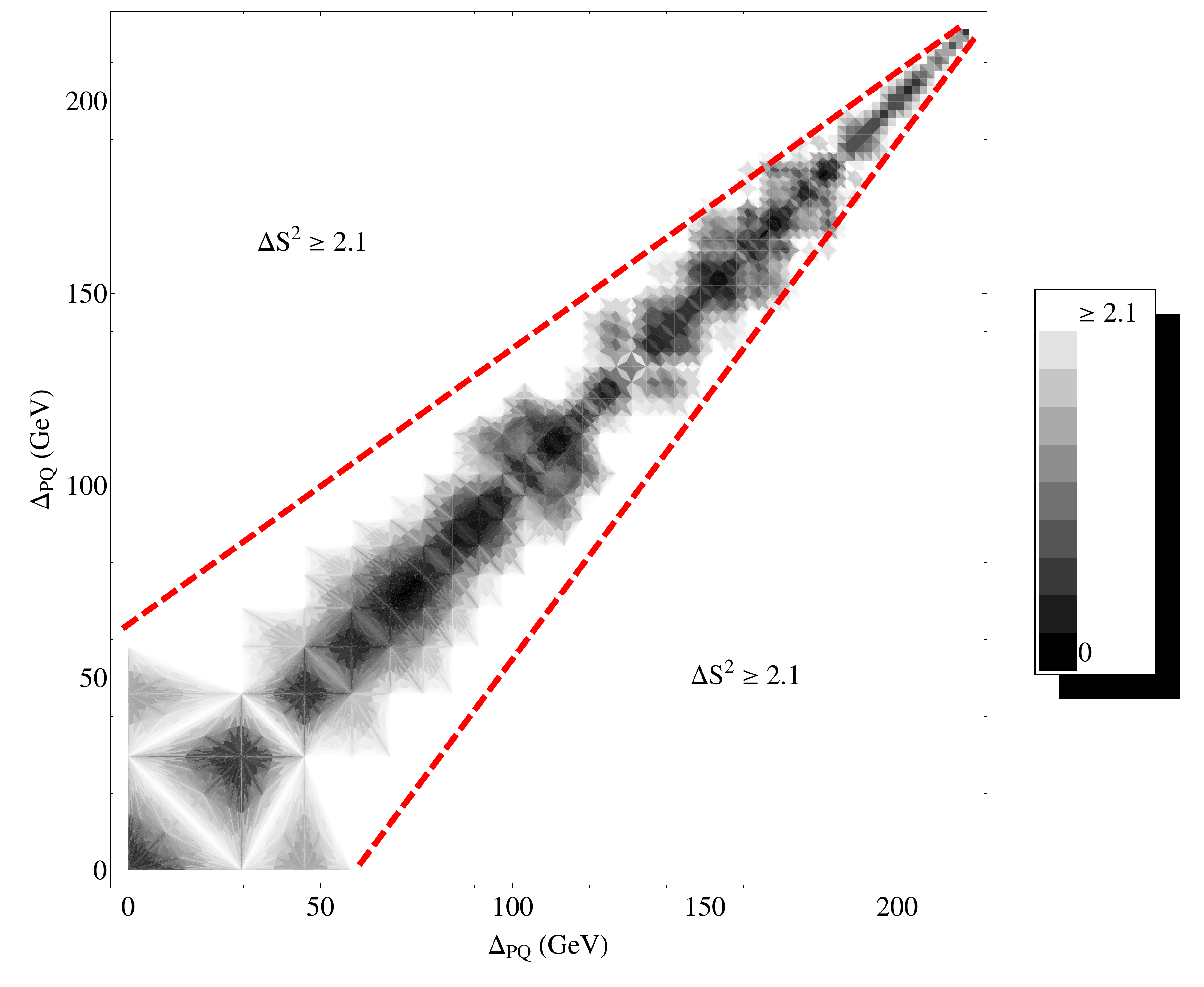}
\end{center}
\caption{Density plot of $\Delta S_{(D)}^{2}$ defined by the signature list
in table \ref{siglist3} of subsection \ref{SIGLIST} at 50 fb$^{-1}$ of simulated data
comparing F-theory GUTs with $N_{5} = 1$ and $\Lambda= 1.28 \times10^{5}$ GeV,
but with varying values of the PQ deformation parameter. Here, we have used a rough notion of distinguishability based
on $99\%$ confidence and $13$ signals so that at $\Delta S_{(D)}^{2}>2.1$ we shall
say that two models are distinguishable. This plot illustrates
that $\Delta S_{(D)}^{2}$ minimizes in a neighborhood of the region $\Delta
^{(1)}_{PQ} = \Delta^{(2)}_{PQ}$. As opposed to the case of 5
fb$^{-1}$ of simulated data shown in figure \ref{FIG:5fbPQ}, at increased
luminosity, it is possible to extract more detailed information about the PQ
deformation.}%
\label{FIG:PQ50FB12E5}%
\end{figure}

In order to determine whether it is possible to extract the value
of $\Delta_{PQ}$, we computed the value of $\Delta S_{(D)}^{2}$
between any two F-theory GUT models with the same values for $N_{5}$
and $\Lambda$. As an illustrative example, we considered in detail
the case of $N_{5}=1$ and $\Lambda=1.28\times10^{5}$~GeV. Plotting
the value of $\Delta S_{(D)}^{2}$ as a function of the PQ deformation
parameter of the two models, $\Delta_{PQ}^{(1)}$ and $\Delta_{PQ}^{(2)}$,
we find that at $5$~fb$^{-1}$ of simulated data, $\Delta S_{(D)}^{2}$
achieves a minimum over a somewhat broad range of values. See figure
\ref{FIG:5fbPQ} for a plot of $\Delta S_{(D)}^{2}$ at $5$~fb$^{-1}$
of integrated luminosity. This figure also illustrates that although this
range is somewhat broad at low values of $\Delta_{PQ}$, with an error
on the order of $\sim \pm100$ GeV, at larger values of $\Delta_{PQ}$,
the region with $\Delta S_{(D)}^{2}<2.1$ occurs over a smaller range of
values. We therefore conclude that it is indeed possible to distinguish
between models with very low $\Delta_{PQ}$ and very large $\Delta_{PQ}$.
Increasing the luminosity allows for further refinements in the determination
of $\Delta_{PQ}$. Figure \ref{FIG:PQ50FB12E5} illustrates that as a
function of the two PQ deformations, the value of $\Delta S_{(D)}^{2}$
at $50$ fb$^{-1}$ allows us to determine the value of $\Delta_{PQ}$
on the order of $\sim \pm10$\ GeV.\footnote{For simplicity, we have estimated the SM background
at $50$ fb$^{-1}$ by simply scaling up by a factor of $10$ the SM background at 5 fb$^{-1}$
of integrated luminosity.}

\section{Conclusions \label{CONCLUSIONS}}

In this paper we have analyzed the extent to which with a
relatively low integrated luminosity of $5$~fb$^{-1}$ how well the
LHC can distinguish F-theory GUTs from other scenarios with an
MSSM spectrum. Focussing on the case of F-theory GUTs with a bino NLSP, we have
used a footprint analysis to determine signatures of potential
interest. We have found that collider signatures with multijets
plus missing energy, as well as b-jets and taus are typically able
to distinguish this scenario from mSUGRA\ models with small
A-terms and from low scale minimal GMSB\ scenarios. In addition, single
messenger F-theory GUTs can be distinguished from mSUGRA models with large A-terms.
We have also found that mSUGRA models with large A-terms can potentially mimic
the signatures of multiple messenger F-theory GUTs with a bino NLSP.
With $5$~fb$^{-1}$ of integrated luminosity, we find that these models can only be
distinguished from large A-term mSUGRA models in a small range of F-theory GUT parameter space.

Focusing on just F-theory GUTs, we have also examined how well
collider signatures can be used to determine the extract the
defining parameters of the model, given by the number of
messengers $N_5$, the characteristic mass scale of minimal
gauge mediation $\Lambda$, and the stringy PQ deformation
parameter $\Delta_{PQ}$. We have found that it is typically possible to
distinguish between single and multiple messenger F-theory GUTs, although
due to the similarities in the soft parameters, distinguishing between
two and three messenger models appears more challenging. We have also
seen that once the number of messengers has been fixed, extracting
the value of $\Lambda$ is quite straightforward. More importantly, we have also seen
that a distinctively stringy feature of F-theory GUTs
corresponding to a deformation away from a minimal gauge mediation
scenario has observable consequences. Indeed, with $5$~fb$^{-1}$
of integrated luminosity, it is possible to distinguish the value
of this parameter up to $\sim \pm 80$~GeV, while for $50$~fb$^{-1}$,
this distinguishability typically improves to $\sim \pm 10$~GeV.

One of the advantages of the footprint method is that rather than
performing a global fit to the data, specifying particular
signatures which are sensitive to characteristics of a class of
models provides a relatively simple methodology for distinguishing
between broad classes of models. Using the footprint as a sieve
for various models, we have seen that in tandem with more
quantitative measures, it is possible to distinguish F-theory GUTs
from other well-motivated models with an MSSM spectrum. On the
other hand, it would of course be of interest to analyze a global
chi-square fit to a given model by including a large class of
binned $P_{T}$ distributions, and other similar observables.
Utilizing the entire class of such observables, a given F-theory
GUT will define a vector in a very high-dimensional vector space.
Performing a fit to such models, it would be important to
ascertain the level of resolution expected in parameters such as
the PQ deformation.

In this paper we have primarily focussed on signatures obtained
which can be obtained from the LHC with low integrated luminosity.
Performing a similar study with high luminosity would
certainly be very interesting. First, this would allow us to
distinguish F-theory GUTs from other models more cleanly, with
more available signatures and better statistics. Second, it would also
allow us to study the potential for performing mass reconstruction of
superpartners. This would provide a more direct verification of
the mass spectrum of F-theory GUTs, and would in particular directly probe
the effects of the PQ deformation. Of course, for future study, improved
signal and background simulation is certainly necessary in order
to confront real LHC data.

In addition, while our analysis has primarily focussed on
determining a class of signatures which can distinguish F-theory
GUTs from other models, our explanation for why these particular
signals are effective in discriminating between various models was
based on a somewhat heuristic analysis of the spectrum and
branching fractions. It would be interesting to perform a more
systematic simulation study to determine whether other potentially
interesting signatures can be extracted. For example, the angular
distribution of electrons and muons (or also the somewhat
different case of taus) could potentially provide another
window into the physics of F-theory GUTs.

Although we have focussed on the case of F-theory GUTs with a bino
NLSP, the case of stau NLSP scenarios is also possible, especially
in the case of multiple messenger models. We have focussed on the
bino NLSP case because it does not contain as striking a signature
as a quasi-stable stau. Nevertheless, it would be interesting to
explore in greater detail this regime of F-theory GUTs using the
footprint method.

\section*{Acknowledgements}

We thank T. Hartman, A. Menon and D. Morrissey for helpful discussions. The
work of JJH and CV is supported in part by NSF grant PHY-0244821,
and that of GLK and JS by the US Department of Energy. JJH thanks
the Michigan Center for Theoretical Physics for hospitality during
part of this work. JS thanks the Department of Physics at
Harvard University for hospitality during part of this work.

\appendix
%dummy comment inserted by tex2lyx to ensure that this paragraph is not empty

\section*{Appendix A: Benchmark mSUGRA Scans}

In this Appendix we describe the mSUGRA parameter scan and the mSUGRA models
studied in our comparison with F-theory GUTs. We recall that such models
are specified by the parameters $M_{1/2}$, $m_{0}$, $A$ and a choice
of $\tan\beta$. For simplicity, we have primarily restricted our
interest to models where the colored sparticle sector has similar
masses to the case of F-theory GUTs. We first describe the scan performed
over mSUGRA models with small A-term. We have considered the following range of parameters:
\begin{itemize}
\item {$M_{1/2} = 400-600$}
\item {$m_{0} = M_{1/2}(2.5+\epsilon)$, with $-0.3\lesssim\epsilon\lesssim-0.3$}
\item {$A = 0-400$ }
\item {$\tan\beta = 2-47$.}
\end{itemize}
We note that in the above scan we have correlated the squark mass with the gluino mass,
allowing it to have a small amount of variation parameterized by $\epsilon$.
In this scan, we have generated $180$ mSUGRA models, which we shall refer to as
small A-term mSUGRA models in our comparison with F-theory GUT models.

In the above scan, we have restricted our attention to models with small A-terms. It
is also of interest to study mSUGRA models with large A-terms. When the A-term is large,
there will be some qualitative differences in the mass spectrum. Most important is the
fact that due to the large trilinear couplings, the lightest stau can now have mass
in between that of $\tilde{\chi}_{1}^{0}$ and $\tilde{\chi}_{2}^{0}$, much as in
F-theory GUTs with a bino NLSP. For this reason, we have also performed a scan over mSUGRA models
with large A-term:
\begin{itemize}
\item {$M_{1/2} = 400-600$}
\item {$m_{0} = M_{1/2}(2.5+\epsilon)$, with $-1.5\lesssim\epsilon\lesssim0.5$}
\item {$A = 0-3000$ }
\item {$\tan\beta = 2-50$,}
\end{itemize}
We have generated a total of $13728$ such models, of which $255$ models satisfy
the stau mass condition $m_{\tilde{\chi}_{1}^{0}}<m_{\tilde{\tau}_{1}}<m_{\tilde{\chi}_{2}^{0}}$
and the Higgs mass limit $m_{h}>114$ GeV. Both of these conditions significantly constrain
the allowed parameter space. This subset of mSUGRA models are referred to as the
large A-term mSUGRA models in our comparison with F-theory GUT models.

\section*{Appendix B: Benchmark Low Scale mGMSB With Stau NLSP Scans}

For the low scale mGMSB models, we focus on the case with $\tilde{\tau}$
NLSP. In this case, the lightest stau will decay inside the detector giving rise
to missing $E_T$ signatures. Because of the generic presence of taus in the events,
this class of models can in principal mimic F-theory GUTs with a missing
$E_{T}$ signal, corresponding to the case of F-theory GUTs with a
bino NLSP.

In this Appendix we describe the scan over low messenger scale mGMSB
models we have performed. To generate a scan of such models, we vary the
parameters $N_{5}$, $\Lambda$, $\tan\beta$ of mGMSB scenarios. For simplicity, we
fix the messenger scale $M_{\text{mess}}$ to a low energy scale, since it
only changes the sparticle mass spectrum logarithmically. The scan
we have performed comprises the following range of parameters:
\begin{itemize}
\item {$N_{5}=1$}

\begin{itemize}
\item {$M_{\text{mess}}=2\times10^{5}$~GeV}
\item {$\Lambda=(1.2-1.92)\times10^{5}$~GeV with step size $8\times10^{3}$~GeV}
\end{itemize}
\item {$N_{5}=2$}

\begin{itemize}
\item {$M_{\text{mess}}=1.2\times10^{5}$~GeV}
\item {$\Lambda=(6.5-11)\times10^{4}$~GeV with step size $5\times10^{3}$~GeV}
\end{itemize}
\item {$N_{5}=3$}

\begin{itemize}
\item {$M_{\text{mess}}=1.2\times10^{5}$~GeV}
\item {$\Lambda=(4.5-7.2)\times10^{4}$~GeV with step size $3\times10^{3}$~GeV}
\end{itemize}
\item {for all of these cases, $\tan\beta\sim2-50$ with step size $2$.}
\end{itemize}
In the above scan, the lower limit on $\Lambda$ corresponds to the
Higgs mass bound $m_{h}>114$~GeV, and the upper limit corresponds
to the discovery limit for $5$~fb$^{-1}$. Of the $750$ low scale
mGMSB models generated this way, $433$ of these have a stau NLSP.

\section*{Appendix C: Signature List}

In this Appendix we provide a list of the 103 signatures used in our
footprint analysis. Most of the signatures are based on the inclusive
$2$ jet and $4$ jet plus missing $E_{T}$ selections as listed in table
\ref{selection-1} of subsection 3.1. The counts of events of these two selections are
denoted by $\ge2$jets($P_{T}>150,100$) and $\ge4$jet($P_{T}>100,50,50,50$)
respectively. These two selections are further divided into subsets
according to the number of leptons, taus and b-jets. The selection
criteria are given below:
\begin{itemize}
\item {leptons: $0l$, $\ge1 l$, $\ge2l$, opposite-sign $2l$ (OSDL) and
same-sign $2l$ (SSDL). }
\item {taus: $0\tau$, $\ge1\tau$, $\ge2\tau$, opposite-sign $2\tau$
(OSDT) and same-sign $2\tau$ (SSDT). }
\item {b-jets: $0b$, $\ge1b$, $\ge2b$ and $\ge3b$}
\item {leptons + b-jets: $0l0b$, $0l\ge1b$, $0l\ge2b$, $\ge1l0b$,
$\ge1l\ge1b$, $\ge1l\ge2b$, $\ge2l0b$}
\item {taus + b-jets: $0\tau0b$, $0\tau\ge1b$, $0\tau\ge2b$,
$\ge1\tau0b$, $\ge1\tau\ge1b$, $\ge1\tau\ge2b$, $\ge2\tau0b$}
\item where we used two different $P_{T}$ cuts for each object:
\begin{itemize}
\item {$P_{T}>10$,$40$~GeV for lepton and tau}
\item {$P_{T}>50$,$100$~GeV for b-jet. }
\end{itemize}
\end{itemize}
Additional signatures include some variations of
the selection criteria (if not specified ${\ds}{\not}E_{T}>100$~GeV
and $M_{eff}>1200$), which are listed below:%
\begin{itemize}
\item $\ge6$ jets ($P_{T}>100,100,20,20,20,20$), ${\ds}{\not}E_{T}>200$~GeV
\item $\ge6$ jets ($P_{T}>100,100,100,100,20,20$)
\item $\ge1\tau$($P_{T}>40$), $\ge2$jets($P_{T}>80$)
\item $\ge1\tau$($P_{T}>40$), $\ge4$jets($P_{T}>40$)
\item $\ge1\tau$($P_{T}<40$), $\ge4$jets($P_{T}>40$), ${\ds}{\not}E_{T}>150$~GeV
\item $\ge1\tau$($P_{T}<40$), $\ge4$jets($P_{T}>40$), ${\ds}{\not}E_{T}>150$~GeV
\item $\ge2$jets($P_{T}>350$), ${\ds}{\not}E_{T}>0.2M_{eff}$
\item $\ge4$jets($P_{T}>250,250,150,150$), ${\ds}{\not}E_{T}>0.1M_{eff}$
\item $\ge6$jets($P_{T}>150,150,100,100,50,50$)
\end{itemize}
The last set of signatures are based on \cite{Altunkaynak:2009tg},
where the base cut is given by ${\ds}{\not}E_{T}>150$~GeV and
$M_{eff}>600$~GeV.
\begin{itemize}
\item $0l$($P_{T}>20$),$\le4$jets($P_{T}>50$), $E_{T}^{miss}>500$~GeV
\item $0l$($P_{T}>20$),$\le4$jets($P_{T}>50$), $M_{eff}>1500$~GeV
\item $\ge1l$($P_{T}>20$), $\le4$jets($P_{T}>50$), ${\ds}{\not}E_{T}>0.2M_{eff}$
and $M_{eff}>1400$~GeV
\item $0l$($P_{T}>20$),$\ge5$jets($P_{T}>50$), $M_{eff}>1200$~GeV
\item $0l$($P_{T}>20$),$\ge5$jets($P_{T}>50$), 4th hardest jet $P_{T}>100$~GeV
\item $0l$($P_{T}>20$),$\ge5$jets($P_{T}>50$), $0.15<r_{\rm jet}<0.5$
\item $0l$($P_{T}>20$),$\ge5$jets($P_{T}>50$), $0.5<r_{\rm jet}<1.0$
\item $0l$($P_{T}>20$),$\ge5$jets($P_{T}>50$), $0.05<E_{T}^{miss}/M_{eff}<0.35$
\item $\ge1l$($P_{T}>20$),$\ge5$jets($P_{T}>50$), $M_{eff}>1350$~GeV
\item $\ge1l$($P_{T}>20$),$\ge5$jets($P_{T}>50$), 4th hardest jet $P_{T}>100$~GeV
\item $\ge1l$($P_{T}>20$),$\ge5$jets($P_{T}>50$), hardest lepton $P_{T}>60$~GeV
\item $\ge1l$($P_{T}>20$),$\ge5$jets($P_{T}>50$), $0.05<{\ds}{\not}E_{T}/M_{eff}<0.35$
\item $\ge1\tau$($P_{T}>20$),$\ge5$jets($P_{T}>50$), $M_{eff}>1350$~GeV
\item $\ge1\tau$($P_{T}>20$),$\ge5$jets($P_{T}>50$), 4th hardest jet $P_{T}>100$~GeV
\item $\ge1\tau$($P_{T}>20$),$\ge5$jets($P_{T}>50$), hardest tau $P_{T}>60$~GeV
\item $\ge1\tau$($P_{T}>20$),$\ge5$jets($P_{T}>50$), $0.05<{\ds}{\not}E_{T}/M_{eff}<0.35$
\end{itemize}
For all $103$ signatures, we have imposed the cut on transverse sphericity
$S_{T}>0.2$. In the above signature, the value $r_{\rm jet}$ is given as a function of the jet $P_{T}$'s as:
\begin{equation}
r_{\rm jet}\equiv \frac{P_T^{\rm jet3}+P_T^{\rm jet4}}{P_T^{\rm jet1}+P_T^{\rm jet2}},
\end{equation}
where $P_T^{{\rm jet}i}$ is the transverse momentum of the $i$-th hardest jet in the event.
For this signature we require at least three jets with $P_T >200$~GeV.

\section*{Appendix D: \texttt{PGS} Trigger}

In our simulation, we have used the PGS4 default detector configuration
and cone jet algorithm with a cone size $0.5$. In addition, the PGS
loose b-tagging is used, which corresponds to a tagging probability
near $0.5$ in the central region for high energy jets. To reduce
the number of SM background events, we choose the PGS triggers with
high thresholds:
\begin{itemize}
\item Inclusive isolated lepton ($e$,$\mu$) $180$~GeV
\item Lepton plus jet ($130$~GeV, $200$~GeV)
\item Isolated dileptons ($\mu\mu$,$ee$) $60$~GeV
\item Dileptons($\mu\mu$,$ee$) plus jet ($45$~GeV, $150$~GeV)
\item Isolated dileptons ($e\mu$) $30$~GeV
\item Isolated lepton ($\mu$,$e$) plus isolated tau ($45$~GeV, $60$~GeV)
\item Isolated ditau $60$~GeV
\item Inclusive isolated photon $80$~GeV
\item Isolated diphoton $40$~GeV
\item Inclusive ${\displaystyle {\not}E_{T}}$ $200$~GeV
\item Inclusive single-jet $1000$~GeV
\item Jet plus ${\displaystyle {\not}E_{T}}$ ($300$ GeV, $125$ GeV)
\item Accoplanar jet and ${\displaystyle {\not}E_{T}}$ ($150$ GeV, $80$
GeV, $1<\Delta\phi<2$)
\item Accoplanar dijets ($400$ GeV, $\Delta\phi<2$)
\end{itemize}

\section*{Appendix E: PQ deformation and Ditau Mass Distribution}

In this Appendix we present a general discussion of the prospects of using the ditau
mass distribution as an additional means of probing the PQ deformation
in F-theory GUTs with a bino NLSP. We find that for given $N_5$ and $\Lambda$,
the edge of the ditau mass distribution is sensitive to the PQ\ deformation.
Interestingly, for F-theory GUTs with $N_5=2,3$, measuring the edge of the ditau mass
distribution will uniquely determine the PQ\ deformation. For the case with $N_5=1$,
the PQ deformation can be determined up to a two-fold degeneracy.

\begin{figure}[ptb]
\begin{centering}
\includegraphics[width=16cm]{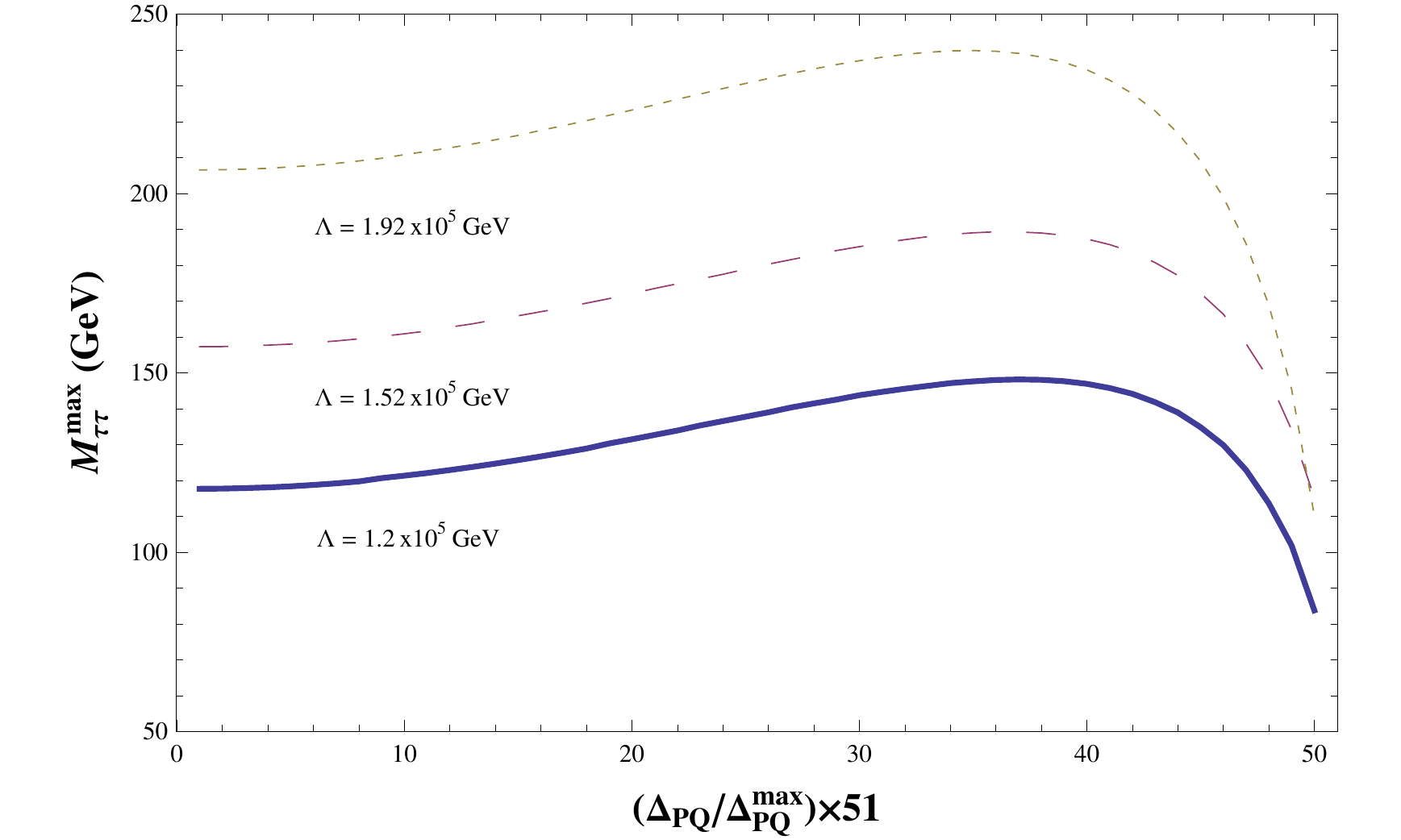}
\includegraphics[width=16cm]{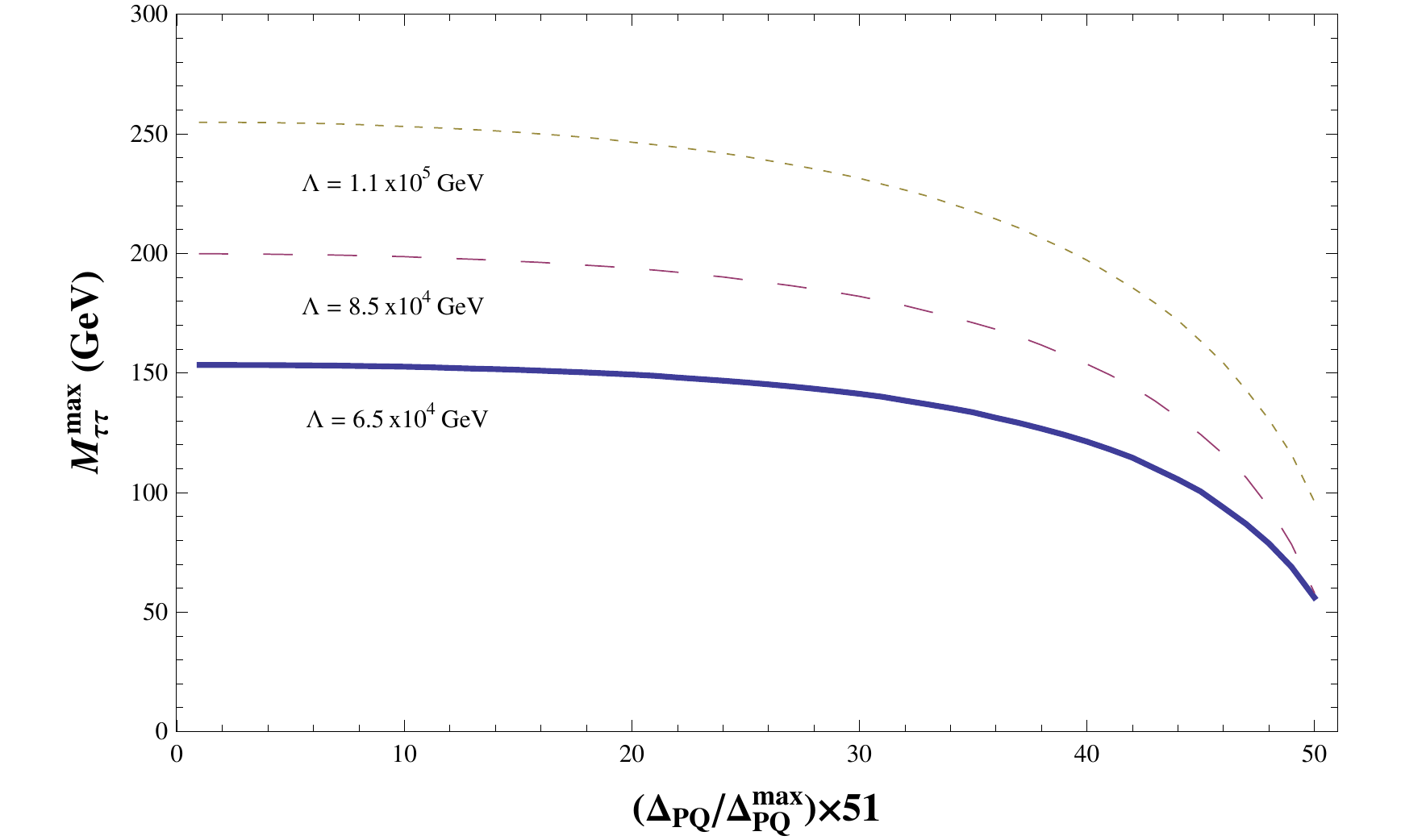}
\par\end{centering}
\caption{Endpoint of the opposite-sign ditau invariant mass distribution as
a function of the PQ deformation parameter $\Delta_{PQ}$ for F-theory
GUTs with a bino NLSP. From top to bottom, the two plots are for
$N_{5}=1,2$ respectively. The case of $N_{5}=3$ is similar to $N_{5}=2$.
Note that in contrast to the case of a single messenger model, for
two and three messengers, the endpoint decreases monotonically as
a function of $\Delta_{PQ}$.}
\label{dilep-edge-thy}
\end{figure}

When the bino is the NLSP, the relative masses of the lightest stau
and two lightest neutralinos obey the relation:
\begin{equation}
m_{\widetilde{\chi}_{1}^{(0)}}<m_{\widetilde{\tau}_{1}}< m_{\widetilde{\chi}_{2}^{(0)}}\text{.}
\label{inequ}
\end{equation}
In this case, the decay chain:
\begin{equation}
\widetilde{\chi}_{2}^{(0)}\rightarrow\widetilde{\tau}_{1}^{\pm}\tau^{\mp}
\rightarrow\widetilde{\chi}_{1}^{(0)}\tau^{\pm}\tau^{\mp}
\end{equation}
will generate events with opposite sign taus. If the $\tau^{+}\tau^{-}$ mass
distribution could be measured directly, it would show a sharp kinematical edge
\begin{equation}
(m_{\tau^{+}\tau^{-}}^{\max})^{2}\equiv\frac{\left(m_{\widetilde{\chi}_{2}^{(0)}}^{2}-m_{\widetilde{\tau}_{1}}^{2}\right)
\left(m_{\widetilde{\tau}_{1}}^{2}-m_{\widetilde{\chi}_{1}^{(0)}}^{2}\right)}{m_{\widetilde{\tau}_{1}}^{2}}\text{.}
\label{ditau-edge}
\end{equation}
However in practice, this edge structure gets smeared out by the missing neutrino from the tau decay.
In addition, there will be a large background under the edge due to the fake taus from jets as well as
taus from the chargino decay $\chi_{1}^{\pm}\rightarrow \tilde \tau^{\pm} \nu$.
In spite of these potential complications, this edge can be experimentally measured with
an estimated $10\%$ error for a few tens of fb$^{-1}$ integrated luminosity \cite{Aad:2009wy}.

With this in mind, we now show that the endpoint of the ditau invariant mass distribution is
is sensitive to the PQ deformation. Scanning over models with different values of $\Delta_{PQ}$,
figure \ref{dilep-edge-thy} shows that $m_{\tau^{+}\tau^{-}}^{\max}$ is a non-trivial
function of $\Delta_{PQ}$. Since the parameter $\Lambda$ and $N_5$ are relatively
easy to determine, we shall take them as fixed in the following discussion.
For the one messenger case, there is one-to-one correspondence between the $m_{\tau^{+}\tau^{-}}^{\max}$
and $\Delta_{PQ}$ in the very large $\Delta_{PQ}$ region, but generally shows a two-fold
degeneracy between small and large $\Delta_{PQ}$. In this case, the value of $m_{\tau^{+}\tau^{-}}^{\max}$
cannot by itself be used to determine the PQ deformation, and requires supplementary observables, such as $P_T$
distributions, to the resolve the degeneracy. However, for the two and three messenger cases,
there is always a one-to-one correspondence between the value of $m_{\tau^{+}\tau^{-}}^{\max}$
and $\Delta_{PQ}$. This makes the ditau invariant mass edge an effective observable for
probing the PQ deformation in F-theory GUTs.

In the following, we shall not focus on the concrete extraction of the ditau
invariant mass edge from the simulated data. Instead, we focus on the analytic
dependence of $m_{\tau^{+}\tau^{-}}^{\max}$ on $\Delta_{PQ}$. As a rough approximation,
the PQ deformation can simply be added as an overall shift to $m_{\widetilde{\tau}_{1}}$. Letting
$\widehat{m}_{\widetilde{\tau}_{1}}$ denote the lightest stau mass
in the absence of the PQ deformation, the endpoint of the ditau invariant
mass distribution is given by the approximation:%
\begin{align}
(m_{\tau^{+}\tau^{-}}^{\max})^{2}  &  \sim\frac{\left(  m_{\widetilde{\chi}%
_{2}^{(0)}}^{2}-\widehat{m}_{\widetilde{\tau}_{1}}^{2}+\Delta_{PQ}^{2}\right)
\left(  \widehat{m}_{\widetilde{\tau}_{1}}^{2}-\Delta_{PQ}^{2}-m_{\widetilde
{\chi}_{1}^{(0)}}^{2}\right)  }{\widehat{m}_{\widetilde{\tau}_{1}}^{2}%
-\Delta_{PQ}^{2}}\\
&  =m_{\widetilde{\chi}_{2}^{(0)}}^{2}+m_{\widetilde{\chi}_{1}^{(0)}}%
^{2}-\frac{m_{\widetilde{\chi}_{2}^{(0)}}^{2}m_{\widetilde{\chi}_{1}^{(0)}%
}^{2}}{\widehat{m}_{\widetilde{\tau}_{1}}^{2}-\Delta_{PQ}^{2}}-\left(
\widehat{m}_{\widetilde{\tau}_{1}}^{2}-\Delta_{PQ}^{2}\right) \\
&  =\widehat{m}_{\widetilde{\tau}_{1}}^{2}\left(  A-\frac{B}%
{1-x}-(1-x)\right)
\end{align}
where we have introduced the parameters:%
\begin{align}
A  &  \equiv\frac{ m_{\widetilde{\chi}_{2}^{(0)}}^{2}+m_{\widetilde{\chi}_{1}%
^{(0)}}^{2} }{\widehat{m}_{\widetilde{\tau}_{1}}^{2}}\\
B  &  \equiv\frac{m_{\widetilde{\chi}_{2}^{(0)}}^{2}m_{\widetilde{\chi}_{1}^{(0)}%
}^{2}}{\widehat{m}_{\widetilde{\tau}_{1}}^{4}}\\
x  &  \equiv\frac{\Delta_{PQ}^{2}}{\widehat{m}_{\widetilde{\tau}_{1}}^{2}}\text{.}%
\end{align}

We now determine the behavior of $(m_{\tau^{+}\tau^{-}}^{\max})^{2}$
as a function of $x$ as $x$ ranges from zero ($\Delta_{PQ}=0$)
to 1 ($\Delta_{PQ}=\widehat{m}_{\widetilde{\tau}_{1}}$). To this
end, we now show that depending on the initial placement of the lightest
stau mass, the value of $(m_{\tau^{+}\tau^{-}}^{\max})^{2}$ can either
increase, or decrease. Note, however, that at $x=1$, the value of
$(m_{\tau^{+}\tau^{-}}^{\max})^{2}$ vanishes. The value of $(m_{\tau^{+}\tau^{-}}^{\max})^{2}$
will increase provided a local maximum is present between $x=0$ and
$x=1$. The function $(m_{\tau^{+}\tau^{-}}^{\max})^{2}(x)$ possesses
a critical point satisfying the condition:
\begin{equation}
x_{\ast}=1-\sqrt{B}=1-\frac{m_{\widetilde{\chi}_{2}^{(0)}}m_{\widetilde{\chi}_{1}^{(0)}}}{\widehat{m}_{\widetilde{\tau}_{1}}^{2}}\text{.}
\end{equation}
Computing the second derivative of $(m_{\tau^{+}\tau^{-}}^{\max})^{2}(x)$
with respect to $x$, it follows that this corresponds to a local
maximum for the function. In other words, for $x<x_{\ast}$, $(m_{\tau^{+}\tau^{-}}^{\max})^{2}(x)$
increases, while for $x>x_{\ast}$, $(m_{\tau^{+}\tau^{-}}^{\max})^{2}(x)$
decreases.

This local maximum will only be attained in the physical range of
interest when $0<x_{\ast}<1$, or:
\begin{equation}
0<\widehat{m}_{\widetilde{\tau}_{1}}^{2}-m_{\widetilde{\chi}_{2}^{(0)}}m_{\widetilde{\chi}_{1}^{(0)}}<\widehat{m}_{\widetilde{\tau}_{1}}^{2}\text{.}
\end{equation}
While the second inequality is indeed always satisfied, the first
inequality requires the geometric mean of $m_{\widetilde{\chi}_{2}^{(0)}}$
and $m_{\widetilde{\chi}_{1}^{(0)}}$ to be less than $\widehat{m}_{\widetilde{\tau}_{1}}$.
When this inequality is not satisfied so that $x_{\ast}<0$, it follows
that $(m_{\tau^{+}\tau^{-}}^{\max})^{2}(x)$ is always a decreasing function
over the physically acceptable range of values for $\Delta_{PQ}$.
In the case where a local maximum is achieved, there is a potential
degeneracy in the location of the endpoint. For large enough values
of the PQ deformation, however, there is typically no degeneracy because
there is a single local maximum and the endpoint must vanish at the
point where the stau and bino masses become degenerate. See figure
\ref{dilep-edge-thy} for plots of the endpoint for representative
F-theory GUTs with $N_{5}=1,3$ vector-like pairs of messengers.
These plots illustrate that the relative spacing between the lightest
stau and the two lightest neutralinos can produce qualitatively different
$\Delta_{PQ}$ dependence.

\section*{Appendix F: Miscellaneous Plots}

In this Appendix we collect additional plots of potential interest.

\begin{figure}[h]
\begin{center}
\includegraphics[
height=4.12in,
width=6.4091in
]{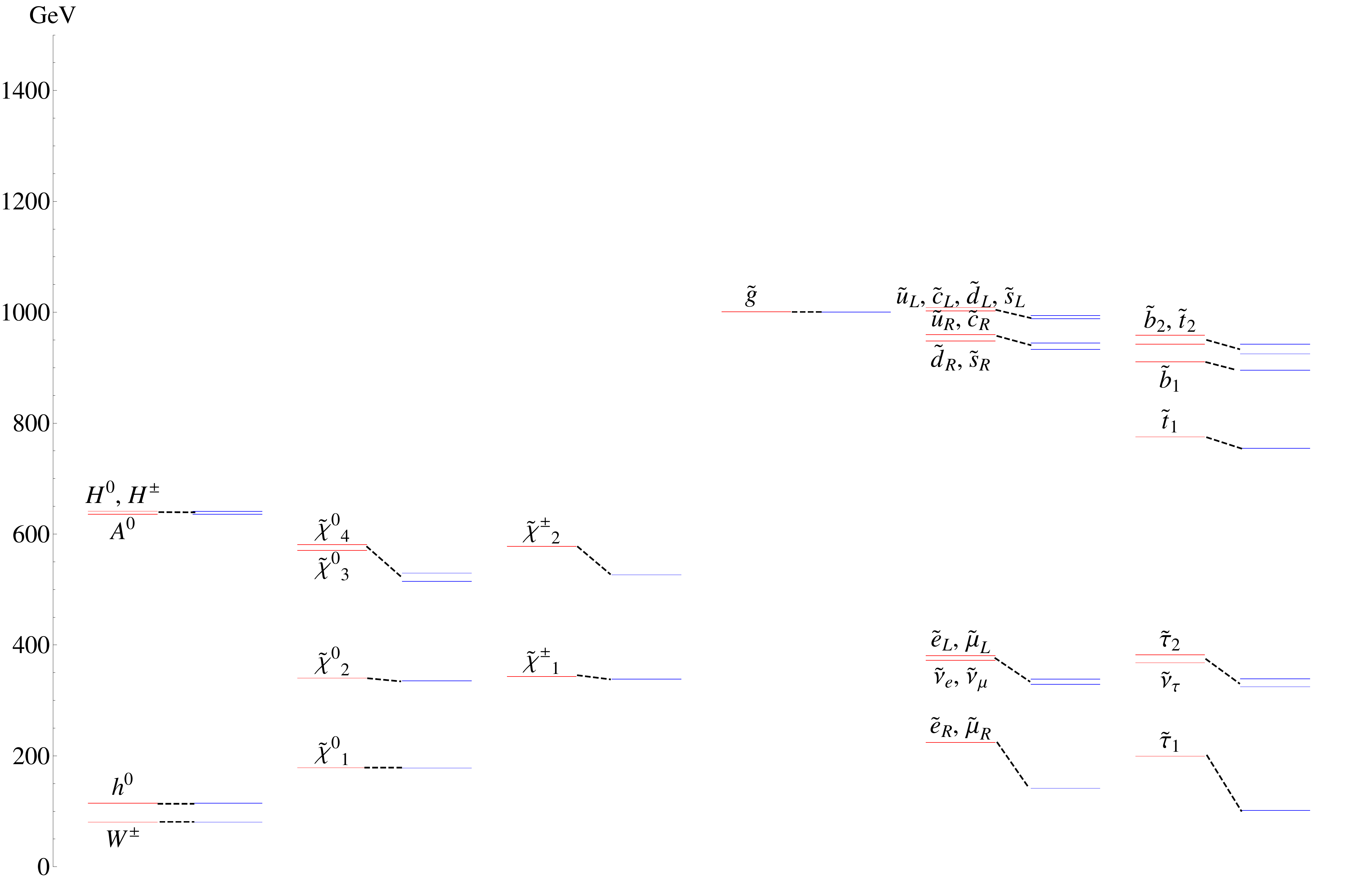}
\end{center}
\caption{Plot of the mass spectrum of an F-theory GUT with $N_{5}=3$,
$\Lambda=4.5\times10^{4}$ GeV, and minimal (red, left part of each column) and maximal (blue, right part of each column) PQ
deformation.}%
\label{FIG:THRMESSSPEC}%
\end{figure}

\begin{figure}[ptb]
\begin{center}
\includegraphics[
height=3.4272in,
width=5.719in
]{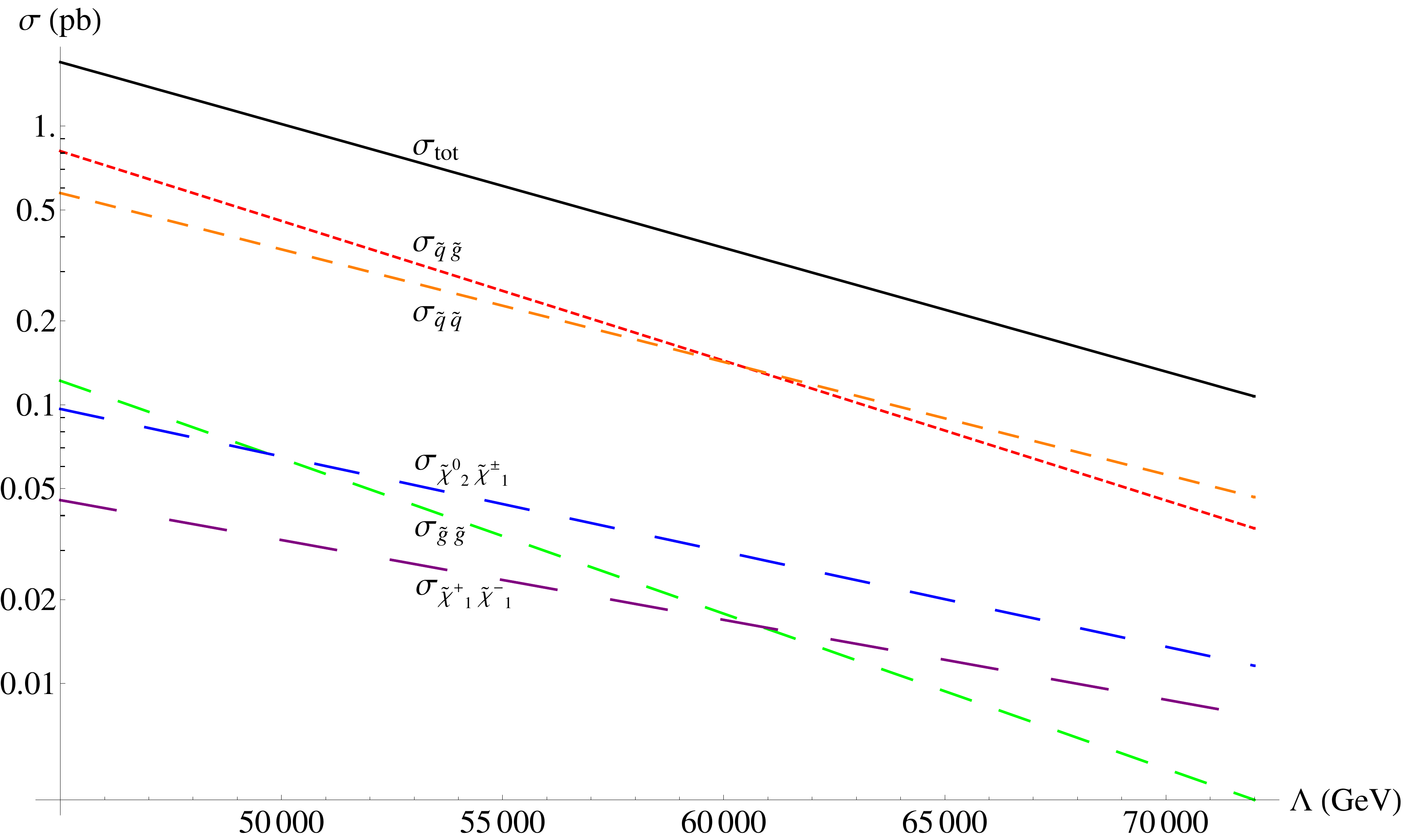}
\end{center}
\caption{Plot of the cross sections at $14$ TeV CM energy for the primary supersymmetric processes as a function of
$\Lambda$ for F-theory GUT models with $N_{5} = 3$ and $\Delta_{PQ} = 0$ GeV. Note
that as $\Lambda$ increases, the cross sections decrease by roughly
an order of magnitude.}%
\label{FIG:ThrL}%
\end{figure}

\begin{figure}[ptb]
\begin{center}
\includegraphics[
height=5.5625in,
width=5.7017in
]{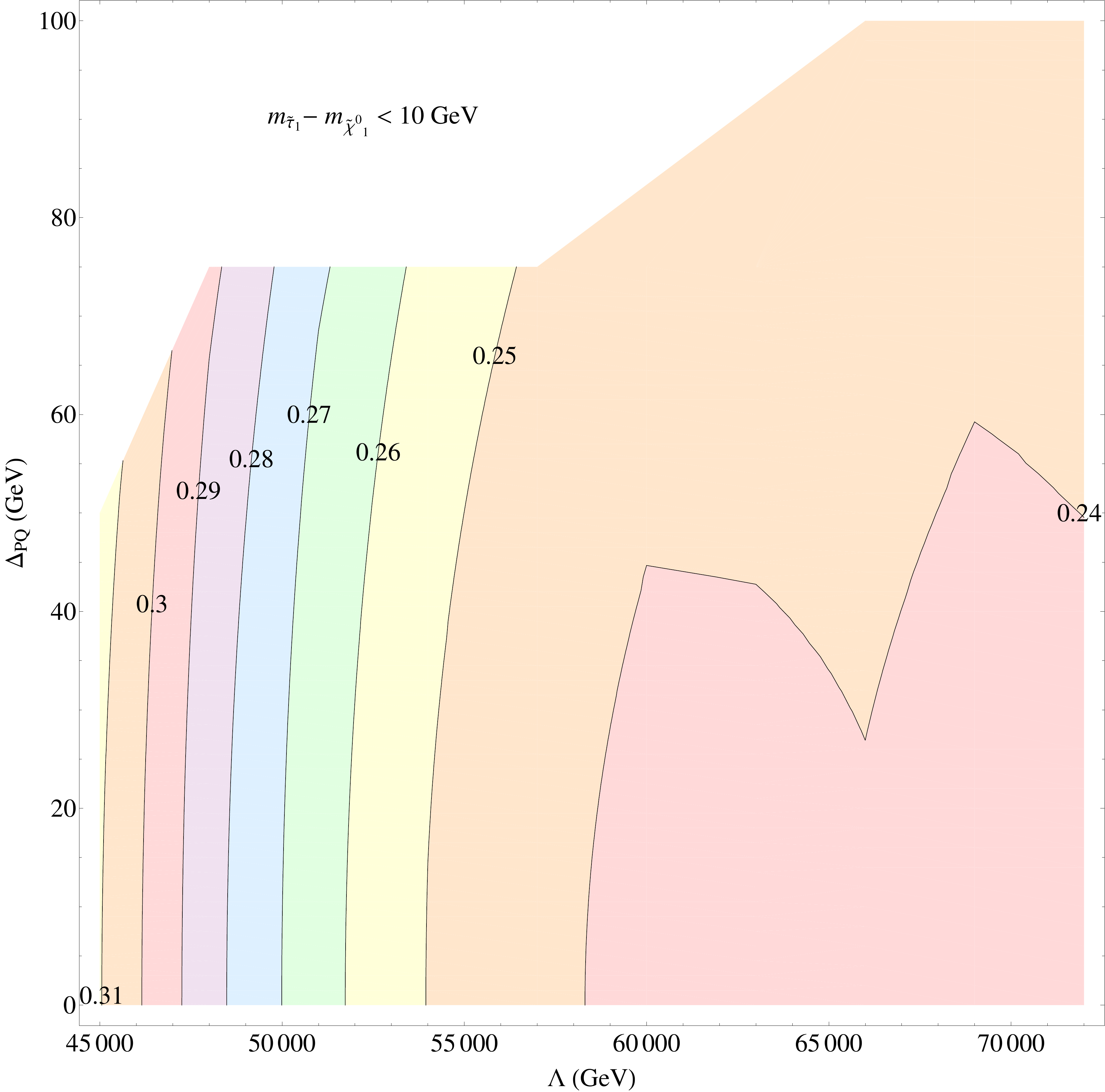}
\end{center}
\caption{Contour plot of the value of $\Delta P^{2}$ obtained by fixing a
particular value of $\Lambda$ and $\Delta_{PQ}$ of an F-theory GUT model with
$N_{5}=3$, and minimizing with respect to all small A-term mSUGRA\ models. We adopt
a rough criterion for theoretical distinguishability specified by the requirement that $\Delta P^2 > 0.01$. By
inspection, $\Delta P^2$ is greater than $0.2$, indicating that
such models are distinguishable at the theoretical level from F-theory GUTs.}%
\label{FIG:softspecthrsugrasa}%
\end{figure}
%EndExpansion

\begin{figure}[ptb]
\begin{center}
\includegraphics[
height=5.0929in,
width=6.2128in
]{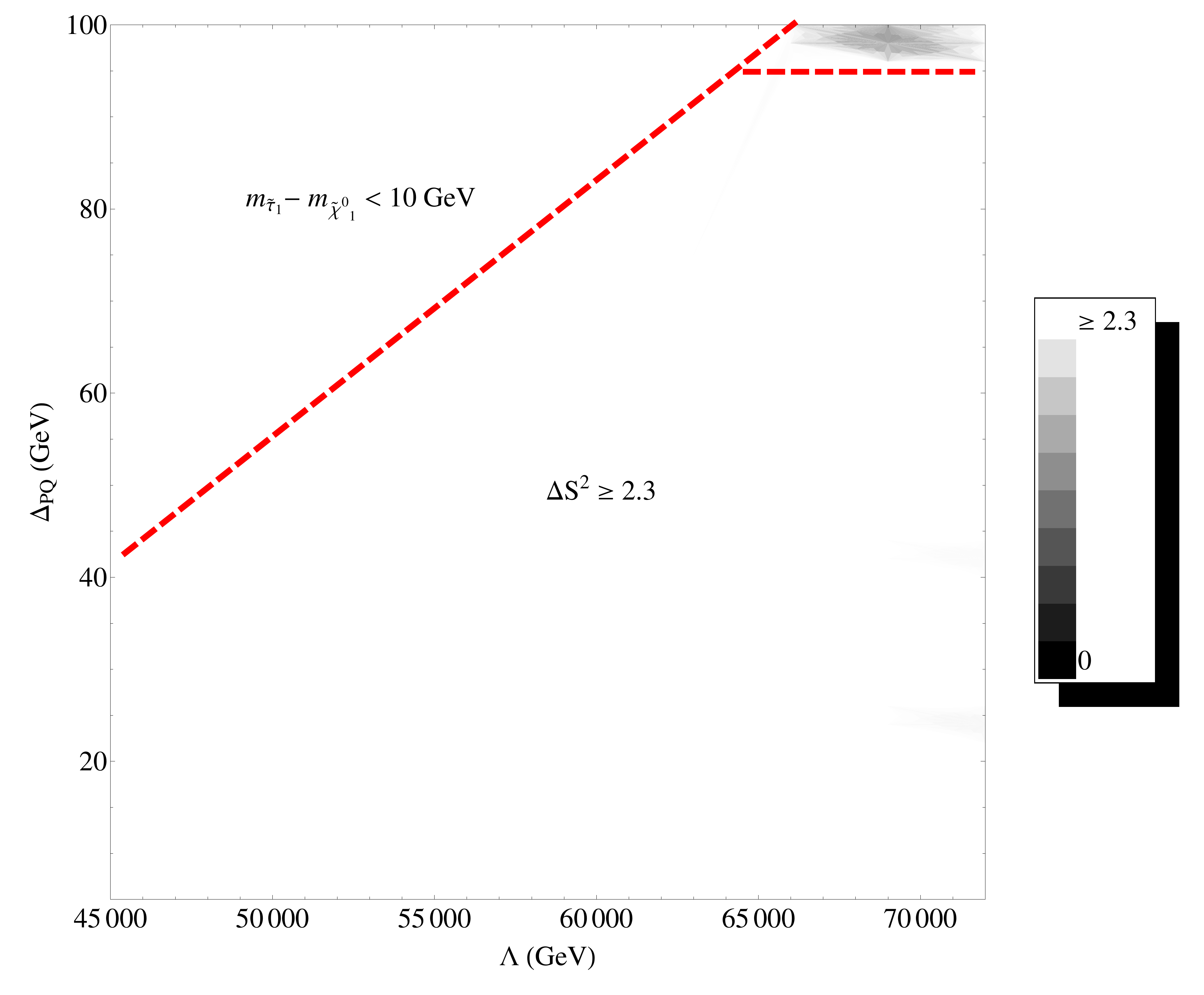}
\end{center}
\caption{Density plot of $\Delta S_{(A)}^{2}$ defined by the signature list
in table \ref{siglist1} of subsection \ref{SIMULATION} comparing the minimal value of a given
$N_{5}=3$ F-theory GUT model with a scan over small A-term mSUGRA models. The signals used are
obtained with $5$ fb$^{-1}$ of simulated LHC data. Here, we have used a rough notion of distinguishability based
on $99\%$ confidence and $10$ signals so that at $\Delta S_{(A)}^{2}>2.3$ we shall
say that two models are distinguishable. By inspection,
for most of the plot, it is possible to distinguish between F-theory GUTs and
mSUGRA models. See figure \ref{FIG:OneSugra} in section \ref{DISTINGUISH} for a similar
plot in the case of $N_{5} = 1$ F-theory GUTs.}%
\label{FIG:ThrSugra}%
\end{figure}

\begin{figure}[ptb]
\begin{center}
\includegraphics[
height=5.5625in,
width=5.7017in
]{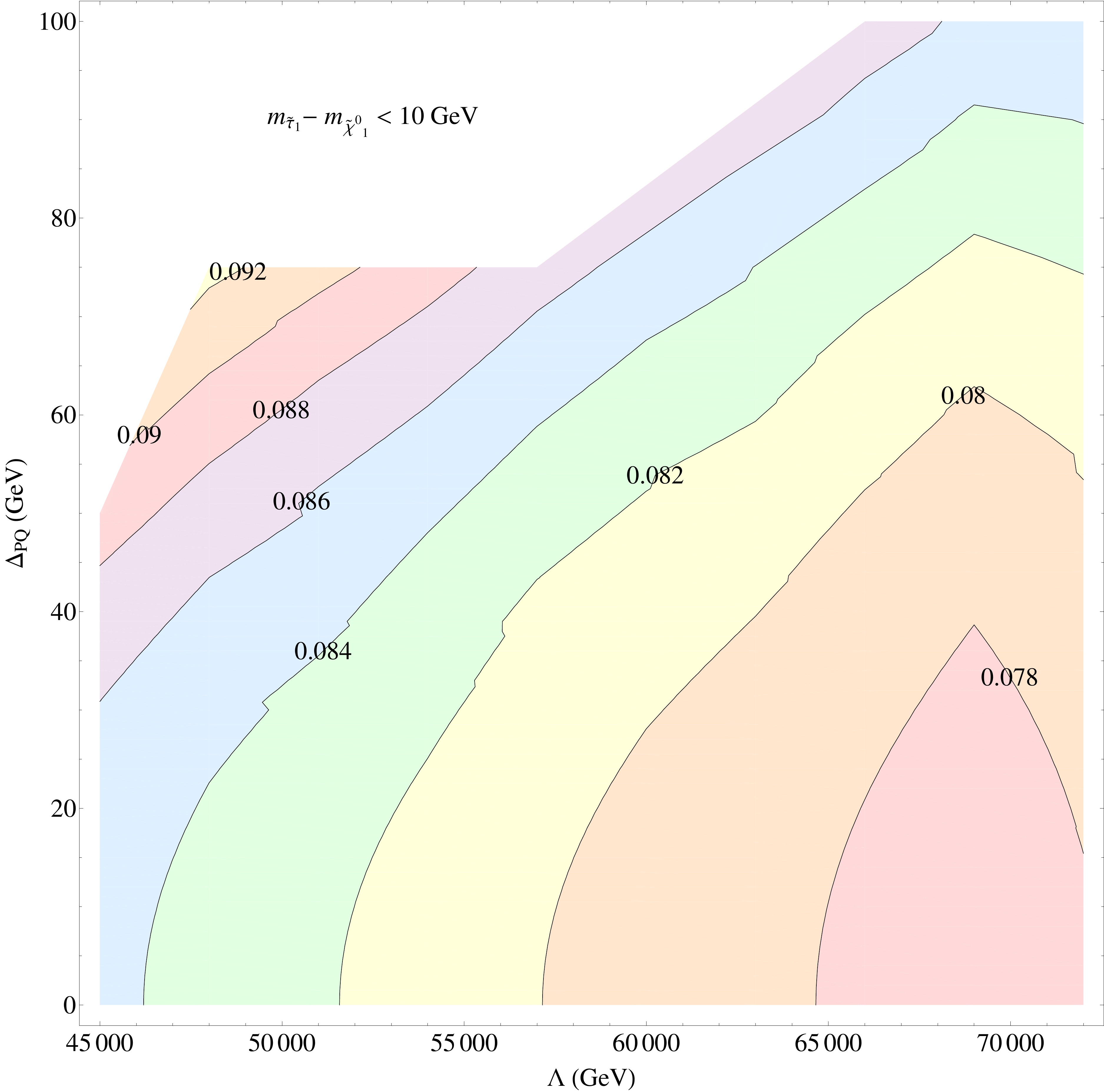}
\end{center}
\caption{Contour plot of the value of $\Delta P^{2}$ obtained by fixing a
particular value of $\Lambda$ and $\Delta_{PQ}$ of an F-theory GUT model with
$N_{5}=3$, and minimizing with respect to all large A-term mSUGRA\ models. We adopt
a rough criterion for theoretical distinguishability specified by the requirement that $\Delta P^2 > 0.01$. By
inspection, $\Delta P^2$ is greater than $0.05$. This is to be
contrasted with the level of distinguishability found for small A-term mSUGRA
models where $\Delta P^{2}$ is greater than $0.15$.}%
\label{softspecthrsugrala}%
\end{figure}

\begin{figure}[ptb]
\begin{center}
\includegraphics[
height=5.0929in,
width=6.2128in
]{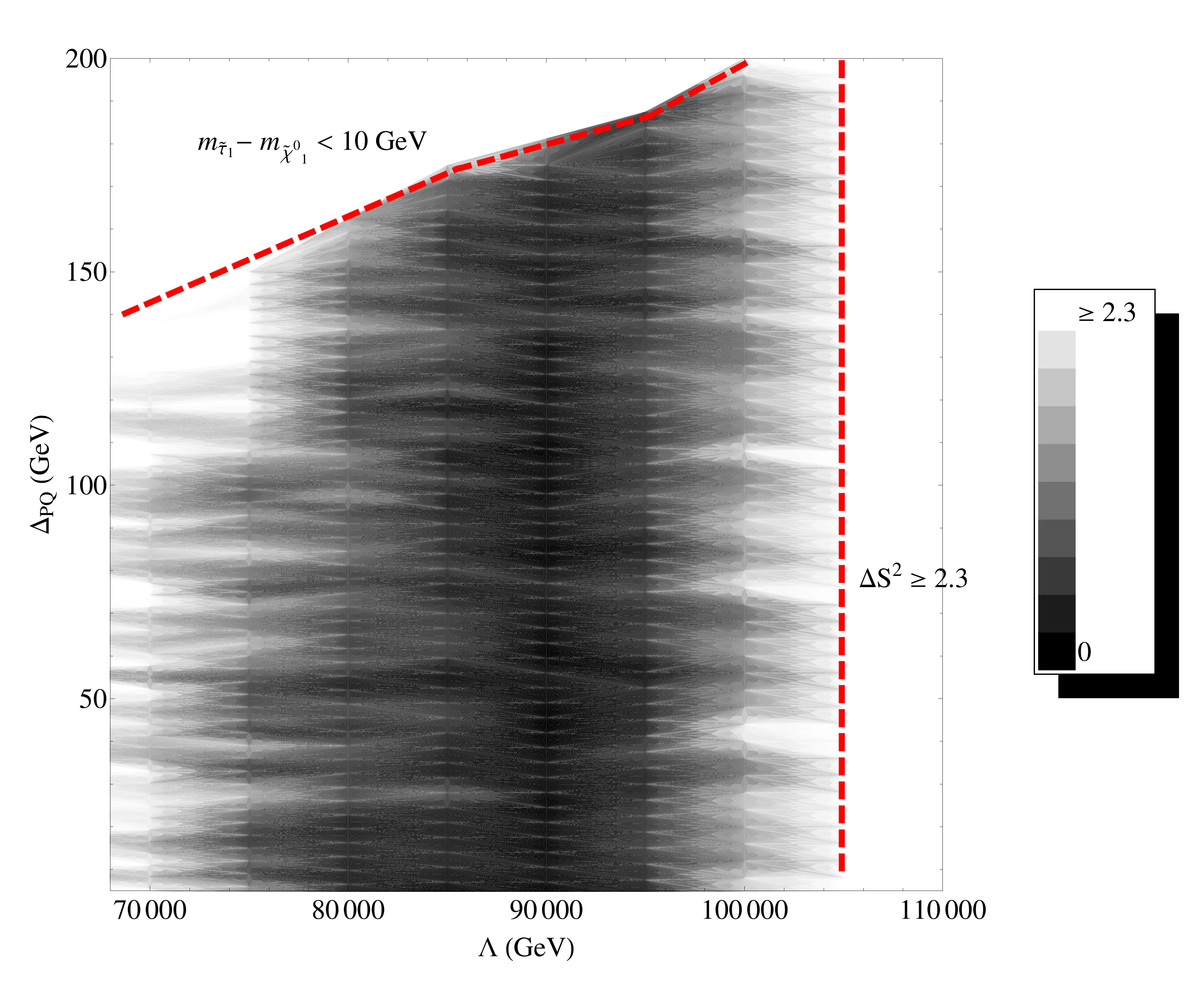}
\end{center}
\caption{Density plot of $\Delta S_{(A)}^{2}$ defined by the signature list
in table \ref{siglist1} of subsection \ref{SIMULATION} comparing the minimal value of a given
$N_{5}=2$ F-theory GUT model with a scan over large A-term mSUGRA models. The signals used are
obtained with $5$ fb$^{-1}$ of simulated LHC data. Here, we have used a rough notion of distinguishability based
on $99\%$ confidence and $10$ signals so that at $\Delta S_{(A)}^{2}>2.3$ we shall
say that two models are distinguishable. The plot shows the
minimal value of $\Delta S^{2}$ for a fixed F-theory GUT\ point. This plot
shows that for this class of signatures, the models are only distinguishable
at larger values of $\Lambda$.}%
\label{twosugrala}%
\end{figure}

\begin{figure}[ptb]
\begin{center}
\includegraphics[
height=5.0929in,
width=6.2128in
]{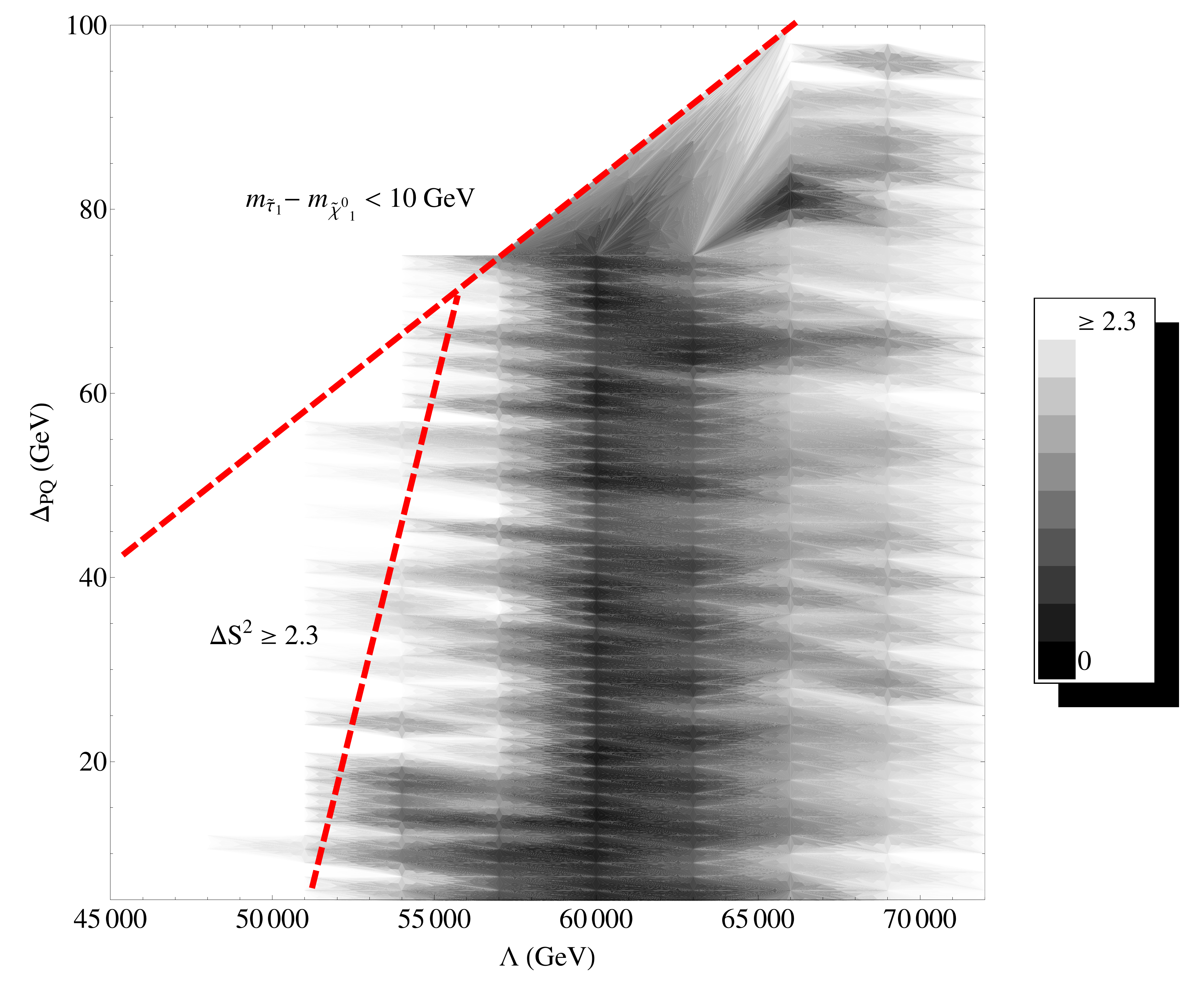}
\end{center}
\caption{Density plot of $\Delta S_{(A)}^{2}$ defined by the signature list
in table \ref{siglist1} of subsection \ref{SIMULATION} comparing the minimal value of a given
$N_{5}=3$ F-theory GUT model with a scan over large A-term mSUGRA models. The signals used are
obtained with $5$ fb$^{-1}$ of simulated LHC data. Here, we have used a rough notion of distinguishability based
on $99\%$ confidence and $10$ signals so that at $\Delta S_{(A)}^{2}>2.3$ we shall
say that two models are distinguishable. This plot
shows that for this class of signatures, the models are only distinguishable
at small values of $\Lambda$.}%
\label{thrsugrala}%
\end{figure}

\begin{figure}[ptb]
\begin{center}
\includegraphics[
height=5.5625in,
width=5.7017in
]{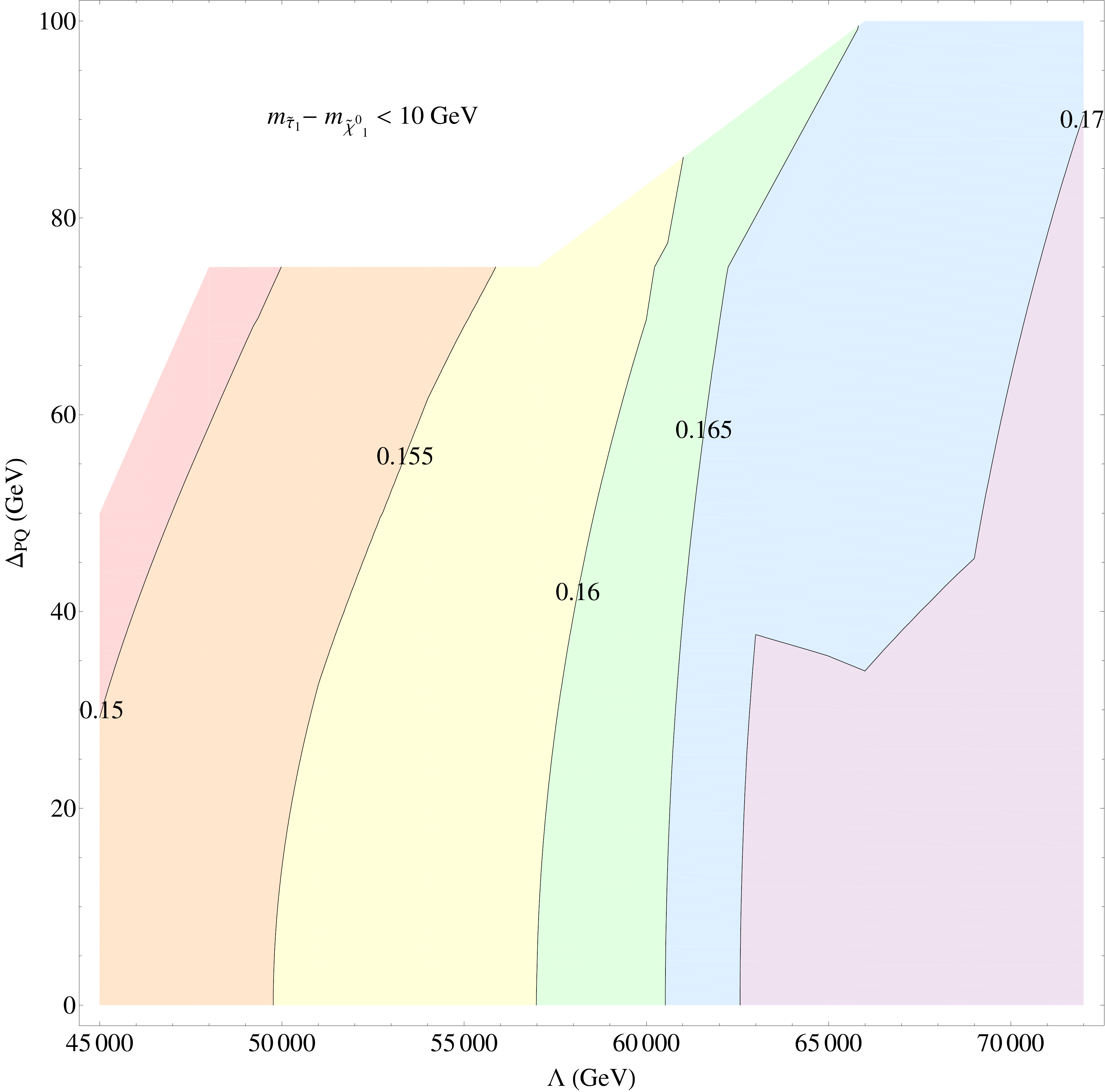}
\end{center}
\caption{Contour plot of the value of $\Delta P^{2}$ obtained by fixing a
particular value of $\Lambda$ and $\Delta_{PQ}$ of an F-theory GUT model with
$N_{5}=3$, and minimizing with respect to all low scale minimal GMSB\ models
with a stau\ NLSP. We adopt a rough criterion for theoretical
distinguishability specified by the requirement that $\Delta P^2 > 0.01$. By inspection, $\Delta P^2$ is greater than
$0.15$, indicating that such models are distinguishable at the theoretical
level from F-theory GUTs.}%
\label{softspecthrgmsbstau}%
\end{figure}

\begin{figure}[ptb]
\begin{center}
\includegraphics[
height=5.5625in,
width=5.7017in
]{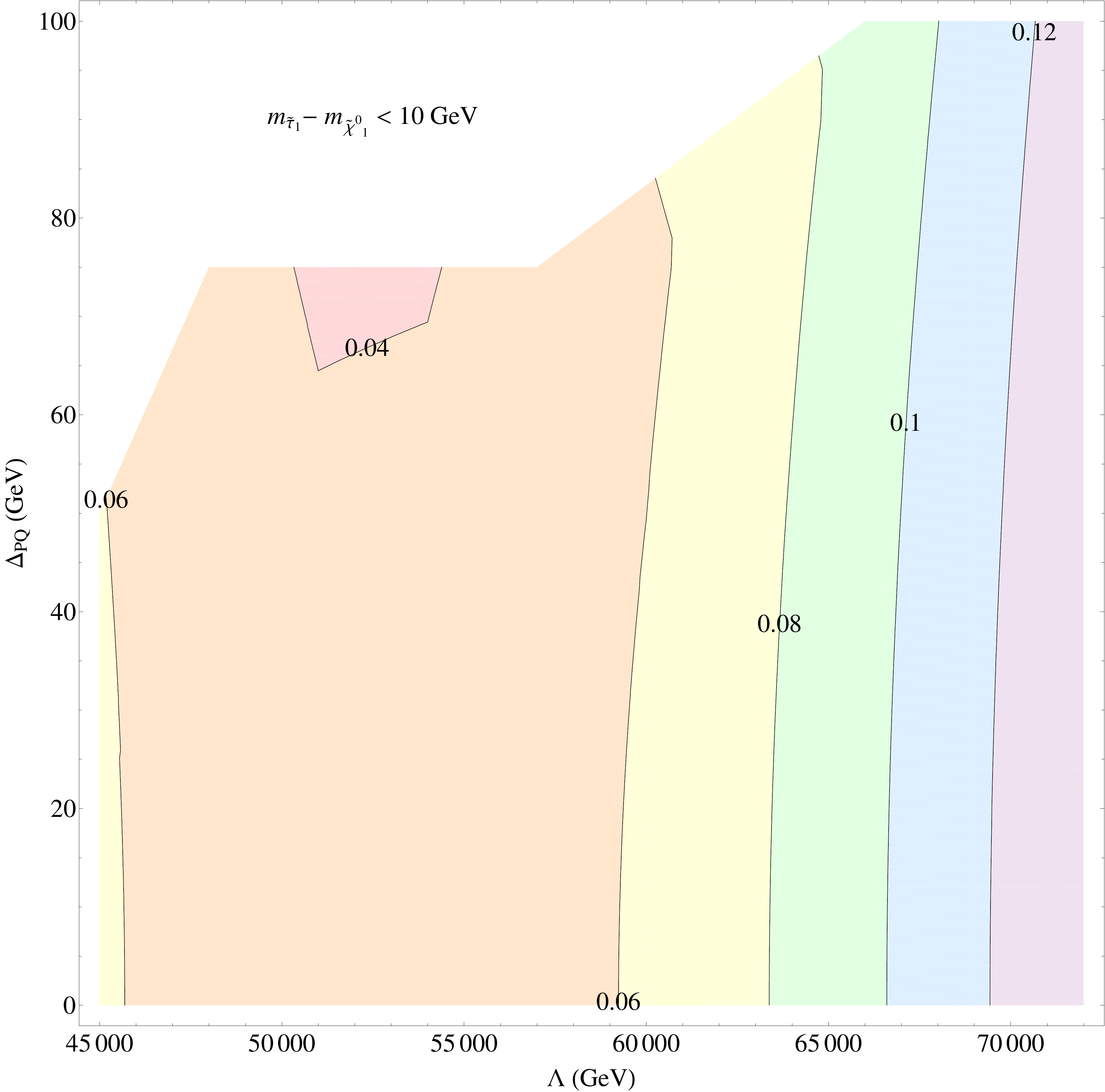}
\end{center}
\caption{Contour plot of $\Delta P^{2}$ between a fixed F-theory GUT\ model with
$N_{5}=1$, $\Lambda=1.28\times10^{5}$ GeV and $\Delta_{PQ}=222$ GeV and
three messenger F-theory GUT models. We adopt a rough criterion for
theoretical distinguishability specified by the requirement that $\Delta P^2 > 0.01$. This figure shows that
$\Delta P^{2}$ minimizes in the vicinity of a small patch. However, it should
be noted that there is no contour for $\Delta P^{2}=0.03$ because the
resulting scan typically saturates above this value. This is in accord
with the fact that the fixed F-theory GUT and the models scanned have
a different number of messenger fields.}%
\label{softspecexampthrfth}%
\end{figure}

\newpage
\bibliographystyle{ssg}
\bibliography{fgutscolliders}

\end{document}